\documentclass{ar2e}

\usepackage{graphicx}
\usepackage{amssymb}

\begin{document}

\date{\today}

\newcommand{\be}{\begin{equation}}
\newcommand{\ee}{\end{equation}}

\jname{Annual Review of Astronomy and Astrophysics}
\jyear{2013}
\jvol{51}
\ARinfo{1056-8700/97/0610-00}

\title{The Dawn of Chemistry}

\markboth{D. Galli, F. Palla}{The Dawn of Chemistry}

\author{Daniele Galli, Francesco Palla
\affiliation{INAF-Osservatorio Astrofisico di Arcetri\\
Largo E. Fermi 5, 50125 Firenze, Italy\\
\bigskip
email: galli,palla@arcetri.astro.it}
{\em To appear in Annual Reviews of Astronomy and Astrophysics Volume 51}}


\begin{abstract}

Within the precise cosmological framework provided by the $\Lambda$-Cold
Dark Matter model and standard Big Bang nucleosynthesis, the chemical
evolution of the pregalactic gas can now be followed with accuracy
limited only by the uncertainties on the reaction rates. Starting
during the recombination era, the formation of the first molecules
and molecular ions containing hydrogen, deuterium, helium, and
lithium was severely hindered by the low density of the expanding
universe, the intensity of the cosmic radiation field, and the
absence of solid catalyzers.  Molecular hydrogen and deuterated
hydrogen, the most abundant species formed in the gas phase prior
to structure formation, played a fundamental role in the cooling
of the gas clouds that gave birth to the first stellar generation,
contributing to determine the scale of fragmentation.  Primordial
molecules also interacted with the photons of the cosmic background
via resonant scattering, absorption and emission. In this review
we examine the current status of the chemistry of the early universe
and discuss the most relevant reactions for which uncertainties
still exist from theory or laboratory experiments. The prospects
for detecting spectral distortions or spatial anisotropies due to
the first atoms and molecules are also addressed.

\end{abstract}

\maketitle

\section{PRESENT CONTEXT AND MOTIVATIONS}
\label{context}

We live in a chemically and dynamically evolving universe. The interplay
between these two domains has shaped the universe in the way we observe
it now. Dynamics, driven by visible and invisible forces, has governed
the transformation of minute density perturbations into structures
of varying degree of complexity (from galaxies to individual stars
within them). Chemistry has been instrumental for modifying a very
simple mixture of nuclei and electrons into complex molecules and solid
particles, the essential ingredients for cooling and fragmentation
of the interstellar gas. This review is an account of the portion of the
chemical evolution that spanned a small fraction of the universe's lifetime,
between the recombination era at redshift $z\approx 1000$ when it was
only 380,000 years old and the transitional phase of cosmic re-ionization
at redshifts $z\approx 15$--6. This epoch is commonly called the ``dark
ages'' since it preceded -- or, more precisely, included the beginning of
-- the formation of the first structures, which were either individual stars
(or groups of stars) of the Population~III, or an entire generation of
stars within pre-galactic gas clouds. The radiation of these stellar
objects lit up the sky and contributed significantly to the re-ionization
of the intergalactic gas, a process that was basically completed by a
redshift of $z\approx 6$ at an age of $\sim 800$ million years.

The first synthesis of the elements, mainly nucleons, occurred in
a very short time interval approximately three minutes after the
initial expansion of the universe, but it took about four hundred
thousand years before conditions were suitable for matter and
radiation to decouple, with the former recombining into neutral
atoms. However, recombination lagged behind expansion and at a
redshift of $z\approx 800$ a minute fraction of electrons and protons
froze out at a level of one part in $\sim 10^4$. The rapidly varying
level of ionization, along with the ever decreasing energy of the
photons of the cosmic microwave background (CMB) radiation, effectively
promoted the formation of the first molecular species, ions and
neutrals, that survived the processes of photoionization/photodissociation
and dissociative recombination. This was, in fact, the ``dawn of
chemistry''.

Since Big Bang nucleosynthesis produced hydrogen and helium (along
with their isotopes), with trace abundances of Li Be B and no solid
particles, chemical reactions in the low density medium of the
expanding universe only occurred in the gas phase. Therefore, the
first molecules and molecular ions were quite simple and limited
to few dominant species, such as H$_2$, HD, HeH$^+$, and LiH (Lepp
\& Shull 1984).  The search for these species prior to structure
formation, either directly in the pregalactic medium or through
their effects on the spectrum of the CMB has proved as yet unsuccessful
in spite of several efforts (de Bernardis et al.~1993; Persson et
al.~2010).  Owing to its potential role in primordial chemistry and
in the cooling of the gas, LiH has also been searched for towards
high redshift QSOs. Recent observations by Friedel, Kemball \&
Fields~(2011) with the CARMA array of the $J=0-1$ transition of
$^7$LiH and $^6$LiH have resulted in a 3$\sigma$-level detection
of $^7$LiH only in the source B0218$+$357 ($z_{\rm abs}=0.685$),
confirming an earlier claim by Combes \& Wiklind~(1998).  Similarly,
a tentative detection of the $J=1-0$ rotational line of HeH$^+$ has
been made towards one of the highest redshift quasars at $z=6.4$
where other molecular lines had been previously detected (Zinchenko, 
Dubrovich \& Henkel~2011).

Another obstacle in probing the physical conditions of the gas
during the dark ages is that measurements towards high redshift
galaxies and the diffuse gas have largely failed to identify clouds
of primordial composition. On the contrary, there appears to be a
systematic contamination by heavy elements at levels of $Z\approx
10^{-3}$ of the solar abundance (e.g., Wolfe, Gawiser \& Prochaska~2005) which
appears to extend up to $z\sim 5$ and which implies enrichment from
the ejecta of earlier generations of stars (Rafelsky et al.~2012).
However, recent studies have identified gas clouds without any trace
of heavy elements or with metallicities consistent with enrichment
by primordial stars (Fumagalli, O'Meara \& Prochaska~2011; Cooke,
Pettini \& Steidel~2011).  Interestingly, the pristine clouds
LLS1134a and LLS0956B with $Z<10^{-4}$~$Z_\odot$ at $z\approx 3$
identified by Fumagalli, O'Meara \& Prochaska~(2011) are located
in regions of density significantly higher than the background, but
unpolluted by heavy elements, indicating that mixing of the metals
in the intergalactic medium is both inhomogeneous and inefficient.

Finally, although a genuine zero-metal star has so far escaped
detection, the search for such objects proceeds unabated due to the
unique information that they carry about the earliest phases of
galactic formation and evolution (e.g., Yong et al.~2012). The most
exciting discovery of the last few years is the low-mass star of
the galactic halo SDSS~J102915$+$172927 with a metallicity $4.5\times
10^{-5}$ times that of the Sun and without enrichment of C, N, O
(Caffau et al.~2011,2012).  Although a number of stars of extremely
low metallicity had been previously detected, they all showed
significant overabundance of the heavy elements, a signature of
pollution by a previous generation of massive stars (Frebel et al.
2008; Christlieb et al. 2002; Norris et al. 2007).  From the analysis
of their sample, Caffau et al.  (2011) conclude that SDSS~J102915
is not an isolated object and that a significant number of similar,
or of even lower metallicity, low-mass stars will be discovered in
the near future. These findings will definitely clarify the issue
on the existence of a transition in the mass scale of the stars
formed in the primordial universe. While a transition from massive
to low-mass fragments based on the cooling properties of atomic
lines of O{\sc i}, C{\sc ii}, Si{\sc ii} and Fe{\sc ii} and on the
existence of a critical value of the C/H and O/H abundance has been
suggested for some time (Bromm \& Loeb 2003, Frebel, Johnson \&
Bromm~2007), the discovery of SDSS~J102915 with metal abundance
below the critical value appears to favor the view of a transition
based on the cooling properties of dust formed in the ejecta of
core-collapse SNe of the first generation (e.g., Schneider et
al.~2012; Klessen, Glover \& Clark~2012).

The exciting developments in the search for primordial molecules,
clouds and stars outlined above make a review of the critical
chemical processes that took place in the post-recombination universe
timely. Until recently, large uncertainties existed not only on the
rates of some key chemical reactions, but also (and, to some extent,
more significantly) on the cosmological model and its fundamental
parameters, and on the yields from Big Bang nucleosynthesis (BBN).
The advent of precision cosmology and a concordance model, along
with significant improvements in standard BBN calculations offer
accurate initial and boundary conditions for the study of the
chemistry of the primordial gas, so that much of the remaining
uncertainty resides in the poor knowledge of some basic reactions.
Therefore, a coherent picture of the chemistry of the main constituents
(hydrogen, deuterium, helium, and lithium) can be provided and,
hopefully, tested.

In this review, we will cover the post-recombination epoch in the
unperturbed universe prior to the development of the first structures,
whether individual stars or galaxies.  Once these objects form,
feedback effects and chemical enrichment become extremely important
and need to be taken into account for a consistent treatment of the
interstellar/intergalactic medium.  Thus, the chemistry will be
very different from that described here and an appropriate discussion
would require a dedicated review. Similarly, the problem of the
fragmentation and collapse of primordial clouds and the characteristic
properties of the first stars will only be addressed in the light
of the chemical properties of the gas. This field of research has
made considerable progress in recent years thanks to the development
of highly sophisticated numerical simulations and excellent summaries
can be found in the literature (e.g., Bromm et al.~2009; Bromm \&
Yoshida~2011; Loeb~2010; Umemura \& Omukai~2013).

The organization of this review is as follows: we sketch in
Section~\ref{earlydevelopments} the early motivations to study the
formation of molecular hydrogen and other molecules in the cosmological
context.  A good recipe for the chemical evolution of the primordial
gas requires the specification of the main ingredients (a cosmological
model, the abundance of the light elements from nucleosynthesis,
the ionization history through recombination) and this will be
provided in Section~\ref{ingredients}.  Section~\ref{chemicalevolution}
will present the results of the calculations of the predicted
abundances of the main molecules and molecular ions, along with a
discussion of the most critical reaction rates. The effects of
chemistry on the cooling of the primordial gas will be highlighted
in Section~\ref{cooling}. Once formed, molecules as well as atoms
can interact with the CMB and possibly leave some imprints both as
spectral distortions and spatial anisotropies and
Section~\ref{interactions} will address this important aspect.
Finally, a summary of the major issues (Section~\ref{summary}) and
some future perspectives (Section~\ref{future}) on both theoretical
and observational studies of primordial chemistry will close the
review.

\section{EARLY DEVELOPMENTS}
\label{earlydevelopments}

The investigation of the role of molecular hydrogen as an important
coolant for the formation of the first structures in the early
universe began in the late 1960s. Saslaw \& Zipoy~(1967) were the
first to point out the important role of H$_2$ molecules for the
thermal and dynamical evolution of pre-galactic gas clouds in the
post-recombination era.  Unlike previous attempts that had ignored
the possible presence of H$_2$ molecules due to the lack of dust
grains and the slow rate of radiative association and three body
reactions, Saslaw \& Zipoy considered the possibility of  radiative
association and charge transfer reactions such as\footnote{They
acknowledge P. Solomon, private communication}
\be
{\rm H} + {\rm H}^+ \rightarrow {\rm H}_2^+ + h\nu
\label{hhp}
\ee
and
\be
{\rm H}_2^+ + {\rm H} \rightarrow {\rm H}_2 + {\rm H}^+ .
\label{h2ph}
\ee
The rate constant for reaction (\ref{hhp}) was taken from a calculation
by Bates~(1951), while the cross section for reaction (\ref{h2ph})
was estimated to have a relatively high value of $\sim 10^{-15}$~cm$^2$,
typical of charge exchange reactions above threshold. Thus, an
H$_2^+$ ion can be converted to H$_2$ as soon as it is formed and
the H$_2^+$ concentration remains very low. In addition, owing to
the low electron abundance, dissociative recombination of H$_2^+$
turned out to be negligible compared with reaction (\ref{h2ph}) and
H$_2$ molecules could then be formed at a significant rate. Using
the radiative cooling function calculated by Takayanagi \&
Nishimura~(1960), Saslaw \& Zipoy~(1967) showed that, under reasonable
assumptions for the initial conditions of the collapsing clouds,
their evolution would depart from pure adiabatic collapse at densities
above $\sim 10^4$~cm$^{-3}$ due to enhanced H$_2$ formation and
cooling.  Cloud contraction could then proceed at high densities
before the onset of optical depth effects and H$_2$ collisional
dissociation, thus allowing the formation of stars of unspecified
characteristics. Saslaw \& Zipoy suggested that the resulting object
might resemble a globular cluster.

Peebles \& Dicke~(1968) also considered the origin of globular
clusters prior to the formation of galaxies. They noticed the
correspondence between the characteristic mass and size of the first
generation of bound systems, given by the Jeans mass and length
($\sim 10^5$~$M_\odot$ and $\sim 5$~pc, respectively), and that of
globular clusters. However, Peebles \& Dicke relied on another
channel for efficient H$_2$ formation and cooling provided by the
presence of negative hydrogen ions. They referred to McDowell~(1961)
who calculated the H$_2$ abundance in the interstellar medium,
following an earlier private communication by
Dalgarno~(1958)\footnote{Pagel~(1959) was the first to point out
the H$^-$ channel for H$_2$ formation in the solar photosphere,
acknowledging A.~Dalgarno for the suggestion. Later, McDowell~(1961)
calculated the H$_2$ fractional abundance in H{\sc i} regions,
quoting Pagel's calculation for the Sun and Dalgarno's estimate for
the rate coefficient. Interestingly, McDowell mentions the fact
that McCrea~(1960) had pointed out that ``consideration of the
problem of star formation in HI regions strongly suggests that the
hydrogen is present in molecular form''. McDowell did not consider
3-body reactions because of the low densities, while McCrea \&
McNally~(1960) suggested that the mechanism for H$_2$ formation
could be that of surface reactions on dust grains.}.  In this scheme,
H$_2$ formation proceeds through the reactions
\be 
{\rm H} + e \rightarrow {\rm H}^- + h\nu 
\label{rad_attach}
\ee
and
\be 
{\rm H}^- + {\rm H} \rightarrow {\rm H}_2 +e .
\label{hm_h2} 
\ee 
The rate coefficient for (\ref{rad_attach}) was calculated from the
cross section given by Chandrasekhar~(1958), while that for reaction
(\ref{hm_h2}) was based on laboratory measurements by Schmeltekopf, Fehsenfeld 
\& Ferguson~(1967).  Once again, since the rate for (\ref{hm_h2}) is
much more rapid than for (\ref{rad_attach}), the formation of a
negative hydrogen is promptly followed by the conversion to a
hydrogen molecule. In addition, this route is faster than that
through H$_2^+$ formation (by a factor of $\sim$100 at 1000~K), so
that the latter was neglected in the calculations.  The efficient
formation of H$_2$, up to a fraction of $\sim 10^{-3}$ by number,
and the enhanced cooling found by Peebles \& Dicke~(1968) led to
fragmentation of a small fraction of the cloud mass, and to collapse
into stars mostly of high mass.

The initial suggestions by Saslaw \& Zipoy~(1967) and Peebles \&
Dicke~(1968) paved the route to more refined calculations of the
role of H$_2$ formation and cooling in collapsing and fragmenting
primordial clouds, mainly with the goal of constraining the typical
masses of the forming objects in pregalactic gas clouds (e.g.,
Hirasawa, Aizu \& Taketani~1969; Takeda, Sato \& Matsuda~1969; Yoneyama~1970;
Hutchins~1976; Silk~1977; Carlberg~1981). The general result was
that even a relatively minor fraction of H$_2$ contributes significant
cooling via rotational and vibrational transitions. However, owing
to the lack of dipole moment, H$_2$ molecules have large radiative
lifetimes, thus are poor radiators.  Nevertheless, these initial
calculations showed that during collapse the gas temperature could
be kept from climbing above $\sim 1000$~K over a large density
range. As a result, the Jeans mass dropped by many decades and the
minimum mass varied between $\sim 20$~M$_\odot$ and $\sim 250$~M$_\odot$
with a maximum fractional abundance of H$_2$ of about $10^{-3}$
before the onset of collisional dissociation.

The inclusion in the chemical network of three body reactions 
\be
{\rm H} + {\rm H} + {\rm H} \rightarrow  {\rm H}_2 + {\rm H}
\ee
and
\be
{\rm H} + {\rm H} + {\rm H}_2 \rightarrow  {\rm H}_2 + {\rm H}_2 .
\ee
modified this picture considerably (Palla, Salpeter \& Stahler~1983).
It was found that, over a wide range of initial conditions, virtually
all the gas can be converted to molecular form by densities of the
order of $10^{12}$~cm$^{-3}$. As a result of significant cooling from
the molecules, the temperature rise during collapse is slowed and
the Jeans mass eventually drops below 0.1~$M_\odot$, thus allowing
the formation of low-mass stars even in clouds of primordial
composition.

Much of the subsequent effort in the field of primordial chemistry
was dedicated to elucidating the atomic and molecular processes,
other than those related to H$_2$, that are important in the early
universe after the recombination epoch and prior to first structure
formation and reionization.  Major studies on this subject include:
Lepp \& Shull (1984), Black~(1990), Puy et al.~(1993), Stancil, Lepp
\& Dalgarno~(1996, 1998), Galli \& Palla~(1998, 2000, 2002), Puy \&
Signore~(2002, 2007), Lepp, Stancil \& Dalgarno~(2002),  Glover \& Abel~(2008),
Schleicher et al.~(2008), Glover \& Savin (2009), Vonlanthen et
al.~(2009), and Gay et al.~(2011).

\section{THE INGREDIENTS OF PRIMORDIAL CHEMISTRY}
\label{ingredients}

\subsection{Cosmological Model}
\label{cosmo}

Over the last few decades, the so-called $\Lambda$--Cold Dark Matter
($\Lambda$CDM) model has emerged as the standard description of
cosmology.  Accordingly, the universe is spatially flat, homogeneous,
and isotropic on large scales with a scale invariant spectrum of
primordial seed perturbations composed of ordinary matter, non-baryonic
dark matter and dark energy. The accurate measurements of the peaks
in the power spectrum of the CMB, along with baryon acoustic
oscillations, the precise estimate of the Hubble constant, of light
element abundances in distant quasars, and the amplitude of mass
fluctuations inferred from clusters and gravitational lensing have
enabled a deeper and more precise understanding of cosmology (e.g.,
Spergel et al. 2003; Komatsu et al.~2011). The essential parameters
(primary and inferred) of such a precision model are listed in
Table~\ref{table_cosmo}. All the values have been taken from the
{\it WMAP} seven-year mean (cf.  Table 1 of Komatsu et al. 2011)
with the exception of the current temperature of the CMB from Fixsen
(2009), and the element abundances from Iocco et al.~(2009).  In
the Table, $z_{\rm eq}$ represents the redshift of equal energy
density between matter and radiation, while $z_{\rm ion}$ is the
redshift of reionization on the assumption that the universe was
re-ionized instantaneously. The fractional abundances of H, He, D
and Li relative to the total number of baryons are given by $f({\rm
H})$, $f({\rm He})$, etc.. These cosmological parameters will be
adopted in the rest of the discussion and will provide the initial
values for the chemical evolutionary models that will be described
in the next Section.  It is expected that the {\it Planck} satellite
by ESA with higher resolution, sensitivity, and frequency coverage
than predecessors will provide an update on the current cosmological
parameter estimates as early as 2013.

\begin{table}[t!]
\begin{center}
\label{table_cosmo}
\caption{Cosmological parameters}
\begin{tabular}{ll}
\hline
Parameter & Value \\
\hline 
$H_0$               & $100\,h$~km~s$^{-1}$~Mpc$^{-1}$ \\
$h$                 & $0.704$ \\
$z_{\rm eq}$        & $3141$ \\
$z_{\rm ion}$   & $10.6$ \\
$T_0$               & $2.725$~K \\
$\Omega_{\rm dm}$   & $0.228$ \\
$\Omega_{\rm b}$    & $0.0455$ \\
$\Omega_{\rm m}$    & $\Omega_{\rm dm}+\Omega_{\rm b}$ \\
$\Omega_{\rm r}$    & $\Omega_{m}/(1+z_{\rm eq})$ \\
$\Omega_\Lambda$    & $0.727$ \\
$\Omega_{\rm K}$    & $1-\Omega_{\rm r}-\Omega_{\rm m}-\Omega_\Lambda$ \\
\hline
$f_{\rm H}$         & 0.924 \\
$f_{\rm He}$        & 0.076 \\
$f_{\rm D}$         & $2.38\times 10^{-5}$  \\ 
$f_{\rm Li}$        & $4.04\times 10^{-10}$ \\ 
\hline
\end{tabular}
\end{center}
\end{table}

\subsection{Standard Big Bang Nucleosynthesis}
\label{SBBN}

According to $\Lambda$CDM cosmology, the formation of the first
elements occurs entirely during the radiation-dominated epoch at a
time $t \sim (T / {\rm MeV})^{-2}$~s, where $T$ is the plasma
temperature. This energy scale corresponds to a time scale of a few
minutes, temperature of $\sim10^9$~K, and redshift of $z\approx
10^8$--$10^9$. The hot and low-density plasma was primarily composed
of free nucleons and electrons that initiated the production of
light elements (e.g., Fields~2011). Owing to the low density, most
of the reactions occur by neutron and proton capture, whereas three
body reactions are effectively suppressed, as well as reactions
with nuclei heavier than $Z_i Z _j \gtrsim 6$.  Thus, at temperatures
$T\sim 100$~keV, deuterium (with binding energy 2.2~MeV) can be
synthesized via $p(n,\gamma)D$ reactions, since blackbody photons
can no longer equilibrate the reverse reaction. The abundance of
deuterium rises sharply, as almost all available neutrons are locked
into D-nuclei. The actual abundance is set by the value of the
baryon-to-photon ratio, parametrized by the quantity
\be
\eta \equiv {n_{\rm b}\over n_\gamma}=2.74\times 10^{-8}\,\Omega_{\rm b} h^2
\ee
where $\Omega_{\rm b}=\rho_{\rm b}/\rho_{\rm cr}$, 
$\rho_{\rm b}$ is the baryon density, $\rho_{\rm cr}$ is the critical density defined by
\be 
\rho_{\rm cr}=\frac{3 H_0^2}{8 \pi G},
\label{rhocr}
\ee
and $h$ is the present-day value of the Hubble constant measured
in units of 100~km~s$^{-1}$~Mpc$^{-1}$.  In the following, we will adopt
$\eta_{10}=10^{-10}\eta$.

The WMAP determination of $\eta_{10}=6.16 \pm 0.15$ implies that
the standard BBN has no free parameter left and can yield precise
predictions of the abundances of D, $^3$He, $^4$He, and $^7$Li
(Serpico, Esposito \& Iocco~2004; Iocco et al.~2009; Coc et al.~2012). These
abundances can then be used as the initial conditions for the
modeling of the chemical evolution of the expanding universe.
Table~\ref{table_sbbn} reports the values of BBN calculations
obtained with two numerical codes: PArthENoPE (Iocco et al.~2009;
Pisanti~2012, priv. comm.), and the nuclear reaction code THALYS
(Coc et al.~2012).  Although the codes can solve the synthesis of
heavy elements (including CNO, $^9$Be, and B), in the following we
only consider the production of light elements since their abundance
is sufficient for the chemistry calculations\footnote{However, the
accurate assessment of the predicted abundances of CNO are of
particular relevance for the evolution of Population III stars whose
properties are significantly affected for values higher than $\sim
10^{-10}$ in mass fraction (Cassisi \& Castellani~1993), or higher
than $10^{-12}$ for low-mass primordial stars (Ekstr\"om et al.~2010).
This is still at least three orders of magnitude above the yields
computed by Coc et al.~(2012) and Iocco et al.~(2009).}.  The
predictions of the detailed models of BBN can be compared with the
observations of light elements at high $z$.  The literature on this
subject is quite extensive and here we briefly summarize the current
status (for a review, see, e.g., Steigman~2007).

\begin{table}[t!]
\begin{center}
\caption{Predicted Abundances from standard BBN}
\begin{tabular}{lcc}
\hline
\hline
Light Element & PArthENoPE & THALYS \\
\hline
$Y_{\rm pi}$  & 0.2467 & 0.2476 \\
D/H        ($\times$10$^{-5}$)  & 2.56 & 2.59 \\
$^3$He/H   ($\times$10$^{-5}$)  & 1.02 & 1.04 \\
$^7$Li/H   ($\times$10$^{-10}$) & 4.60 & 5.24 \\
$^6$Li/H   ($\times$10$^{-14}$) & 1.11 & 1.23 \\
\hline
\end{tabular}
\label{table_sbbn}
\end{center}
\end{table} 

\subsubsection{Deuterium.}

As for deuterium, there is consistency between the inferred primordial
abundance deduced from the angular power spectrum of the CMB, and
the predictions of BBN. Assuming that the measured D-abundance
in low-metallicity environments at redshifts $z\gtrsim 2$--3
corresponds to the true D/H ratio produced by the BBN (see, e.g.,
Kirkman et al. 2003), observations of D{\sc i} and H{\sc i} absorption
lines in damped Ly-$\alpha$ (DLA) systems towards QSOs have provided
the best determinations of deuterium. The mean value derived from
8 QSO absorbers is (Fumagalli, O'Meara \& Prochaska~2011)
\be 
\log \left({\rm D/H}\right)_{\rm p} = (2.78\pm 0.21) \times 10^{-5}.  
\ee
This implies a value of $\Omega_{\rm b} h^2$~(BBN)$=0.0213 \pm
0.0012$, in agreement within the errors with the value of $\Omega_{\rm
b} h^2=0.0222\pm 0.0004$ deduced from the angular power spectrum
of the CMB (Keisler, Reichardt \& Aird~2011).  However, the authors note that
the scatter between different measurements is larger than the
published errors. The preferred explanation of the dispersion is
an underestimate of the systematic errors affecting D/H measurements
(e.g., Pettini et al.~2008)\footnote{Alternative interpretations
involving an early destruction of deuterium have been suggested
(Olive et al.~2012).}.  Recently, Pettini \& Cooke~(2012) have
obtained an accurate measurement towards the metal-poor ($Z\simeq
10^{-2}\,Z_\odot$), DLA at  redshift $z=3.05$ in the $z_{\rm em}\simeq
3.03$~QSO SDSS~J1419$+$0829 with ${\rm (D/H)}_{\rm p}=(2.535\pm
0.05)\times 10^{-5}$ which implies $\Omega_{\rm b}\,h^2$~(BBN)$=0.0213
\pm 0.0012$. Similarly, Noterdaeme et al.~(2012) have derived a
primordial abundance of (D/H)$_{\rm p} =2.8^{+0.8}_{-0.6}\times
10^{-5}$, consistent with the predictions of BBN using the constraint
on $\Omega_{\rm b} h^2$ from WMAP7.

\subsubsection{Helium.}

The situation for helium is less favorable\footnote{Since $^3$He
is both produced and destroyed in stars and since it can be observed
only in the ISM of the Milky Way (e.g., Bania, Rood \& Balser~2010), its
abundance is affected by large uncertainties and cannot be used as
a test for BBN predictions.}. Since $^4$He is produced in stars,
the determination of its primordial value requires extrapolation
from the measured abundances in metal-poor galaxies (extragalactic
H{\sc ii} regions), a procedure which is affected by systematic
uncertainties (e.g., Olive \& Skillman~2004, Izotov \& Thuan~2010).
Different analyses have yielded values that differ systematically
from each other, with a weighted mean of
\be
Y_{\rm p}=0.2566 \pm 0.0023,
\ee
(Aver, Olive \& Skillman~2010), that shows a slight discrepancy
with the predicted BBN value reported in Table~\ref{table_sbbn}.
We note that helium has the most accurately predicted abundances
and it is quite insensitive to the precise value of the baryon
density $\Omega_{\rm b}\,h^2$. In a recent re-analysis of the dataset
of Izotov, Thuan \& Stasinska~(2007), Aver, Olive \& Skillman~(2012)
using Markov Chain Monte Carlo techniques have obtained an extrapolated
primordial value of $Y_{\rm p}=0.2534 \pm 0.0083$, in agreement
with the WMAP7 result of $0.2487 \pm 0.0002$.

\subsubsection{Lithium.}

Finally, of all the light elements, lithium (in both isotopic
versions, $^7$Li and $^6$Li) is most problematic, owing to the well
established discrepancy between the predicted abundance of $^7$Li
and the value measured in metal-poor halo dwarf stars.  In fact,
the observations extended over a time span of about 30 years have
confirmed the existence of a discrepancy between the measured value
of $^7$Li/H$=(1.58 \pm 0.31) \times 10^{-10}$ -- the so-called Spite
plateau -- and that inferred from WMAP measurements, larger by a
factor of $\sim 3$ (Iocco et al.~2009). The realization of a dramatic
scatter of the $^7$Li abundance at very low metallicities, below
$[\mbox{Fe/H}] \approx -2.8$, has complicated the interpretation
of the evolution of lithium in low-mass stars even further (see
Sbordone et al.~2010 and Bonifacio, Sbordone \& Caffau~2012 for a
recent analysis). A surprising, yet very encouraging, result has
been reported by Howk et al.~(2012) who have measured the $^7$Li
abundance in the interstellar medium of the Small Magellanic Cloud.
Unlike the case for old metal-poor stars, the derived abundance of
$^7$Li/H$=(4.8\pm 1.7)\times 10^{-10}$ is almost equal to the BBN
predictions, thus providing a precious hint that the ``lithium
problem" posed by stellar measurements may not be representative
of the primordial value. Since inferring the Li abundance in the
ISM in not straightforward and requires several assumptions, these
initial results must be substantiated by further analysis.

BBN predicts that $^7$Li must be definitely more abundant than its
isotopic version $^6$Li by several orders of magnitude (cf.
Table~\ref{table_sbbn}).  High resolution observations of halo stars
have been able to detect $^6$Li with ratios as high as $\sim 0.5$
(Asplund et al.~2006; Garc\'ia P\'erez et al.~2009). However, the
observations remain controversial, as well as the evidence for a
tendency of $^6$Li to follow the same Spite plateau. Note that the
observations towards the Small Magellanic Cloud give a limit to the
isotopic ratio of $^6{\rm Li}/^7{\rm Li}<0.28$ (Howk et al.~2012).

The standard BBN model rests on the assumption of homogeneity and
isotropy of all constituent species, a hypothesis supported by the
observed low amplitude of the CMB temperature fluctuations. However,
alternative models of {\em inhomogeneous} nucleosynthesis have been
developed, mainly in order to provide a solution to the $^7$Li
problem. In this case, the the baryon-to-photon and the neutron-to-proton
ratios are allowed to vary on small scales allowing for a large
variation of the element abundances (see, e.g., Iocco et al. 2009
for a general discussion). Here, it suffices to underline the
potential impact that these models have on the yields of CNO and
heavier elements as a distinctive diagnostic between standard and
inhomogeneous models.  Vonlanthen et al.~(2009) have investigated
the effects of the higher CNO abundances resulting from non-standard
BBN and their implications on molecule formation.

\subsection{Cosmological Recombination}
\label{recombination}

The importance of cosmological recombination for studies of primordial
chemistry and structure formation was already emphasized in the
seminal works of Peebles~(1968) and Peebles \& Dicke~(1968). According
to Peebles~(1968), ``one also would like to know the residual
ionization in the matter, because the electrons and ions mediate
the formation of molecular hydrogen and so help determine the rate
of energy radiation by clouds that formed at the Jeans limit''.
{\bf Figure~1} shows the cosmological recombination of H, D, He and Li.
Of all the nuclei of interest, Li$^{3+}$ is the first to recombine
at redshift $z\approx 14,000$,
\be
{\rm  Li}^{3+} + e \rightarrow {\rm Li}^{2+} + h\nu, 
\ee
then Li$^{2+}$ recombines at $z\approx 8600$,
\be
{\rm Li}^{2+} + e \rightarrow {\rm Li}^+ + h\nu, 
\ee
followed by the recombination of He$^{2+}$ at $z\approx 6000$,
\be
{\rm He}^{2+} + e \rightarrow {\rm He}^+ + h\nu, 
\ee
and of He$^+$at $z\approx 2500$,
\be
{\rm He}^+ + e \rightarrow {\rm He} + h\nu, 
\ee
By $z\approx 2000$ all these reactions have run to completion, owing
to the high density of the matter and the large availability of
electrons as long as H is still ionized.  Finally, at $z\approx
1300$ hydrogen recombines, along with deuterium:
\be
{\rm H}^+ + e \rightarrow {\rm H} + h\nu.
\ee
Notice that the cosmological recombination of Li,
\be
{\rm Li}^+ + e \rightarrow {\rm Li} + h\nu, 
\ee
essentially never occurs, owing to the photoionization of Li due
to ultraviolet photons produced by the recombination of H and He
(Switzer \& Hirata~2005) and the five orders of magnitude drop of
electron abundance following H recombination.  Notice also that the
residual fraction of D$^+$, unlike those of H$^+$ and He$^+$, does
not reach a freeze-out because D$^+$ is efficiently removed at low
redshift by the near-resonant charge-transfer reaction
\be
{\rm D}^+ + {\rm H} \rightarrow {\rm D} + {\rm H}^+,
\ee
which is exothermic by 3.7~meV (43~K).

\begin{figure}[t!]
\begin{center}
\resizebox{8cm}{!}{\includegraphics{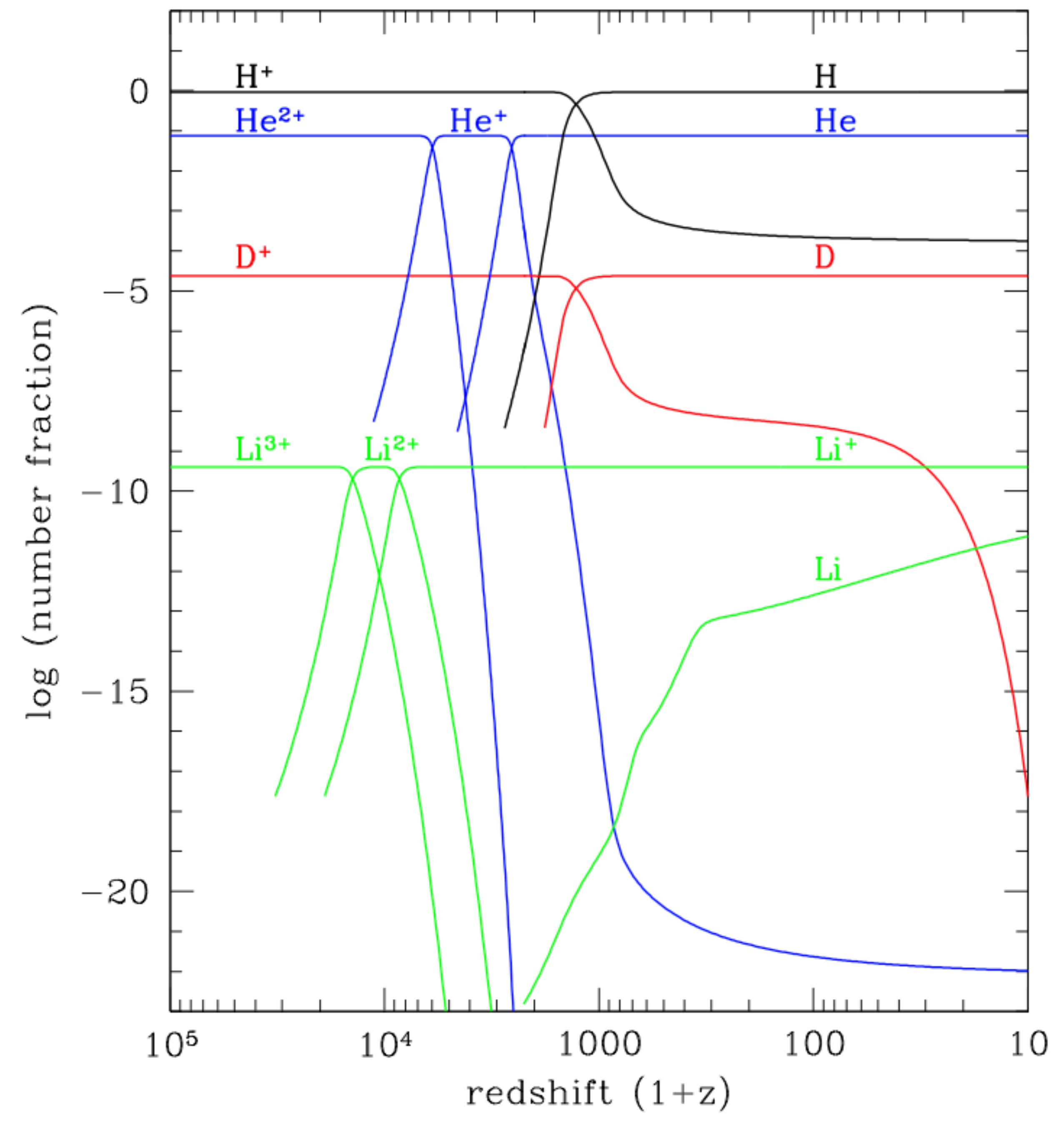}}
\caption{Cosmological recombination of H, D, He, and Li. Note the
small residual fraction of H$^+$ (and He$^+$), while D$^+$ is
completely removed at low redshifts by charge exchange with H.
Unlike all other elements, lithium remains largely ionized even at
the lowest redshifts.}
\end{center}
\label{fig_rec}
\end{figure}

\subsubsection{Recombination of Hydrogen.}
\label{hydrogenrecombination}

The basic physics of cosmological H recombination was fully understood
in the late 1960s (Peebles~1968; Zel'dovich et al.~1968; see
Peebles~1993). Direct recombination to the $n=1$ ground state of H
or He produces photons which can easily ionize neighboring neutral
atoms, because the primordial gas has a large optical depth ($\sim
10^7$) to ionizing photons.  Therefore, cosmological H and He
recombination proceeds by capture of electrons into excited $n\ge
2$ states followed by decay to the ground state.
However, electrons in excited states cannot cascade instantaneously
down to the ground state, because of the large reservoir of photons
in the CMB with energies sufficient to produce reionizations out
of these states ($E>3.4$~eV).  In fact, the population of these
excited states is, with good approximation, in thermal equilibrium
with the CMB. In addition, transitions to the ground state produce
Lyman photons that can excite other neutral atoms allowing them to
be easily re-ionized.  Unlike the present-day ISM, where Lyman
photons are eventually absorbed by dust, the only way they can
escape reabsorption in the conditions of the early universe is
through two processes of low efficiency: by redshifting out of the
line because of the cosmological expansion, or by two-photon forbidden
emission from the $n=2$ level.  The ``bottleneck'' caused by slow
decay from the $n=2$ to the $n=1$ level, combined with the density
drop due to the expansion, prevents H recombination from being completed
and results in a residual fraction of electrons and protons.  On
the basis of a simple two-level approximation ($n=1$, $n\geq 2$)
plus continuum, Peebles~(1968) and Zel'dovich et al.~(1968) found
a residual ionization $2\times 10^{-5}$--$2\times 10^{-4}$ depending
on the present-day value of the mass density, in agreement with
previous estimates of Novikov \& Zel'dovich~(1967).  Many refinements
and additional effects were later added to the original formulation
(e.g. Matsuda, Sato \& Takeda~1971; Jones \& Wise~1985; Krolik~1990; Sasaki
\& Takahara~1993) without however changing the basic concepts
summarized above. Still in 2000, the major uncertainty on the
residual electron fraction was due to the poor knowledge of the
cosmological parameters $\Omega_{\rm b}$ and $h$.

As a result of the exquisite accuracy of experiments such as WMAP,
several detailed multilevel calculations, supplemented by computer codes, are
now available to determine with precision better than $0.1$\% the
evolution of the electron fraction down to $z\approx 500$:
RECFAST\footnote{http://www.astro.ubc.ca/people/scott/recfast.html} (Seager, Sasselov
\& Scott~1999,2000; updated by Wong, Moss \& Scott~2008 and Scott \& Moss~2009);
RICO\footnote{http://cosmos.astro.uiuc.edu/rico} (Fendt et al.~2009);
COSMOREC\footnote{http://www.cita.utoronto.ca/$\sim$jchluba/
{\em Science}$_-$Jens/Recombination/CosmoRec.html} (Switzer \& Hirata~2008ab;
Ali-Haimoud \& Hirata~2010; Chluba, Vasil \& Dursi~2010; Grin \&
Hirata~2010; Rubi{\~ n}o-Mart{\'{\i}}n et al.~2010; Chluba \& Thomas~2011);
HyRec\footnote{http://tapir.caltech.edu/$\sim$yacine/hyrec/hyrec.html}
(Ali-Ha{\"i}moud \& Hirata~2011);
ATLANT\footnote{http://www.ioffe.ru/astro/QC/CMBR/atlant/atlant.html}
(Kholupenko et al.~2011).

\subsubsection{Recombination of Helium.}

Cosmological He recombination was first computed by Matsuda, Sato 
\& Takeda~(1969). An updated and detailed review of the processes involved
can be found in Switzer \& Hirata~(2008a,b,c) and Chluba, Fung \&
Switzer~(2012).  In currently available computer codes like RECFAST,
H and He recombinations are computed simultaneously, with He$^+$
and He$^{2+}$ treated as multilevel atoms.  The recombination of
He$^{2+}$ presents no $n=2$ ``bottleneck'' because of the high
two-photon decay rate.  Consequently, the abundance of He$^{2+}$
is well approximated by the Saha formula. This is not true for
He$^+$ recombination, which follows essentially a hydrogenic-like
case B recombination and is found to be slow essentially for the
same reasons that H recombination is slow (see, e.g., Seager, Sasselov
\& Scott~2000).

\subsubsection{Recombination of Lithium.}
\label{lithiumrecombination}

The calculation of Li recombination has not been straightforward.
It was first computed by Palla, Galli \& Silk~(1995), and later refined by
Stancil, Lepp \& Dalgarno~(1996), Bougleux \& Galli~(1997) and Galli \&
Palla~(1998).  Defining the recombination redshift $z_{\rm rec,
Li}$ as the redshift when ${\rm Li}/({\rm Li}+{\rm Li}^+)=0.1$,
Stancil, Lepp \& Dalgarno~(1996) and Galli \& Palla~(1998) found $z_{\rm rec,
Li}\approx 400$--450, and a residual Li ionization fraction ${\rm
Li}^+/{\rm Li}\approx 0.5$--0.7 as $z\rightarrow 0$.  However, it
was later realized by Switzer \& Hirata~(2005) that the recombination
history of Li is strongly affected by UV photons generated by the
recombination of H and He at $z < 1000$ via Ly-$\alpha$ and two-photon
decay (see Section~\ref{hydrogenrecombination}). These nonthermal
photons are sufficient to reduce the neutral Li fraction by about
3 orders of magnitude at $z=100$.  The resulting evolution with
redshift in Figure~1 shows that Li never recombines completely and
${\rm Li}^+$ is still $\sim 98$\% of the total Li at $z=10$ (cf.
Table~\ref{table_abund}).  As discussed in Section~\ref{opticaldepth},
the recombination history of Li has important cosmological consequences.

\section{CHEMICAL EVOLUTION AFTER RECOMBINATION}
\label{chemicalevolution}

The full network of chemical reactions for the early universe  has
become increasingly complex over the last decade, from the $\sim
90$ reactions of Galli \& Palla~(1998) to the $\sim 250$ reactions
of Gay et al.~(2011). The number of species formed from nuclei up
to Li is limited to $\sim 30$,  but for state-resolved chemistry
each state must be considered as a separate species.  Thus, for
example, H$_2^+$ and H$_2$ require 19 and 15 coupled equations,
respectively, one for each vibrational state of the fundamental
electronic level (Coppola et al.~2011).  The network has been
extended to molecular species produced by elements heavier than Li,
like F (Puy et al.~2007) and C, N, O (Vonlanthen et al.~2009), but
their abundances are extremely small even if non-standard BBN models
are adopted. Ignoring the very small change in particle number due
to chemical reactions, the total baryon density $n_{\rm b}$ evolves
with redshift as
\be 
n_{\rm b}=\frac{\Omega_{\rm b} \rho_{\rm cr}}{\mu m_{\rm H}} (1+z)^3
\ee
where $\rho_{\rm cr}$ is the critical density defined by
equation~(\ref{rhocr}), $m_{\rm H}$ is the mass of the hydrogen
atom, and $\mu=4/(4-3Y_{\rm p})$ the mean atomic weight of the
primordial gas. Numerically, $n_{\rm b}\approx 2.2\times
10^{-7}(1+z)^3$~cm$^{-3}$.

Once formed, molecules can be photodissociated by the CMB photons
with rates that depend on its spectrum, a blackbody characterized
by a radiation temperature $T_{\rm r}=T_0(1+z)$, with $T_0=2.725$~K
(see Table~\ref{table_cosmo}). In addition, distortions of the CMB
produced by photons emitted during the recombination of H and He
cannot be neglected, as they affect significantly, and in some cases
dramatically, the rates of several photodestruction processes, such
as H$^-$ photodetachment (see Section~\ref{globaltrends}) and Li
recombination (see Section~\ref{lithiumrecombination}).  

For applications to the homogeneous universe, the chemical reaction
network must be supplemented by a routine that provides the ionization
fraction as function of redshift for the adopted cosmological model,
and an equation for the evolution of gas temperature $T_{\rm g}$,
\be 
\frac{dT_{\rm g}}{dt}=-2T_{\rm g}\frac{\dot R}{R}
+\frac{2}{3k n_{\rm b}}[(\Gamma-\Lambda)_{\rm Compton}
+(\Gamma-\Lambda)_{\rm mol}+\Gamma_{\rm chem}].
\ee
In this equation, the first term represents the adiabatic cooling
of the gas due to the expansion of the universe, $R$ being the scale
factor; the second and the third the net transfer  of energy (heating
$\Gamma$ minus cooling $\Lambda$) from the radiation to the gas via
Compton scattering of CMB photons on electrons, and via excitation
and de-excitation of molecular transitions, respectively; the fourth
term represents chemical heating of the gas (see e.g.  Glassgold, Galli 
\& Padovani~2012). The ionization fraction is sufficient to maintain
$T_{\rm g}\approx T_{\rm r}$ up to $z\approx 300$ via Compton
heating of the gas. Because of the small molecular abundances, the
term $(\Gamma-\Lambda)_{\rm mol}$ is very small and can raise the
gas temperature by only a few degrees (Khersonskii~1986; Galli \&
Palla~1989; Puy et al.~1993).  The chemical heating term $\Gamma_{\rm chem}$
is also negligible (Puy et al.~1993). Consequently, at redshift
$z\lesssim 100$, the gas cools almost adiabatically, with $T_{\rm
g} \approx 0.02 (1+z)^2$~K.  Additional sources of heating for the
gas that could affect its chemical evolution are dark matter particles
decays and annihilations (Ripamonti, Mapelli \& Ferrara~2007) and ambipolar
diffusion of a cosmological magnetic field (Schleicher et al.~2009).

Finally, the chemical network is completed by the equation for the redshift,
\be
\frac{dt}{dz}=-\frac{1}{(1+z)H(z)},
\ee
where $H(z)$ is the Hubble parameter
\be
H(z)=H_0[\Omega_{\rm r}(1+z)^4+\Omega_{\rm m}(1+z)^3
+\Omega_{\rm K}(1+z)^2+\Omega_\Lambda]^{1/2}.
\label{hz}
\ee
The meaning of these quantities has been discussed in Section~\ref{cosmo} 
and their numerical values are listed in Table~\ref{table_cosmo}.

\subsection{Overview of Global Trends}
\label{globaltrends}

\begin{figure}[t!]
\begin{center}
\resizebox{8cm}{!}{\includegraphics{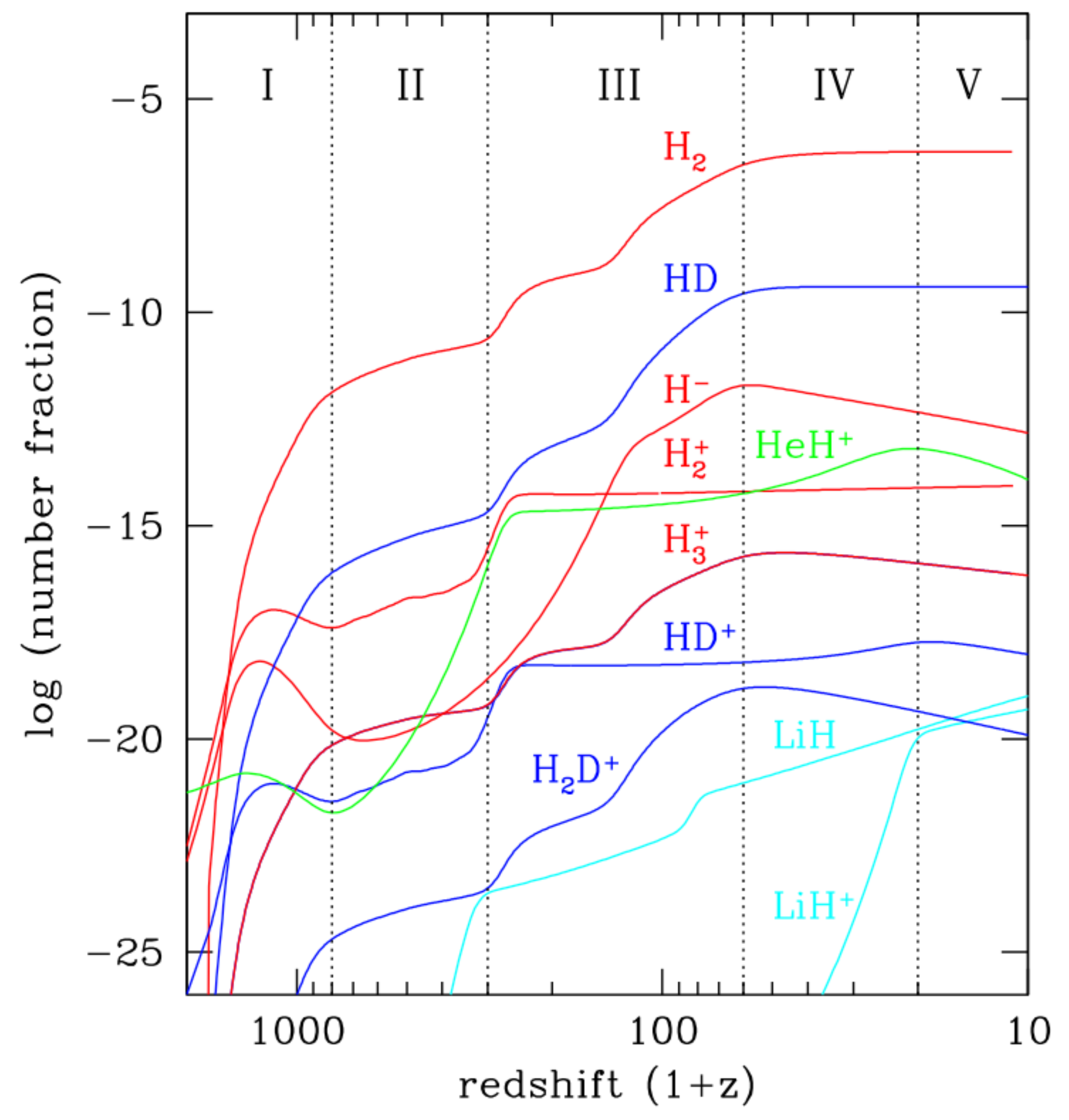}}
\caption{Fractional abundances (relative to the total number of baryons) 
of the main molecules and ions formed
in the early universe as function of redshift. The vertical dotted
lines indicate the boundaries of the five main evolutionary phases
described in the text.}
\end{center}
\label{fig_global}
\end{figure}

The evolution of the main molecular species using the cosmological
parameters and initial element abundance described before is displayed
in {\bf Figure~2} as fractional abundances relative to the total number
of baryons.  It can be conveniently described following the five
main periods during which  some common process can be identified
that dominates the chemical kinetics of the gas. As a function of
decreasing redshift, the sequence is as follows:

{\bf I.} $2000\lesssim z\lesssim 800$.
At $z\sim 1000$ the ionization
fraction is still $\sim 10$\%, while at $z=800$ it drops to $\sim
1$\%. While hydrogen is still mostly ionized,  the availability of
protons promotes the initial steep rise of all molecular species.
The first reaction to occur is the radiative association of H$^+$
and He
\be
{\rm H}^+  + {\rm He} \rightarrow  {\rm HeH}^+  + h\nu ,
\ee
quickly followed by the sequence of radiative association of  H and
H$^+$ (reaction \ref{hhp}) that leads to H$_2$ formation via charge
exchange (reaction \ref{h2ph}). The formation of H$_2^+$ via reaction
(\ref{hhp}) is more efficient than the conversion of HeH$^+$ to
H$_2^+$ by reaction with hydrogen atoms. In this phase, the main
destruction process is through energetic photons from the CMB.  As
shown in Figure~2, the formation of H$_2$, HD, H$_3^+$, and H$_2$D$^+$
proceeds quite rapidly and then levels off at a redshift of $z\sim
1000$.  However, the HeH$^+$, H$_2^+$, and HD$^+$ abundances 
display a turnover followed by a decrease to a minimum
at $z\approx 800$ due to recombination with electrons.  

{\bf II.} $800\lesssim  z \lesssim 300$. As the ionization fraction
freezes out, all the molecular species start forming at a moderate
pace, sustained by the decreased importance of photodissociation
processes by the CMB.  The major limiting factor to the increase
of H$_2$ molecules is the importance of deviations from local
thermodynamic equilibrium (LTE) in the H$_2^+$ level population and
the highly efficient photodissociation by CMB photons (Hirata \&
Padmanabhan 2006; Coppola et al.~2011). Thus, the behavior of HD,
H$_3^+$, and H$_2$D$^+$ follows exactly that of the H$_2$ molecules.
In principle, the kinetics of HeH$^+$ formation should also be
followed including the effects of non-LTE analysis of the level
populations, but this calculation has not been done yet. One would
then expect that the steep rise shown in Figure~2 could just reflect
the assumption of a LTE level population and that the actual abundance
should be considerably reduced. At redshifts $z\lesssim 400$, LiH
begins to form by radiative association, but its growth is limited
by efficient photodestruction.

{\bf III.} $300\lesssim z \lesssim 60$. The abundance of most species
rises and reaches a peak value at $z\sim 60$. As the energy density
of the CMB photons keeps decreasing, H$_2^+$ forms more effectively
through radiative association of H and H$^+$, but the charge exchange
reaction \ref{h2ph} limits its abundance. At redshifts below $\sim
100$, the formation of H$^-$ ions starts becoming important as the
main route for H$_2$ formation. H$^-$ ions (binding energy 0.754~eV)
are no longer effectively photodetached by the CMB and promote the
formation of H$_2$ molecules by radiative attachment (reaction
\ref{hm_h2}).  However, as first computed by Hirata \& Padmanabhan~(2006),
the presence of a significant tail of non-thermal photons of the
CMB produced by hydrogen and helium recombination slows down the
rise of H$^-$ abundance and, hence, of H$_2$ and H$_3^+$.

{\bf  IV.} $60\lesssim z \lesssim 20$. The abundance of H$_2$ and
HD reaches freeze-out, while that of H$^-$ steadily drops due to
its conversion into H$_2$ molecules.  Freeze-out is achieved when
the formation time becomes longer then the Hubble expansion time. 
At redshift $z\lesssim 40$ the marked increase in the HeH$^+$ and 
HD$^+$ abundance results from the exponential drop in the reaction
\be
{\rm HeH}^+  +  {\rm H}  \rightarrow {\rm He}  +  {\rm H}_2^+
\label{hehph}
\ee
below $T_{\rm g}=100$~K (Bovino et al.~2011c). Some of the HeH$^+$
participates in the enhanced formation of HD$^+$ via
\be
{\rm HeH}^+  +  {\rm D} \rightarrow  {\rm HD}^+  +  {\rm H} ,
\ee
which drives further production of H$_2^+$ and HD$^+$. Below $z\approx
40$, LiH$^+$ is readily formed due to the decrease of the intensity
of the CMB, and its abundance reaches that of LiH.

{\bf V.} $z\lesssim 20$. As first structure formation begins to be
important, the steady decrease of some species is accounted for by
the relevant role played by dissociative recombinations.  Interestingly,
only H$_2$ and HD have reached their steady state abundances, whereas
the rest remain out of equilibrium. This fact introduces some
arbitrariness in the initial conditions for the calculations of the
dynamical evolution of the density perturbations, even though the
minute fractional abundances of most chemical species do not affect
the thermal properties of the primordial gas.

The chemical evolution of each element will now be described in some detail.

\begin{table}[t!]
\begin{center}
\caption{Elemental and molecular abundances relative to hydrogen at $z=10$}
\begin{tabular}{cccccc}
\hline
H$_2$/H & H$^-$/H & H$_2^+$/H & H$_3^+$/H & H$_3^-$/H & \\
$6.3\times 10^{-7}$ & $1.8\times 10^{-13}$ & $9.2\times 10^{-15}$ & $8.0\times 10^{-17}$ & $5.5\times 10^{-33}$ & \\
\hline 
HD/H & HD$^+$/H & D$^+$/H & H$_2$D$^+$/H & & \\
$4.2\times 10^{-10}$ & $1.2\times 10^{-18}$ & $6.1\times 10^{-19}$ & $1.6\times 10^{-20}$ & & \\
\hline
HeH$^+$/H & He$^+$/H & He$_2^+$/H & & & \\
$1.7\times 10^{-14}$ & $9.8\times 10^{-23}$ & $9.3\times 10^{-36}$ & & & \\
\hline
Li$^+$/H & Li/H & LiH/H & LiH$^+$/H & Li$^-$/H & LiHe$^+$/H \\
$4.3\times 10^{-10}$ & $8.0\times 10^{-12}$ & $9.0\times 10^{-20}$ & $4.6\times 10^{-20}$ & $1.7\times 10^{-22}$ & $3.0\times 10^{-23}$ \\
\hline
\end{tabular}
\end{center}
\label{table_abund}
\end{table}

\subsection{The Chemistry of Hydrogen}
\label{hydrogenchemistry}

\subsubsection{H$_2$ formation through excited states.}

During the recombination era, the formation of H$_2$ molecules can
proceed either via radiative association of excited hydrogen atoms
(Latter \& Black~1991)
\be
{\rm H}(n=2) + {\rm H}(n=1) \rightarrow {\rm H}_2(X\,^1\Sigma_g^+) + h\nu
\label{rad_ass_hh}
\ee
or via Raman association (the inverse Solomon process; Dalgarno \& van
der Loo 2006; Alizadeh \& Hirata 2011)
\be
{\rm H} + {\rm H} + h\nu \rightarrow {\rm H}_2(B\,^1\Sigma_u^+, C\,^1\Pi_u)
\rightarrow {\rm H}_2(X\,^1\Sigma_g^+) + h\nu,
\label{raman}
\ee
where $X\,^1\Sigma_g^+$ is the ground state of H$_2$, and
$B\,^1\Sigma_u^+$ and $C\,^1\Pi_u$ are the excited electronic states
with energies $\sim$10~eV.  The potential curves of the $B\,^1\Sigma_u^+$
and $C\,^1\Pi_u$ with allowed transitions to the $X\,^1\Sigma_g^+$
ground state of H$_2$ correlate with atomic hydrogen in $n=2$ (both
$2s$ and $2p$) and $n=1$ levels, respectively.

In the first process, as pointed out by Latter \& Black~(1991), the
population of the $n=2$ excited level is maintained over the LTE
value as a result of resonant scattering by Ly-$\alpha$ photons
(Peebles~1968; Zel'dovich et al.~1968), thus allowing reaction
(\ref{rad_ass_hh}) to occur rapidly at a typical rate of few
$10^{-14}$~cm$^3$~s$^{-1}$ over a temperature interval of 50--30000~K.
Latter \& Black showed a pronounced increase of H$_2$ molecules due
to this reaction in the redshift interval between 2500 and 1500.

Recombination of H$_2$ by Raman association is facilitated owing
to the abundance of ultraviolet photons, both blackbody CMB photons
(Dalgarno \& van der Loo 2006) and spectral distortion photons
(Alizadeh \& Hirata 2011). The resulting excited H$_2$ molecule
reemits the photon and can decay either to the bound state
$X\,^1\Sigma_g^+$ or to the continuum of the $X$ level leading to
dissociation. Dalgarno \& van der Loo~(2006) computed the H$_2$
production rates via Raman association including all relevant
rovibrational resonances and found them to be more dominant than
the radiative association channel of reaction (\ref{rad_ass_hh}).
In these calculations, most of the contribution to H$_2$ formation
occurs at a redshift $z\approx 1600$.

On the other hand, Alizadeh \& Hirata~(2011) have demonstrate the
importance of the distortion photons of the CMB using a code that
included the absorption and emission of Lyman-series photons in the
H{\sc i} lines. This contribution starts to be dominant at redshifts
$z\lesssim 1400$ when the photon density deviates significantly
from the Planck spectrum. The overall effect is a {\it net} reduction
of the H$_2$ abundance due to efficiency of the inverse of
reaction~(\ref{rad_ass_hh}). In the redshift interval 1300--900,
the H$_2$ fractional abundance remains well below the value of
$10^{-16}$, several orders of magnitude lower than the value computed
neglecting this process. At lower redshifts, however, the process
becomes rapidly negligible as the spectral distortion is redshifted
at energies lower than the Lyman-Werner bands. Interestingly,
Alizadeh \& Hirata~(2011) find that the H$_2$ abundance is too low
to affect the recombination process or to introduce secondary
anisotropies in the CMB: in principle, absorption or emission of
H$_2$ molecules of photons above the Lyman-Werner bands could add
or remove photons that would otherwise excite a hydrogen atom and
alter the recombination history.

\subsubsection{State-resolved formation of H$_2$ and H$_2^+$.}

At redshifts lower than $z\approx 800$, the abundance of H$_2$
molecules is determined by the balance between charge transfer with
H$_2^+$ ions, reaction~(\ref{h2ph}) and the reverse of reaction
(\ref{hhp}). Other minor routes involve reactions of HD molecules
with either H or H$^+$. In spite of its simplicity, the kinetics
of the main reactions for H$_2$ is quite complex due to the presence
of a large number of quantum states and the evaluation of the exact
vibrational distribution function is indeed necessary (see, e.g.,
Lepp, Stancil \& Dalgarno~2002). This aspect has proved particularly important
in the case of the photodissociation rates of H$_2^+$ ions, the
main destruction route in competition with conversion into H$_2$
molecules, which have usually been computed either assuming LTE
level populations of the rovibrational levels or with all the H$_2^+$
ions in the ground state. Detailed calculations of the fully
level-resolved populations of H$_2^+$ have been performed by Hirata
\& Padmanabhan~(2006) and Coppola et al.~(2011; see also Longo et
al.~2011). The resulting behavior is shown in {\bf Figure~3a}. The main
conclusion of these studies is that full thermalization of the
vibrational levels is never achieved since the vibrational relaxation
processes are too slow to balance the strong vibrational selectivity
of formation rates and radiative processes affect both the vibrational
distribution function and the overall fractional abundances. As a
consequence of the non-LTE level population, the formation of H$_2$
molecules is greatly reduced mainly because the H$_2^+$ ions are
photodissociated before they can either decay to the ground state
or undergo charge transfer to become H$_2$ molecules.

\subsubsection{H$_2$ formation via H$^-$ ions.}

The last step in the formation of H$_2$ molecules is via associative
detachment, reaction~(\ref{hm_h2}). Recently, the cross section for
this simple reaction has been accurately measured in experiments
with merged beams apparatus at collision energies of cosmological
interest (Kreckel et al.~2010; Miller et al.~2011).  Once formed,
H$^-$ ions are readily destroyed by photons of the CMB owing to the
rather low photodetachment threshold energy (0.754~eV).  The major
improvement on the precise magnitude of this process has been the
inclusion of the spectral distortions in the CMB (Hirata \& Padmanabhan
2006). The rate coefficient for photodetachment can be written as
\be
k_{\rm nonth}= n_{\rm H} c \int_{\nu_0}^{\infty} r_\nu \sigma_\nu {d\nu\over\nu}
\ee
where $r_\nu$ is the number of recombination photons per H atom
and per logarithmic frequency range, $\sigma_\nu$ is the photodetachment
cross section and $h\nu_0=0.754$~eV. The nonthermal photons become
the dominant process at redshifts below $z\approx 130$ when the
photon energy is $\sim 1$~eV, corresponding to the peak of the
photodetachment cross section. The effect on the H$^-$ abundance
is clearly visible in {\bf Figure~3b} where the sharp peak at $z\approx 100$
(dotted line) is replaced by a shallow increase and a maximum at
$z\approx 60$.  This behavior accounts for the evolution of H$_2$
molecules, also shown in Figure~3b. Unlike the case without non-thermal
photons, the fractional abundance of H$_2$ reaches the freeze-out
value only at $z\lesssim 60$, when the nonthermal photons have been
redshifted below the threshold. The freeze-out value of H$_2$
molecules at $z=10$ is $6\times 10^{-7}$.

\subsubsection{Minor species: H$^-$, H$_2^+$ and H$_3^+$.}

As shown in Figure~3b and Table~\ref{table_abund}, the
fractional abundance of these species at $z=10$ is very small,
varying between $\sim 10^{-13}$ for H$^-$ and $\sim 10^{-16}$ for
H$_3^+$. The reason why H$^-$ ions do not reach equilibrium at the
lowest redshifts is that they are constantly removed by the radiative
association with H atoms to form H$_2$ molecules.  Similarly, the
H$_3^+$ ion displays a steady decrease from its peak abundance of
$\sim 2.5\times 10^{-16}$ at $z\approx 50$ due to efficient destruction
by dissociative recombination with electrons. According to the
experimental results of McCall et al.~(2004) and Kreckel et al.~(2005),
the dissociative reaction rate is in general agreement with the
theoretical predictions and the branching ratios tend to favor the
channel H$+$H$+$H over the H$+$H$_2$ channel.  The global evolution
of H$_3^+$ follows quite closely that of H$_2$ as the main formation
process is via radiative association with H$^+$ at all redshifts.
Due to its potential importance as a coolant of the collapsing gas,
we shall return to H$_3^+$ in the discussion on the rates
(Section~\ref{criticalrates}).

As for H$_2$, the behavior of H$_2^+$ is strongly affected
by the inclusion of the non-LTE level population in the calculation
of the photodissociation rate, which is the main destruction process
of this ion.  The latter ceases to be important at redshifts below
$z\approx 300$ and that explains the flattening of the abundance
shown in Figure~3a.  At lower redshifts, formation by radiative
association is balanced by charge exchange reaction into H$_2$,
accounting for a nearly constant fractional abundance of $\sim
10^{-14}$. Finally, at redshifts smaller than $z\approx 60$, the
formation of H$_2^+$ is enhanced by the HeH$^+$ channel via the
reaction with ${\rm H}_2^+$ whose rate drops significantly at
temperatures below $T_{\rm g}\approx 100$~K (Bovino et al. 2011c),
as will be discussed in Section~\ref{heliumchemistry}.

We also mention the possibility of the formation of H$_3^-$, the
negatively charged counterpart of H$_3^+$, which is predicted to
be stable by about 0.013 eV (St\"arck \& Meyer~1993). Recently,
Ayouz et al.~(2011) have developed the theory of its formation in
the ISM by radiative association of H$_2$ and H$^-$ and derived the
relevant cross sections. Using the reaction rate by Ayouz et
al.~(2011), we have verified that the resulting abundance is
negligible ($\lesssim10^{-30}$) at all redshifts.

\begin{figure}[t!]
\begin{center}
\resizebox{6.5cm}{!}{\includegraphics{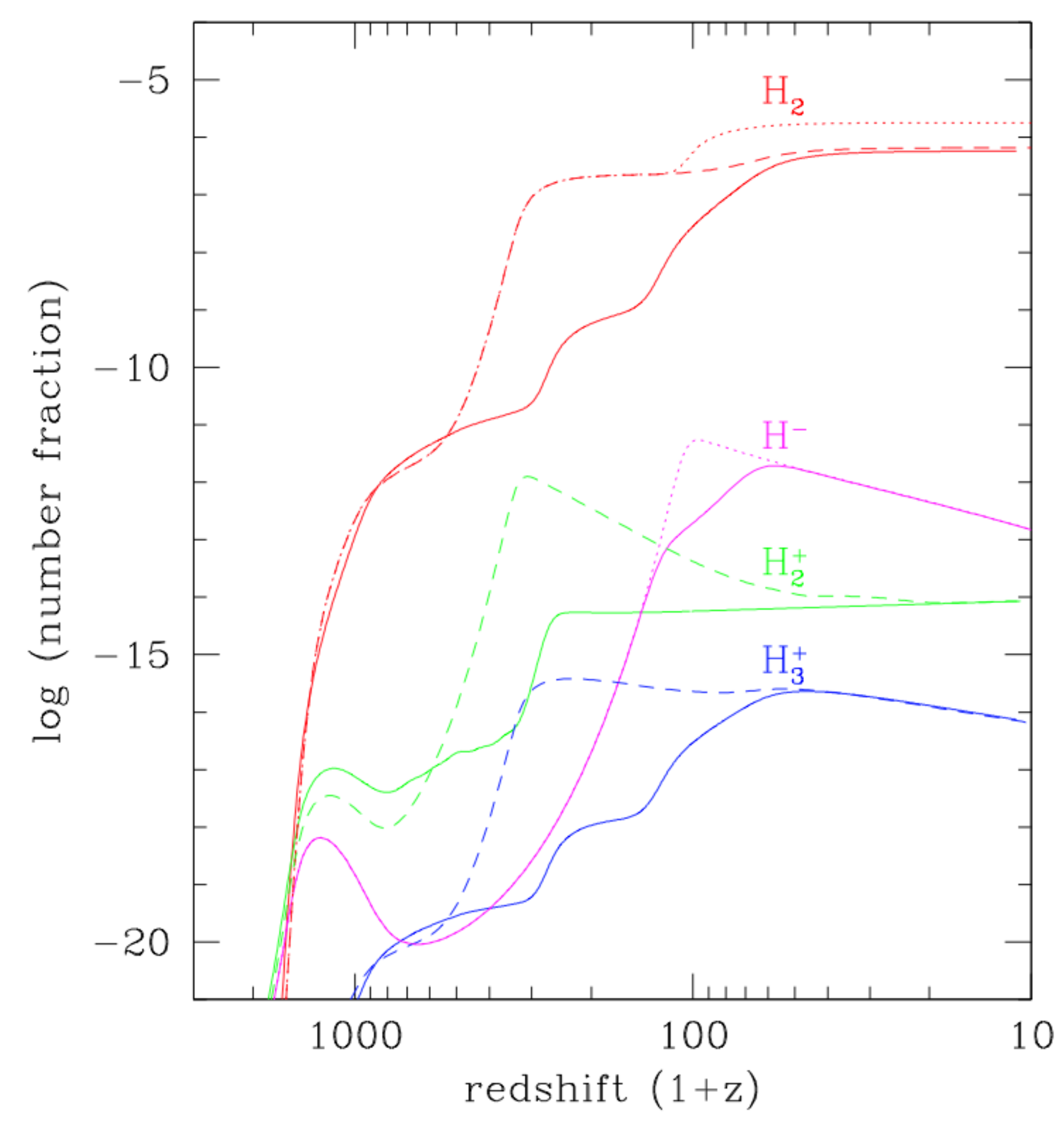}}
\resizebox{6.5cm}{!}{\includegraphics{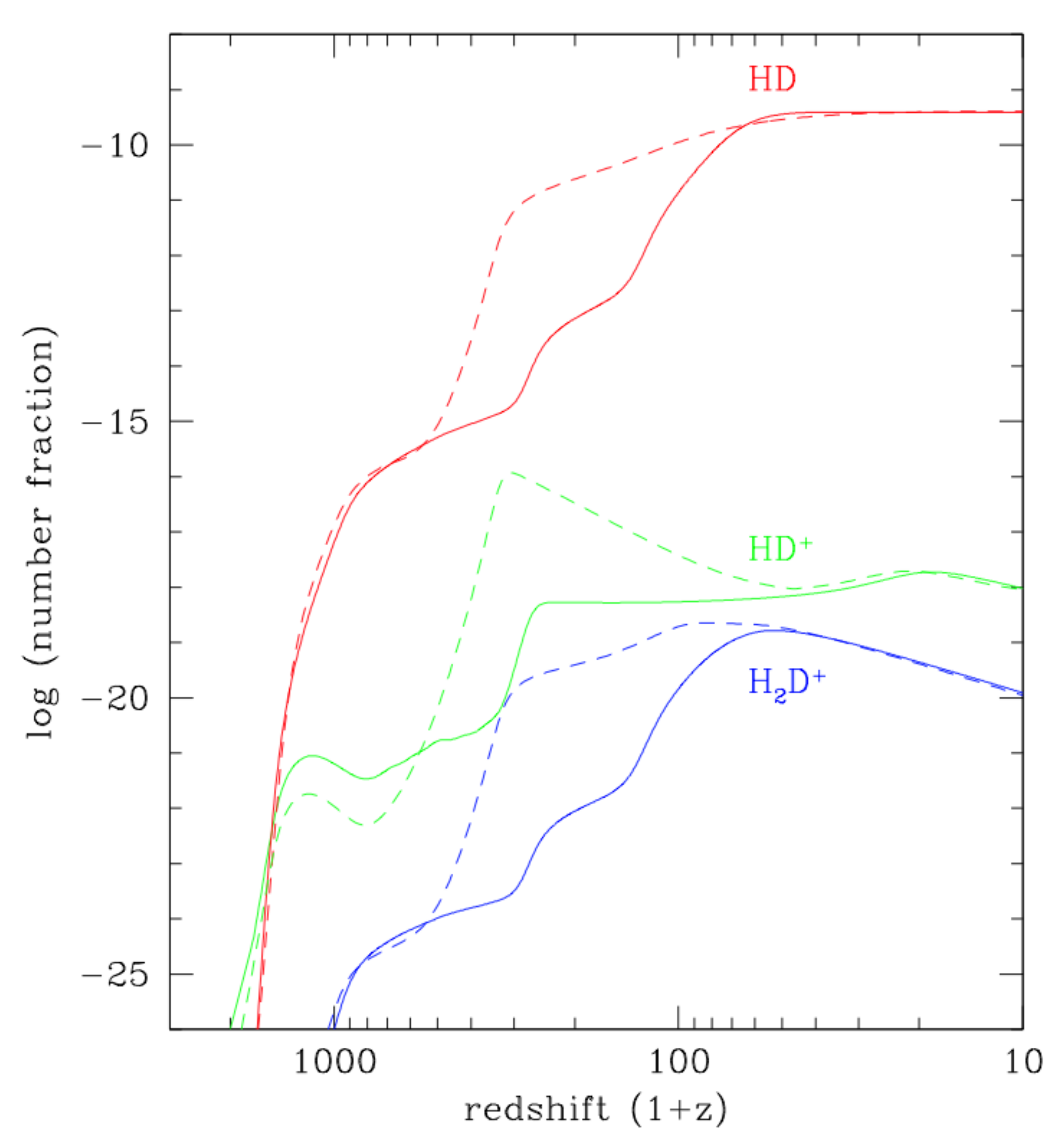}}
\resizebox{6.5cm}{!}{\includegraphics{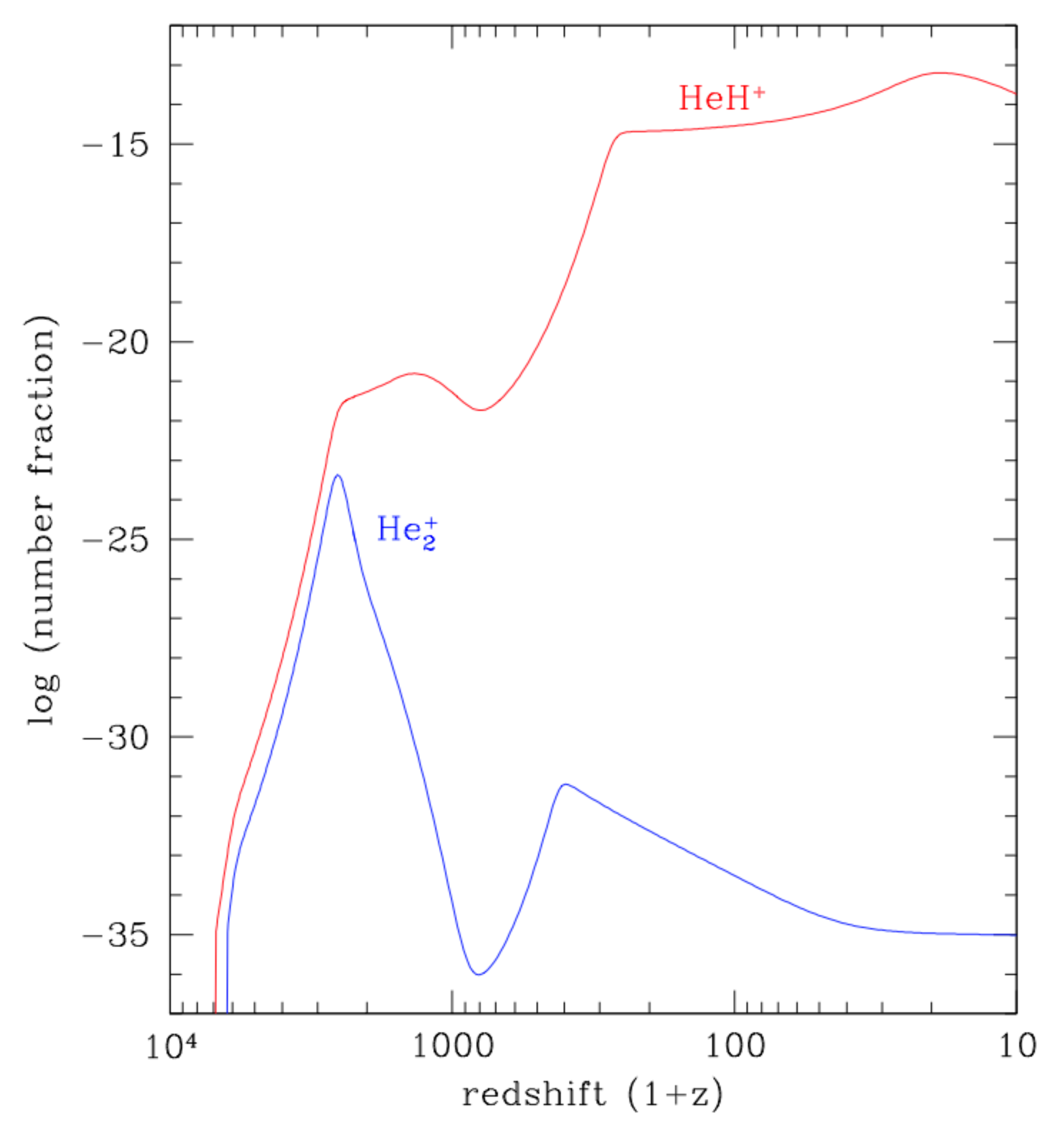}}
\resizebox{6.5cm}{!}{\includegraphics{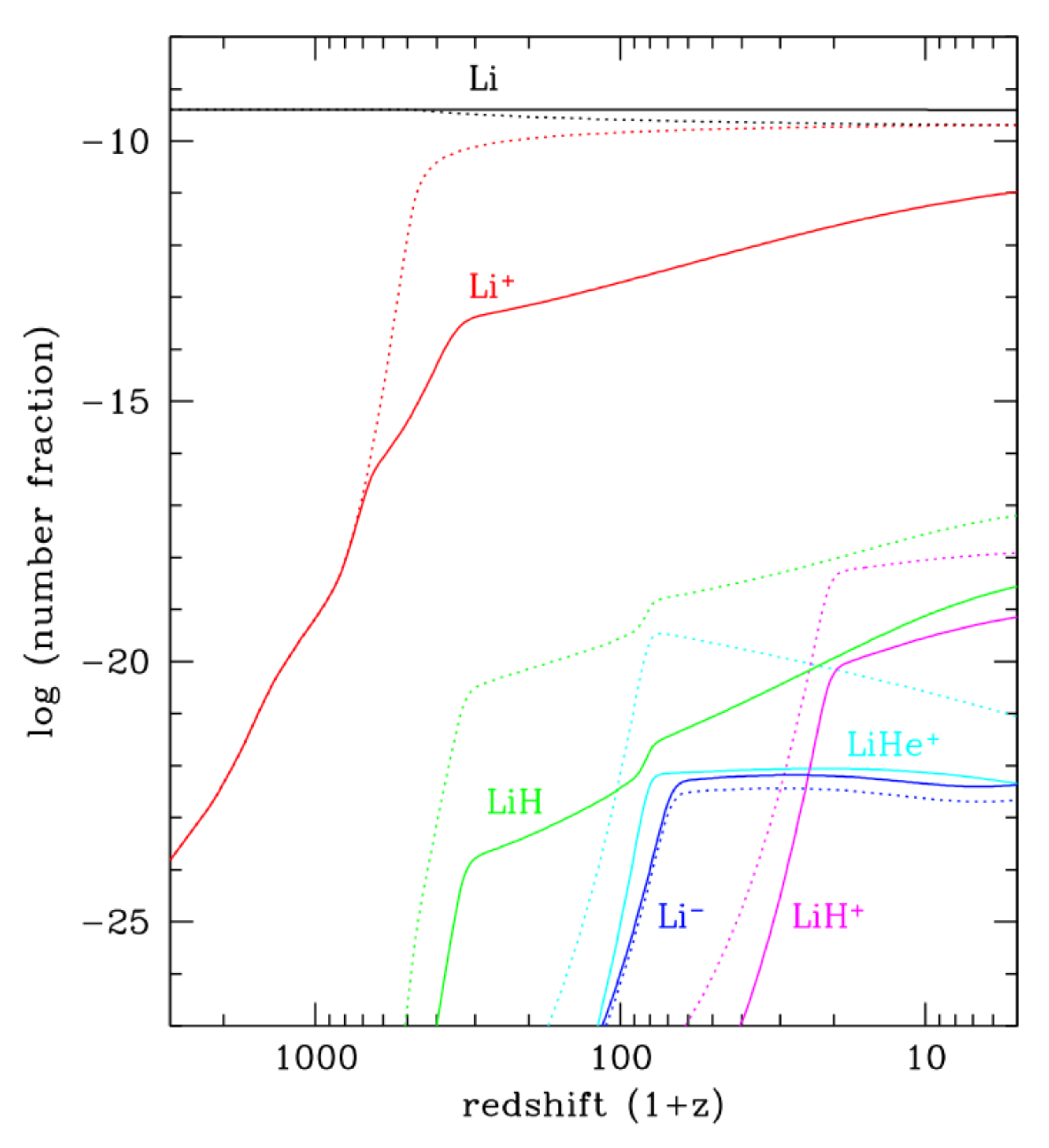}}
\caption{({\em a}\/) Hydrogen chemistry. The solid curves show
the case with state-resolved formation of H$_2$ and
H$_2^+$, while the dashed lines correspond to the
evolution obtained by ignoring this effect. The dotted lines display
the evolution of H$_2$ and H$^-$ without the contribution of the
non-thermal photons of the CMB to the photodissociation of H$^-$.
(Adapted from Coppola et al.~2011). ({\em b}\/) Deuterium chemistry.
The solid and dashed curves have the same meaning as in the case{\em {\em Astron. Astrophys.}}
of the hydrogen chemistry shown in Figure~3a. (Adapted from Galli
\& Palla~2002). ({\em c}\/) Helium chemistry. Evolution of the
main He-bearing molecular ions, HeH$^+$ and He$_2^+$. (Adapted from
Bovino et al.~2011c). ({\em d}\/) Lithium chemistry.  Evolution of
Li-bearing molecules and ions with (solid curves) and without (dotted
curves) the effects of non-thermal photons on Li recombination. In
the latter case, Li never recombines, molecular abundances decrease
significantly at all redshifts, with the exception of Li$^-$, and
LiH remains the most abundant species. (Adapted from Bovino, Tacconi
\& Gianturco~2011b and Bovino et al.~2011d).}
\end{center}
\end{figure}

\subsection{The Chemistry of Deuterium}
\label{deuteriumchemistry}

Deuterium in the early universe is important for a variety of
reasons.  First, owing to its larger mass, deuterated hydrogen has
a spacing of the rotational levels smaller than that of hydrogen
molecules. Thus, cooling through HD dipole radiation can effectively
drive the primordial gas to substantially low temperatures, provided
a sufficient abundance is obtained.  Thus, the chemistry of deuterium
has been computed by numerous groups following the initial study
of Lepp \& Shull (1984), such as  Dalgarno \& Lepp (1987), Latter
(1989), Puy et al.  (1993), Palla, Galli \& Silk~(1995), Stancil, 
Lepp \& Dalgarno~(1998), Galli \& Palla (1998, 2002), Lepp, 
Stancil \& Dalgarno~(2002), and Glover
\& Abel (2008).  Recently, Gay et al. (2011) have presented a
thorough analysis of the chemistry of highly deuterated species,
including D$_2$, D$_2^+$, D$_2$H$^+$, and D$_3^+$. The formation
of HD follows two main routes involving a deuteron exchange with
H$_2$
\be
{\rm D} + {\rm H_2} \rightarrow {\rm HD} + {\rm H} ,
\label{d1}
\ee
and
\be
{\rm D^+} + {\rm H_2} \rightarrow {\rm HD} + {\rm H^+}. 
\label{d2}
\ee
The thermal rate constant for reaction~(\ref{d1}) has been measured
in numerous experiments over a wide range of temperatures (e.g.,
Mitchell \& LeRoy~1973; Michael \& Fisher 1990); theoretical
calculations (e.g., Mielke et al.~1994; Charutz, Last \& Baer~1997) show
very good agreement with each other and with the experimental data.
Reaction~(\ref{d2}) represents the major source of HD in diffuse
interstellar clouds (Dalgarno, Weisheit \& Black~1973). Its rate coefficient
is almost constant and close to the Langevin value $2.1\times
10^{-9}$~cm$^3$~s$^{-1}$ (e.g., Gerlich~1982).  Finally, at $z
\lesssim 10$, HD formation takes place via charge exchange (Karpas, Anicich 
\& Huntress~1979)
\be
{\rm HD}^+ + {\rm H} \rightarrow {\rm HD} + {\rm H}^+.   
\ee

The destruction of HD occurs mainly via the reverse reactions 
of (\ref{d1}) and (\ref{d2})
\be
{\rm HD} + {\rm H} \rightarrow {\rm D} + {\rm H_2}   
\label{d3}
\ee
and
\be
{\rm HD} + {\rm H^+} \rightarrow {\rm D^+} + {\rm H_2}.   
\label{d4}
\ee
The rate coefficient for the first reaction has been computed by
Shavitt~(1959) using a semi-empirical H$_3$ energy surface.  As for
reaction (\ref{d4}), owing to its endothermicity by 0.0398 eV
(462~K), the removal of HD at low temperatures is reduced by a
factor $\exp(-462~{\rm K}/T_{\rm g})$, leading to significant
fractionation. At freeze-out, the ratio HD/H$_2 \approx 7\times
10^{-4}$, with an enhancement factor of $\sim 25$ over the initial
D/H abundance.

The relative abundance of D and D$^+$, 
is governed by the charge exchange reactions
\be
{\rm H}^+ + {\rm D} \rightarrow {\rm H} + {\rm D}^+ ,
\label{d5}
\ee
and
\be
{\rm H} + {\rm D}^+ \rightarrow {\rm H}^+ + {\rm D}. 
\label{d6}
\ee
Since reaction (\ref{d5}) has a threshold of 43~K, the rate coefficient
for reaction (\ref{d6}) is usually obtained by multiplying that of
reaction (\ref{d5}) by $\exp(43~\rm{K}/T_{\rm g})$. Savin~(2002)
has computed accurate rate coefficients for both reactions, improving
on the early estimates by Watson~(1976) and Watson, Christensen \& Deissler~(1978)
that have been widely used in studies of deuterium chemistry.

Figure~3b highlights the close similarity of the evolution with
redshift of the HD abundance with that of H$_2$ and the dramatic
effect of including the state resolved levels in the calculations.
Generally, the chemistry of HD is dominated by the ion-neutral
reactions (\ref{d2}) and (\ref{d4}) in a gas of low density (e.g.,
prior to the formation of the first structures), whereas the
neutral-neutral reactions (\ref{d1}) and (\ref{d3}) become more
important in conditions of high density and temperature (as during
cloud collapse or in shocked gas).

Figure~3b also displays the evolution of the main deuterated cations
HD$^+$ and H$_2$D$^+$. The case of H$_2$D$^+$ is of interest due
to the suggestion of Dubrovich~(1993) and Dubrovich \& Lipovka~(1995)
that its abundance could reach significant values (up to
H$_2$D$^+$/H$_2=10^{-5}$), in which case specific rovibrational
transitions could leave detectable imprints on the spectrum of the
CMB. In principle, such an abundance could result from the complete
conversion of H$_3^+$ into H$_2$D$^+$ following deuterium fractionation
at low temperatures ($T_{\rm g} \ll 50$~K). However, the actual
abundance of H$_3^+$ in the primordial gas is quite low due to fast
dissociative recombination. Although Dubrovich \& Lipovka~(1995)
claim observable features in the CMB for abundances of H$_2$D$^+$
as low as 10$^{-8}$, Figure~3b shows that this value is never
achieved in the post-recombination universe where H$_2$D$^+$ reaches
a peak of only $\sim 2\times 10^{-19}$ at $z\sim 80$. On the other
hand, the isotopic ratio is enhanced with respect to the primordial
D/H abundance by a factor of $\sim 8$.  As in the case of H$_3^+$,
the main removal of H$_2$D$^+$ occurs via dissociative recombination.

Finally, the chemistry of HD$^+$ is directly related to that of
H$_2^+$ since its abundance is determined by the equilibrium between
the forward and reverse reaction
\be
{\rm H_2^+} + {\rm D} \rightarrow {\rm HD^+} + {\rm H}. 
\ee
HD$^+$ would maintain the freeze-out value of $\sim 5\times 10^{-19}$,
reached at a redshift of $\sim 250$ if not for the enhancement due
to the production from HeH$^+$ via
\be
{\rm HeH^+} + {\rm D} \rightarrow {\rm HD^+} + {\rm He} .
\ee
Comparison of Figure~3a and Figure~3b visualizes the similar behavior
of HD$^+$ and H$_2^+$, as well as H$_2$D$^+$ and H$_3^+$.

\subsection{The Chemistry of Helium}
\label{heliumchemistry}

The helium hydride ion (HeH$^+$) and the molecular helium ion
(He$_2^+$) contend for the role of the first molecular species
formed in the universe.  Searches for HeH$^+$ in space, either in
gaseous nebulae (Moorhead et al.~1988, Liu et al. ~1997, Dinerstein
\& Geballe~2001) or in high-redshift absorbers (Zinchenko, Dubrovich 
\& Henkel~2011), have only given tentative detections.

In the early universe, HeH$^+$ and He$_2^+$ are formed at $z\approx
7000$ by radiative association of He with H$^+$ and He$^+$,
respectively. The rate of the former process is well determined
theoretically (Saha, Datta \& Barua~1978; Flower \& Roueff~1979; Roberge \&
Dalgarno~1982; Basu \& Barua~1984; Kimura et al.~1993; Dumitriu \&
Saenz~2009; Sodoga et al.~2009) and the absolute photodissociation
cross section has been measured by the free electron laser FLASH
(Pedersen et al.~2007), although at high photon energy (38.7~eV).
Some studies have also addressed the enhancement of the photodissociation
rate produced by CMB photons (Ju{\v r}ek, {\v S}pirko \& Kraemer~1995;
Zygelman, Stancil \& Dalgarno~1998). The rate of the He$_2^+$ formation has a
temperature dependence similar to that of the radiative association
of H and H$^+$, but is about one order of magnitude smaller 
(Stancil, Babb \& Dalgarno~1993).

The history of helium chemistry is displayed in {\bf Figure~3c}.
Since He$^+$ recombines rapidly at $z\approx 2500$, the abundance
of He$_2^+$ reaches a maximum of $\sim 10^{-23}$ around this redshfit,
and then is readily destroyed by photodissociation and dissociative
recombination. In contrast, the incomplete recombination of H$^+$
keeps the abundance of HeH$^+$ steadily rising to $\sim 10^{-13}$
at $z\approx 20$.  The main reactions responsible for the bumpy
behavior of the evolution of HeH$^+$ and that limit its final
abundance are photodissociation for $z \gtrsim 300$, collisions
with H for $20\lesssim z \lesssim 300$ (Linder, Janev \& Botero~1995, Bovino
et al.  2011c), and dissociative recombination for $z \lesssim 20$
(Guberman~1994). As mentioned in Section~\ref{hydrogenchemistry},
destruction of HeH$^+$ by collisions with H atoms is an important
source of H$_2^+$ and accounts for its enhancement at the smallest
redshifts. Finally, at redshifts below $z\approx 800$, photodissociation
processes become inefficient, and the abundance of He$_2^+$ rises
again by about 5 orders of magnitudes, but its growth is limited
by efficient destruction by collisions with H for $z\lesssim 400$.

In spite of the potential importance of HeH$^+$ molecules for
efficiently scattering the CMB photons (see Section~\ref{interactions}),
an accurate calculation of photodissociation rates is still lacking.
Miyake, Gay \& Stancil~(2011) have obtained cross sections for all rotational
transitions from the vibrational levels $v^{\prime\prime}=0$--11
of the $^1X\Sigma^+$ state for temperatures between 500 and 12,000~K,
including both $X\leftarrow X$ and $A\leftarrow X$ transitions.
These cross sections can then be used to directly obtain the
photodissociation rate due to the CMB radiation.  Miyake et al.
present some initial estimate for different radiation temperatures
and find that at redshifts $\gtrsim 1000$ the $A\leftarrow X$
transitions (which are usually neglected) dominate the photodissociation,
while at lower redshifts the $X\leftarrow X$ transition become more
important.  Thus, one would expect that the initial rapid rise of
the HeH$^+$ abundance shown in Figure~3c will be further reduced
by inclusion of this photodestruction channel. Future work is needed
in order to clarify the magnitude of this effect.

\subsection{The Chemistry of Lithium}
\label{lithiumchemistry}

The chemistry of lithium in the early universe has been reviewed
by Dalgarno, Kirby \& Stancil~(1996), Stancil, Lepp \& Dalgarno~(1996), Bougleux
\& Galli~(1997), Galli \& Palla~(1998), Puy \& Signore~(2002) and
Bodo, Gianturco \& Martinazzo~(2003).  After the recombination of
Li$^{3+}$ and Li$^{2+}$ (see Section~\ref{recombination}), the main
remaining species Li and Li$^+$ react with the residual electrons,
H, D, and He atoms and ions to form Li$^-$, LiH, LiH$^+$, LiD,
LiD$^+$, LiHe$^+$ and Li$^-$.  The chemical network of lithium is
complex (for updated chemical reaction rates see Bovino, Stoecklin
\& Gianturco~2010; Bovino et al.~2010; Bovino, Tacconi \&
Gianturco~2011b; Bovino et al.~2011d; Bovino et al.~2012), but the
molecular abundances are very small.  In addition, as these species
are easily photodissociated, their formation in significant amounts
is delayed to relatively low redshifts, $z\lesssim 200$.  To
complicate matters, the evolution is also extremely sensitive to
the effects of nonthermal photons from the recombination of H and
He (cf. Switzer \& Hirata 2005).  The latter can easily maintain
Li ionized down to the lowest redshifts, whereas in their absence
lithium would recombine at the same level of Li$^+$. This behavior
is displayed in {\bf Figure~3d} where the dramatic effect of the
inclusion of non-thermal photons can be appreciated by comparing
the evolution in the two cases.  In addition to the neutral and
singly ionized species, some Li$^-$ ions can also form at redshifts
below $z\sim 100$. Their evolution follows closely that of H$^-$,
both anions being removed at low temperatures mainly by reactions
of mutual neutralization with H$^+$, with a rate typically proportional
to $T_{\rm g}^{-1/2}$.  As a result, the abundance of Li$^-$ at
$z=10$ is $\sim 2\times 10^{-22}$.

The first Li-bearing molecule to form is LiH mainly via radiative
association, 
\be 
{\rm Li} + {\rm H} \rightarrow {\rm LiH} + h\nu,
\ee 
with a contribution for $20 \lesssim z \lesssim 80$ from Li$^-$
via the reaction 
\be 
{\rm Li}^- + {\rm H} \rightarrow {\rm LiH} + e.  
\ee 
Note, however, that the rate for the latter reaction is only
estimated.  Radiative association takes place first from the Li($^2P$)
state, then from the Li($^2S$) state. Since the $^2P$ level lies
about 1.85~eV above the $^2S$ level, the formation of LiH molecules
via electronically excited levels is important only at $z\gtrsim
800$, when the temperature is greater than some thousand degrees
K. Early estimates for the radiative association rate from the $^2S$
state resulted in relatively large values, and therefore the abundance
of LiH in the early universe was expected to be a substantial
fraction of the Li abundance (Lepp \& Shull~1984).  However, the
situation was clarified by refined quantum mechanical calculations
that reduced significantly the magnitude of the cross sections
(Dalgarno, Kirby \& Stancil~1996; Gianturco \& Gori Giorgi~1996).
As for destruction, LiH is easily photodissociated by CMB photons
for $z\gtrsim 300$, whereas at lower redshifts it is mainly destroyed
by collisions with H, a process for which quantal calculations are
available (Bovino, Wernli \& Gianturco~2009).

Owing to the small binding energy (0.14~eV), the main production
of LiH$^+$ via radiative association of Li with H$^+$ is delayed
until redshifts below $z\approx 40$, although a minute contribution
can result at $z\gtrsim 700$ from reactions of Li$^+$ ions with H.
The reaction rates for both channels have been computed with fully
quantal methods (Dalgarno, Kirby \& Stancil~1996; Gianturco \& Gori Giorgi~1997).
As in the case of LiH, photodissociation is the dominant destruction
mechanism of the molecular ion, with an uncertain contribution from
collisions with H and a contribution from dissociative recombination
({\v C}urik \& Greene~2007,2008) at very low redshifts ($z\lesssim
20$). Interestingly, at $z=10$ the abundance ratio of LiH/LiH$^+$
is $\sim 2$, either including or ignoring the non-thermal photons
of the CMB.

Finally, LiHe$^+$ is formed by spontaneous and stimulated radiative
association of Li$^+$ with He (Bovino, Tacconi \& Gianturco~2011a,
2012).  At $z\gtrsim 100$, LiHe$^+$ is destroyed by photodissociation,
while at lower redshift by collisions with H and by dissociative
recombination (Bovino et al.~2012). However, the abundance of
LiHe$^+$ at $z=10$ remains exceedingly small, $\sim 4\times 10^{-23}$.

\subsection{Critical Rates}
\label{criticalrates}

As stressed in the Introduction, the evolution described so far
depends more on the uncertainties of the chemical model than on the
initial conditions provided by the cosmological models. The abundances
of the various species are only as accurate as the rates given as
input. The literature is rich in studies that critically examine
the uncertainties of the reaction rates and their effects on the
chemistry (and cooling) of the primordial gas. Most of the important
reactions involved in the chemistry of the major species have well
determined rate coefficients. Below, we give a brief overview of
the rates for the most relevant reactions or that still present
significant differences either from theory or from experiments.
Extensive and updated compilations of reaction rates can be found
in Glover \& Abel~(2008), Schleicher et al.~(2008),  Glover \&
Savin~(2009), Bovino et al.~(2011d), and Gay et al.~(2011).

\subsubsection{Associative detachment of H$^-$.}

Surprisingly enough, in spite of many years of research, there has
been considerable uncertainty on both the magnitude and temperature
dependence of the rate of this important reaction that affected all
the calculations of the chemical and thermodynamical evolution of
Population~III stars (e.g., Glover, Savin \& Jappsen~2006).  The
series of merged beams measurements carried out by Bruhns et
al.~(2010), Kreckel et al.~(2010), Miller et al.~(2011) and the ion
trap study by Gerlich et al.~(2012) have given results in agreement
with the theoretical predictions (e.g., Sakimoto~1989, C{\'{\i}}zek,
Hor\'acek \& Domcke~1998).  {\bf Figure~4} illustrates the good
match (to within $\pm$25\%) between experimental data of Gerlich
et al.  (squares), Bruhns et al. (solid line) and theory (dashed
curve) over a wide temperature interval and highlights the departure
from the Langevin value that had been previously adopted in chemical
models. The impact of the newly computed rates on the collapse and
fragmentation of primordial clouds has been discussed by Kreckel
et al.~(2010) who have shown that the uncertainty on the predicted
Jeans mass of the fragments is reduced from a factor of 20 to only
a factor of 2 (see Savin et al.~2012; Savin~2013).

\begin{figure}[t!]
\begin{center}
\resizebox{12cm}{!}{\includegraphics{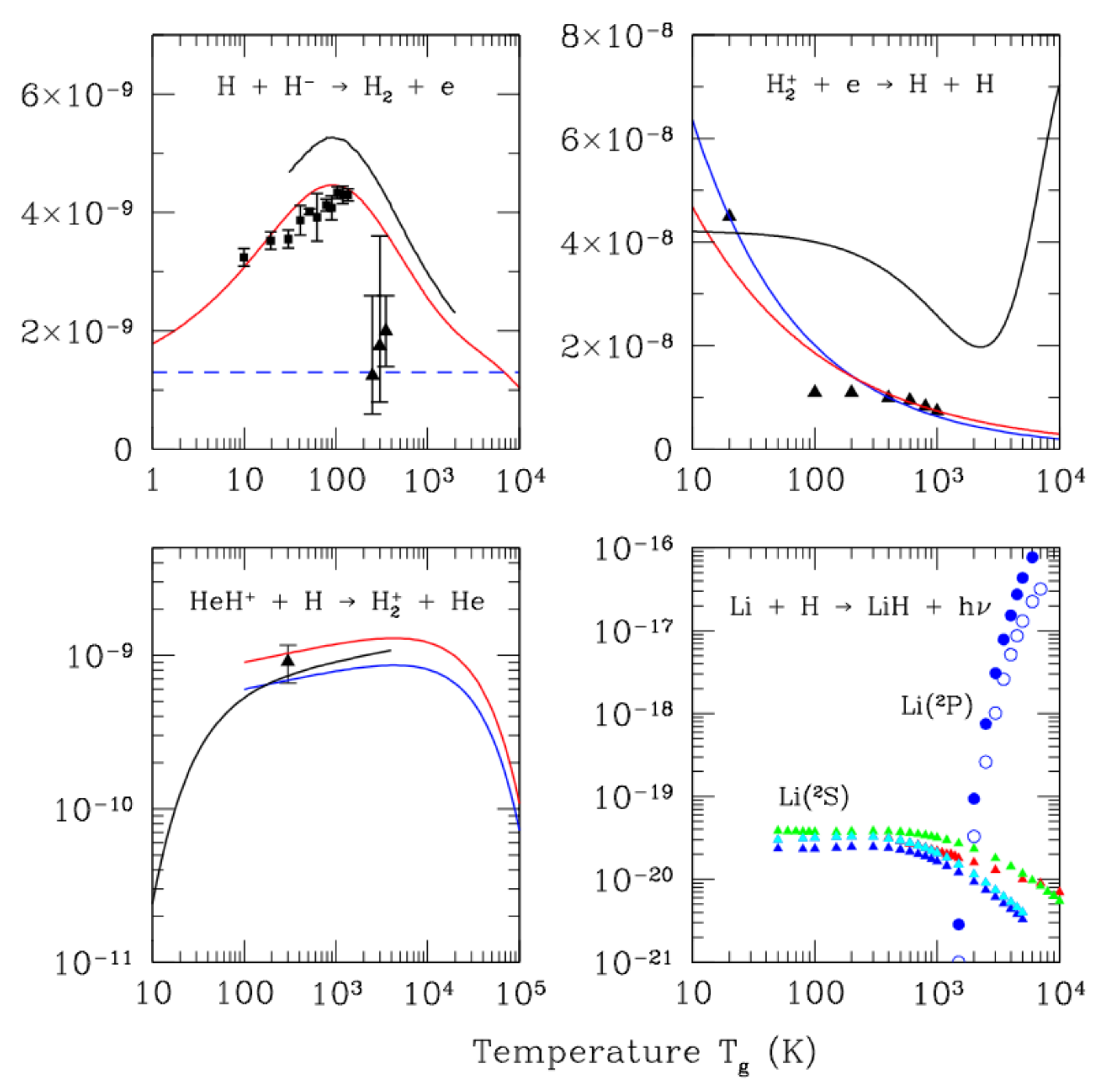}}
\caption{Rate coefficients (in cm$^3$~s$^{-1}$) for some critical
reactions.  {\em Upper left panel:} Filled squares: data
from Gerlich et al.~(2012) with 1$\sigma$ statistical uncertainty.
Black line: results from the merged-beam experiments
of Bruhns et al.~(2010) and Miller et al.~(2011). Red curve: 
calculation by C{\'{\i}}zek et al.~(1998).  Triangles: measurements
at room temperature by Schmeltekopf, Fehsenfeld \& Ferguson~(1967), Fehsenfeld,
Howard \& Ferguson~(1973), and Martinez et al.~(2009, 
slightly shifted for better readability). Short-dashed
line: Langevin value.  {\em Upper right panel:} Blue and
red curves: fits by Galli \& Palla~(1998) and Stancil, Lepp \& 
Dalgarno~(1998) of the rate by Schneider et al.~(1994) for the $v=0,
J=0$--20 levels (triangles). Black curve: LTE rate computed
by Coppola et al.~(2011) from the state-resolved cross sections of
Takagi~(2002).  {\em Lower left panel:} Triangle: experimental result at
$300$~K (Karpas, Anicich \& Huntress~1979). Red and
blue curves: reaction rate obtained by Stancil, Lepp \& Dalgarno~(1998)
and Schleicher et al.~(2008) from the experimental
cross sections of Linder, Janev \& Botero~(1995). Black curve:
result of the quantum calculations by Bovino et al.~(2011c).  {\em
Lower right panel:} Li$(^2S)$: data from Dalgarno, Kirby
\& Stancil~(1996, cyan), Gianturco \& Gori Giorgi~(1996, blue),
Bacchus-Montabonel \& Talbi~(1999, red), Bennett et al.~(2003, 2008,
green). Li$(^2P)$: data from Gianturco \& Gori
Giorgi~(1997, blue) for the $B^1\Pi\rightarrow X^1\Sigma^+$ (filled
dots) and the $A^1\Sigma^+\rightarrow X^1\Sigma^+$ (empty dots),
respectively.}
\end{center}
\label{fig_4panels}
\end{figure}

\subsubsection{Mutual neutralization of H$^-$ and H$^+$.}

Although this process is irrelevant in the chemistry of the expanding
universe, it becomes important when the gas gets reionized and
recombines out  of equilibrium during the gravitational collapse
phase (see, e.g., Glover, Savin \& Jappsen~2006). The older experiment
by Moseley, Aberth \& Peterson~(1970) reports cross sections that are consistently
higher at energies above 3~eV than the most recent ones by Szucs
et al.~(1984) and Peart \& Hayton~(1992). The latter are in very
good agreement with the theoretical calculations (see, e.g., Stenrup, Larson 
\& Elander~2009).  Experiments at energies below $\sim 3$~eV are required
in order to remove the remaining uncertainty (Urbain et al.~2010).

\subsubsection{Radiative association of H$_2^+$.}

This reaction which dominates the H$_2^+$ kinetics is important at
all temperatures and redshifts. The rate coefficient calculated by
Stancil, Babb \& Dalgarno~(1993) is in excellent agreement with the older
results of Ramaker \& Peek~(1976). Below $\sim 10$~K, the rate is
basically constant at the value of $\sim 2\times 10^{-20}$~cm$^3$~s$^{-1}$
which accounts for the flat portion of the curve of the H$_2^+$
abundance at low redshifts (see Figure~3a).  We note that the fitting
formulae given by Stancil, Lepp \& Dalgarno~(1998) and Gay et al.~(2011) become
inaccurate below $T_{\rm g}\sim 10$~K, possibly affecting the resulting
H$_2^+$ abundance at the lowest redshifts.

\subsubsection{Photodissociation of H$_2^+$.}

This is the main destruction process for H$_2^+$ which determines
the history of H$_2$ molecules down to redshifts $z\sim 300$.
State-resolved cross sections have been calculated by Dunn~(1968),
Stancil, Babb \& Dalgarno~(1993), Lebedev, Presnyakov \& Sobel'Man~(2000), and
Hirata \& Padmanabhan~(2006). However, only the early results of
Dunn~(1968) are available in tabular form and have been adopted in
the work by Coppola et al.~(2011) on which the discussion of
Section~\ref{hydrogenchemistry} is based.

\subsubsection{Charge exchange of H$_2$ with H$^+$.}

This endoergic reaction has a threshold of 1.83~eV. The most complete
theoretical work on this chemical reaction is that of Krsti\'c~(2002).
The thermal rate coefficient has been computed by Savin et al.~(2004)
for H$_2(v=0,J=0)$. Glover \& Abel~(2008) discuss the large
uncertainties on this rate that exist in the literature, quoting
as extreme values the prescription by Shapiro \& Kang~(1987) based
on detailed balance from the reverse reaction, and the much lower
rate of Abel et al.~(1997), based on cross sections by Janev, Langer
\& Evans~(1987).  However, the latter is valid only at temperatures
higher than $\sim 10^4$~K, and should not be used at lower temperatures
where it exponentially drops to unrealistic values.  At temperatures
above $\sim 10^3$~K, the rate coefficient by Savin et al.~(2004)
is in excellent agreement with the fit by Galli \& Palla~(1998)
based on cross sections of Holliday, Muckerman \& Friedman~(1971).
At lower temperatures, the exponential drop due to the energy
threshold makes the uncertainty in the rate coefficient of no
consequence for primordial chemistry and collapse calculations.

\subsubsection{Charge exchange of H$_2^+$ with H.}

Also for this exoergic reaction, state-resolved cross sections have
recently been computed by Krsti\'c~(2002) for collision energies
in the range 0.6--9.5~eV, but need to be extrapolated to lower
energies for applications to primordial chemistry.  The extrapolation
is usually made in such a way as to reproduce the value of the
thermal rate of this reaction measured by Karpas, Anicich \& Huntress~(1979) at
$T_{\rm g}=300$~K (Coppola et al.~2011). Clearly, specific calculations
at low energy (and additional experimental measurements) are needed.

\subsubsection{Dissociative recombination of H$_2^+$.}

A wealth of theoretical and experimental data is available for the
dissociative recombination of H$_2^+$ and its isotopologues (see,
e.g., Schneider \& Suzor-Weiner~2002).  The level of experimental
accuracy made possible by the advent of storage-ring experiments
has encouraged a continuous refinement of the computational techniques
and the inclusion of previously neglected physical processes. Using
multi-quantum defect theory, Takagi~(2002) has calculated state-resolved
dissociative recombination cross sections that include the rotational
structure and all excited dissociative states. A relevant characteristic
is the presence of many resonances at low energies, and the sensitivity
of the cross section to the vibrational quantum number $v$, especially
at low energies.  As a result, the reaction rate computed by Coppola
et al.~(2012) in LTE has a non-monotonic behavior significantly
different from the usual power-law approximations commonly adopted
in the literature (see Figure~4).

\subsubsection{Three-body H$_2$ formation.}

In the low density gas of the expanding universe, three-body reactions
cannot take place.  However, in the collapsing clouds of the
primordial halos where the first generation of stars is thought to
occur, these reactions effectively transform the atomic gas into
fully molecular hydrogen at densities above $\sim 10^9$~cm$^{-3}$.
In addition, since each reaction liberates 4.48 eV of energy,
corresponding to the binding energy of H$_2$, into the gas, three-body
reactions represent an important energy source for the dynamical
evolution of the clouds. The main reaction for H$_2$ formation
involves atomic hydrogen as the third body, but a number of other
channels are also possible. Although there is consensus on the small
value of the rate coefficient of the reaction, the numerical values
quoted by different authors vary by up to two orders of magnitude
at the low temperatures of interest for cosmological simulations,
while there is substantial agreement (to within a factor of $\sim$2)
at temperatures greater than $\sim 2000$~K  (see, e.g., Glover \&
Savin~2009).

Current estimates of the three body reaction rates are based on
laboratory measurements of the reverse reaction, collisional
dissociation of H$_2$ by hydrogen atoms, using the principle of
detailed balance (e.g., Palla, Salpeter \& Stahler~1983; Flower \& Harris 2007).
Similar, although smaller, differences exist in the rate coefficients
for reactions where the third body is an H$_2$ molecule. The source
of the discrepancy is attributed by Flower \& Harris~(2007) to
different assumptions made for the equilibrium constant.  However,
including the nuclear degeneracy, as well as the electron spin
degeneracy, in the calculation removes the discrepancy and reduces
the rate to the values adopted by Palla, Salpeter \& Stahler~(1983) (F.~Esposito,
priv. comm.).  An even lower value of the rate was proposed by Abel
et al.~(2002), based on the calculations of Orel (1987) at temperatures
$<300$~K and assuming an inverse temperature dependence at higher
temperatures. From the theoretical viewpoint, Esposito \&
Capitelli~(2009) have presented a unifying approach that includes
the contributions from quasibound and unbound states by means of
the classical orbiting resonance theory (e.g., Orel 1987) and from
continuum states through directly calculated dissociation with
detailed balance application.  Their total recombination rate shows
a dependence on the temperature to pressure ratio, but it is in
fairly good agreement with the estimates of Abel, Bryan \& Norman~(2002)
at low temperatures.

Performing direct experiments of three-body association has proved
extremely difficult in the laboratory due to the high densities and
temperatures required.  Savin~(2013) gives an account of the
challenges of such experiments and concludes that the situation is
not too promising in the short term. Turk et al.~(2011) have performed
an analysis of the effects of the current uncertainty in the rate
using 3D simulations of the collapse of primordial clouds in the
high density regime. The results clearly show how sensitive the
outcome is to the precise value and temperature dependence on the rate,
concluding that ``the uncertainty represents a major limitation on
our ability to accurately simulate the formation of the first stars
in the universe''.  Clearly, this is a field that urgently requires
some basic technological or methodological breakthrough.

\subsubsection{Radiative association of H$_3^+$.}

Although the reaction
\be
{\rm H}_2+{\rm H}^+\rightarrow {\rm H}_3^+ + h\nu
\ee
is the main route for H$_3^+$ formation, the rate coefficient has
an uncertainty of four orders of magnitude. The study of Gerlich
\& Horning~(1992) based on an ion trap measurement and a classical
trajectory analysis recommends the constant value of $1\times
10^{-16}$~cm$^3$~s$^{-1}$.  This value has been adopted in most
chemical models (e.g., Schleicher et al.~2008; Glover \& Savin~2009). 
However, Stancil, Lepp \& Dalgarno~(1998), on the basis of quantal
calculation of other diatomic molecules, argue that the rate should
be much smaller and indicate a value of $1\times 10^{-20}$
~cm$^3$~s$^{-1}$ also adopted in the recent compilation of Gay et
al.~(2011). The evolution of H$_3^+$ is clearly affected by such
a large uncertainty. If instead of the large rate, the value suggested
by Stancil, Lepp \& Dalgarno~(1998) had been adopted in the chemical model
discussed in Section~\ref{hydrogenchemistry}, the resulting abundance
of H$_3^+$ would decrease by three orders of magnitude with respect
to that shown in Figure~3a. Thus, this reaction needs to be further
investigated before the effective role of H$_3^+$ can be elucidated.

\subsubsection{Dissociative recombination H$_3^+$ and H$_2$D$^+$.}

The ion storage ring measurements of McCall et al.~(2004) and Kreckel
et al.~(2005) have settled the issue of the rate coefficient of
this most relevant reaction. Previously, a large uncertainty existed
between the results of merged beam experiments with typical values
of $\sim 10^{-7}$~cm$^3$~s$^{-1}$ at room temperatures (e.g.,
Sundstr\"om, Mowat \& Danared~1994) and of flowing afterglow experiments with
values about ten times smaller (e.g., Smith \& {\v S}panel~1993).
More recent experiments (e.g., Glos\'ik et al. 2008), as well as
refined theoretical calculations have converged towards the results
of McCall et al.~(2004) with the remaining disagreement below a
factor of 2.  In the evolution presented here, we have followed
McCall et al.~(2004) using a branching ratio of 0.66 for the
recombination into ${\rm H}+{\rm H}+{\rm H}$ and of 0.34 into ${\rm
H}_2 + {\rm H}^+$. As for H$_2$D$^+$, the experiments of Datz et
al.~(1995) and Larsson et al.~(1996) still remain the main source
for the determination of the rate coefficient with branching ratios
of 0.73 for the channel ${\rm H}+{\rm H}+{\rm D}$, 0.20 for ${\rm
HD}+{\rm H}$, and 0.07 for ${\rm H_2}+{\rm D}$.

\subsubsection{Isotopic exchange of HD.}

In spite of being the most important channel for the destruction
of HD, only sparse laboratory data exist for reaction (\ref{d3}).
In addition, all the experiments are limited to temperatures greater
than $\sim 200$~K (e.g., Galli \& Palla~ 2002), thus confirming the
plea made long ago by Shavitt ~(1959) for more experimental work
on such a basic reaction.

\subsubsection{Transfer reaction of HeH$^+$.}

Reaction~(\ref{hehph}) is the main destruction channel of HeH$^+$
in the redshift interval $300\lesssim z \lesssim 10$. Prior to
Bovino et al.~(2011c), it was not studied with sufficient accuracy
from a theoretical point of view.  The crossed-beam data for energies
higher than 0.2~eV (Rutherford \& Vroom~1973) were fitted by Linder, Janev 
\& Botero~(1995) and adopted in the chemical models of Schleicher et
al.~(2008).  Bovino et al.~(2011c) present new ab initio quantum
calculations for this process for collision energies from $10^{-4}$~eV
to 1~eV.  The reaction rate is displayed in Figure~4. The marked
drop of the thermal rate below $T_{\rm g}\approx 100$~K is due to
the non-Langevin behavior of the cross sections at energies below
$\sim 10^{-2}$~eV.

\subsubsection{Radiative association of ${\rm LiH}$.}

The cross sections and rates for the radiative association from the
Li($^2P$) state have been computed by Gianturco \& Gori Giorgi~(1997).
As for the $^2S$ state, using semi-classical arguments, Lepp \&
Shull~(1984) estimated a radiative association rate $\sim
10^{-17}$~cm$^3$~s$^{-1}$, a value confirmed by Khersonskii \&
Lipovka~(1993).  Subsequent quantal calculations performed by two
groups on the basis of independent sets of potential curves and
wavefunctions (Dalgarno, Kirby \& Stancil~1996; Gianturco \& Gori
Giorgi~1996) provided a value of the rate smaller than the previous
results by about three orders of magnitude.  Semiclassical calculations
by Talbi \& Bacchus-Montabonel~(1998) and Bacchus-Montabonel \&
Talbi~(1999) were found in agreement with the quantal results. The
inclusion of quasi-bound levels with radiative widths exceeding
their tunneling widths (Bennett et al.~2003) resulted in an increase
of the rate by a factor of $\sim 2$.  The rates of radiative
association of LiH from the $^2S$ and $^2P$ states are shown in
Figure~4 as function of temperature.

\section{COOLING OF THE PRIMORDIAL GAS}
\label{cooling}

Cooling and thermal balance of the primordial gas are key ingredients
in the formation of the first structures, as they determine the
characteristic fragmentation scale of the clouds and the magnitude
of the accretion flow onto the growing protostars (e.g., Abel et
al~2002; Yoshida et al.~2006).  While chemistry determines the
abundances of the main molecular constituents of the gas, the
microphysics of each individual component defines the cooling
efficiency of the gas. In the case of the homogeneous expanding
universe, the thermal balance of the gas is simply set by adiabatic
expansion since molecules do not form in sufficient fraction to
effectively couple the matter and radiation fields (see
Section~\ref{chemicalevolution}).  However, in the collapsing gas
of the first minihalos or larger halos, the density and temperature
increase considerably allowing the excitation of more molecular
degrees of freedom. Thus, a detailed knowledge  of the main cooling
agents is needed to determine the thermal state of the gas.

\subsection{Molecular Cooling}

In order to evaluate the cooling function at each redshift, the
radiative and collisional transition probabilities and the frequencies
of the rovibrational transitions must be specified for each molecule.
These data can be found in the literature not only for the most
relevant molecules, like H$_2$ and HD, but also for the other
species, such as H$_2^+$, H$_3^+$, HD$^+$, H$_2$D$^+$, HeH$^+$,
LiH, and LiH$^+$. Then, for each molecule, one must solve the level
populations from the rate equations. The studies by Glover \&
Abel~(2008) and Glover \& Savin~(2009) provide detailed discussions
for each species. More recently, Coppola, Lodi \& Tennyson~(2011)
has computed updated radiative cooling functions valid in the
high-density regime where LTE conditions prevail. One should notice
that accurate collisional coefficients are available only in limited
cases, and for the minor species simple approximations are used in
the evaluation of the cooling functions. We summarize in {\bf
Figure~5} the individual cooling functions for the most relevant
species in the low-density case ({\em a}\/) and in LTE conditions
({\em b}\/), respectively.

\begin{figure}[t!] 
\begin{center}
\resizebox{6.6cm}{!}{\includegraphics{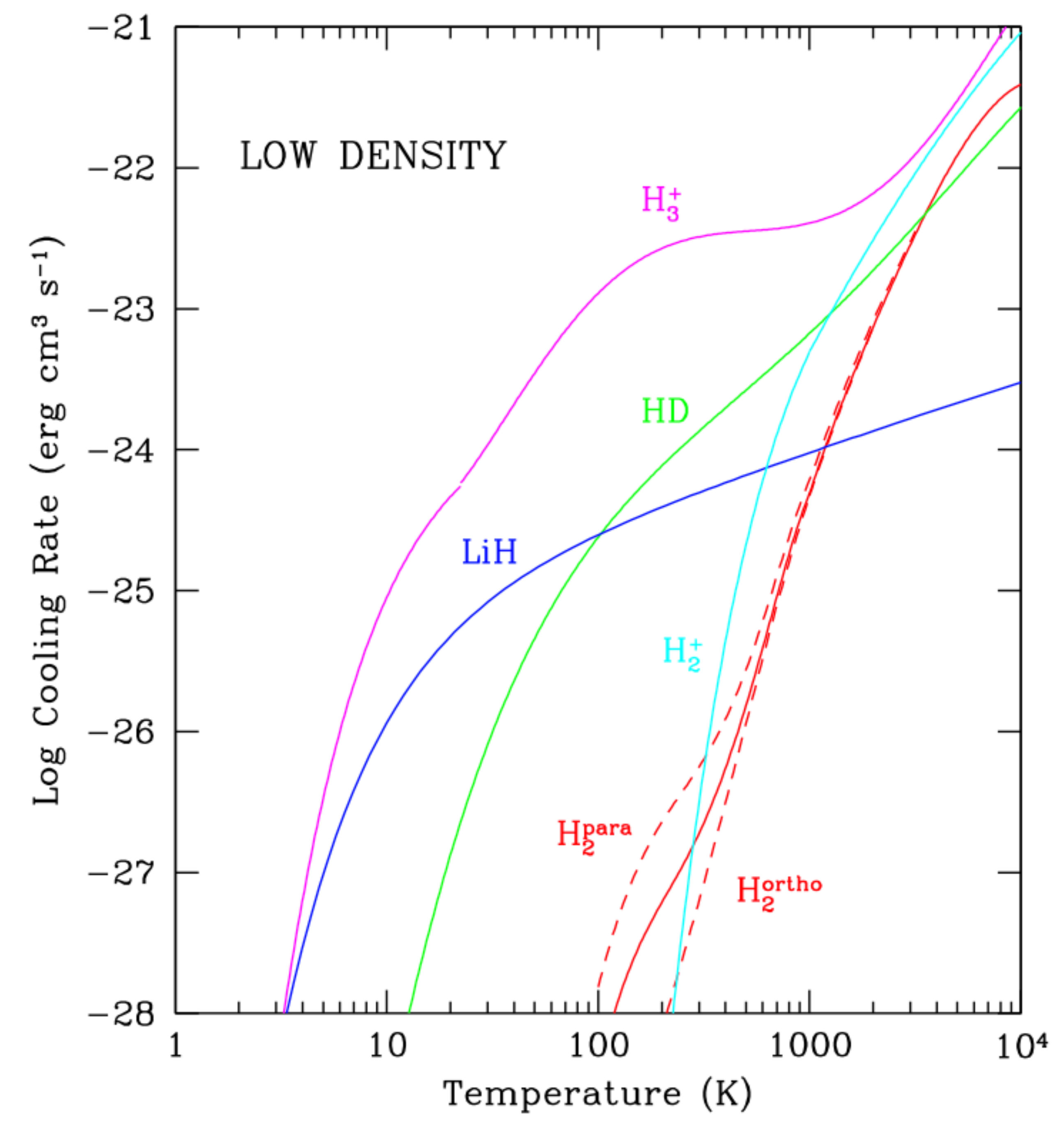}} 
\resizebox{6.6cm}{!}{\includegraphics{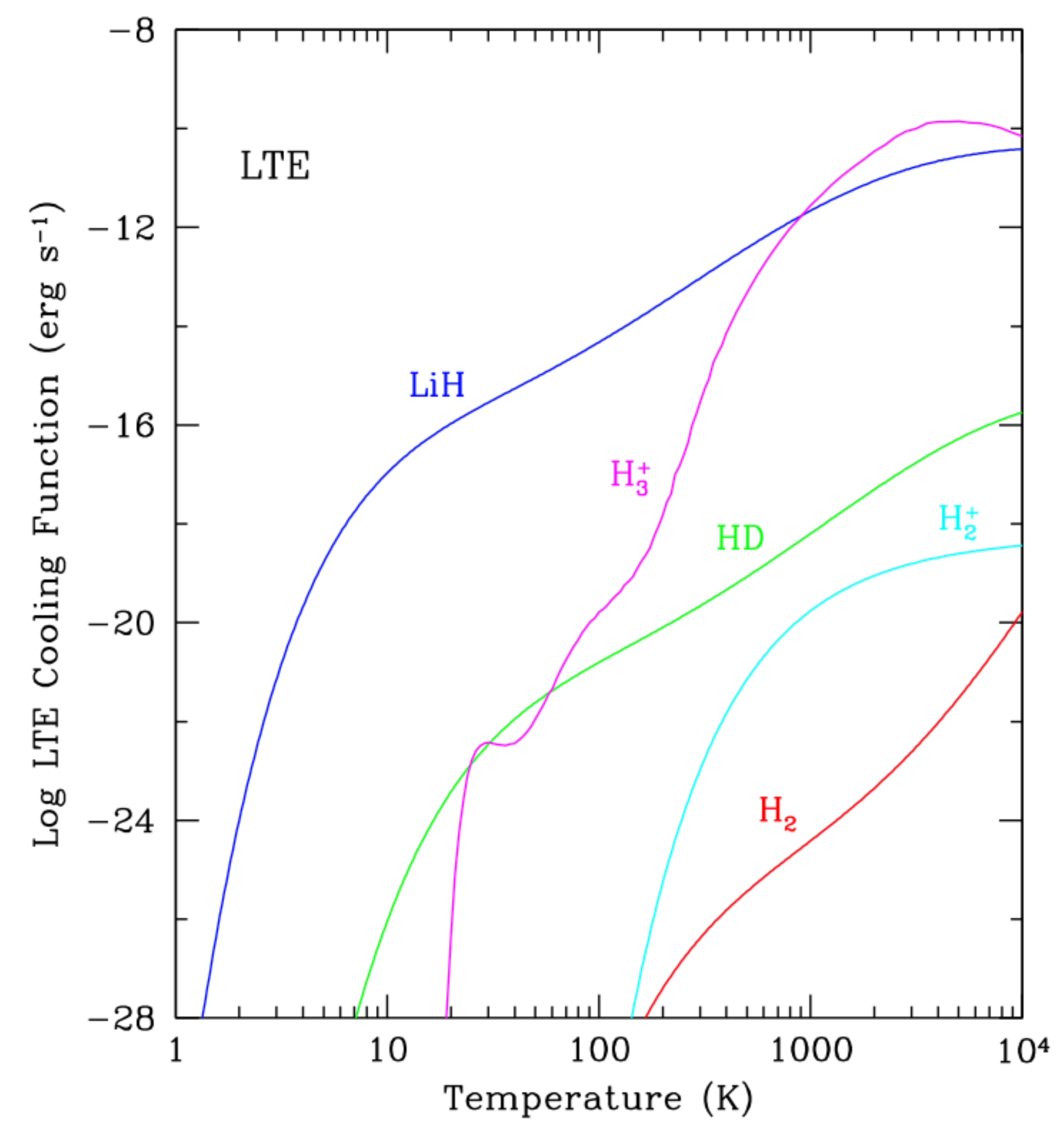}} 
\caption{Cooling rate/function for H$_2$, H$_2^+$, H$_3^+$, HD, and
LiH in the low-density ({\em a}\/) and LTE ({\em b}\/) limits. In
the low-density case, we plot the cooling rate of ortho-H$_2$,
para-H$_2$ (dashed curves), and total H$_2$ (solid curve) assuming
an ortho/para ratio of 3:1. Sources for the individual low-density
cooling rates are: Glover \& Abel~(2008) for H$_2$; Suchkov \&
Shchekinov~(1978) and Glover \& Savin~(2009) for H$_2^+$; Glover
\& Savin~(2009) for H$_3^+$; 
Lipovka, N\'u\~nez-L\'opez \& Avila-Reese~(2005) for HD;  Bougleux
\& Galli~(1998) for LiH.  Sources for the LTE cooling functions
are: Coppola et al.~(2012) for H$_2$; Glover \& Abel~(2008) for
H$_2^+$; Glover \& Savin~(2009) for H$_3^+$; Coppola et al.~(2011)
for HD and LiH. }
\end{center} 
\label{fig_cool} 
\end{figure}

\subsubsection{H$_2$ cooling.}

In spite of its dominant role, H$_2$ has two basic limitations: the
excited ro- and vibrational transitions have small radiative
transition probabilities (radiative lifetimes $\gtrsim 10^6$~s) and
collisional deexcitation becomes competitive with spontaneous
transitions at rather low densities ($n\sim 10^4$~cm$^{-3}$). Thus,
as the density increase the cooling rate of H$_2$ reaches the LTE
value which depends only on the transition probabilities and the
energy levels. Since the former are small, the cooling rate is also
small.  The second effect is the onset of optical depth in the cores
of the rovibrational lines that effectively reduces the cooling
efficiency and requires the solution of the appropriate radiative
transfer equations. Detailed calculations have shown that the cooling
by H$_2$ molecules is effectively suppressed at densities $n\gtrsim
10^{10}$~cm$^{-3}$ (e.g., Yoshida et al.~2006). Another important
aspect is the H$_2$ ortho/para ratio, since the equilibrium value
of 3:1 is achieved only at relatively high temperature and densities.
Indeed, in the expanding low-density universe, the H$_2$ formation
process can decrease the LTE value for abundances $f_{{\rm H}_2}
\lesssim 10^{-6}$. Glover \& Abel~(2008) have shown that for
H$^+$/H$\gtrsim 10^{-4}$ and temperatures lower than several hundred
K, the ortho/para ratio is controlled by collisions with protons,
so that the cooling rate is mainly due to para-H$_2$ ($J=2$--0,
$\Delta E/k=510$~K), although most of the gas contains ortho-H$_2$
($J=3$--1, $\Delta E/k=845$~K).

In the low-density limit, the total cooling is the sum of individual
contributions due to the collisional partners of H$_2$: H, H$_2$,
He, electrons, and protons. Glover \& Abel~(2008) provide a thorough
discussion and empirical prescriptions of the various terms. Some
recent improvements concern the H$_2$--H$_2$ collisions,
discussed by Lee et al.~(2008, see also Balakrishnan et al.~2011)
who provide updated collisional coefficients based on the potential
energy surface of Diep \& Johnson~(2000).

\subsubsection{HD cooling.}

Due to small but non zero dipole moment ($8.3\times 10^{-4}$~debyes),
HD has radiative lifetimes that are a factor of $\sim 100$ shorter
than those of H$_2$, so that the cooling rate reaches its LTE limit
only at densities $n\sim 10^6$~cm$^{-3}$. In addition, unlike H$_2$
molecules, transitions with $\Delta J=1$ are permitted, allowing
cooling to occur through the $J=1$--0 transition (corresponding to
$\Delta E/k=128$~K). Finally, due to the exothermicity of
reaction~(\ref{d2}) by 462~K, HD is highly fractionated at low
temperatures, thus boosting the relevance of this molecule as an
important coolant for the gas at the lowest temperatures.  The
cooling function computed by Lipovka, N\'u\~nez-Lop\'ez 
\& Avila-Reese~(2005) for HD-H collisions
for $1 \lesssim n \lesssim 10^8$~cm$^{-3}$ and $100 \lesssim T_{\rm
g} \lesssim10^4$~K, still remains the main reference. In fact, it
is in very good agreement with the results of Wrathmall, Gusdorf \& Flower~(2007)
who have improved the interaction potential and included a larger
set of rovibrational levels. In the LTE limit, Coppola, Lodi \&
Tennyson~(2011) find a slightly larger (factor of $\sim 2$) cooling
rate for temperatures $T_{\rm g}\gtrsim 700$~K due to the contribution
of highly excited rovibrational levels.  Collisions of HD with He
and electrons have been studied by Nolte et al.~(2012) and Yoon et
al.~(2010), respectively. However, to our knowledge, the corresponding
cooling functions have yet to be evaluated.

\subsubsection{${\rm LiH}$ and ${\rm LiH}^+$ cooling.}

The very large dipole moment (5.89~debyes) of LiH makes this molecules
an excellent radiator, and potentially important as a coolant of
the primordial gas.  However, its low abundance renders its
contribution to the total cooling rate insignificant in the expanding
universe while it may play a role in collapse calculations. The
cooling function for collisions with H has been computed by Bougleux
\& Galli~(1997, fitted by Galli \& Palla~1998). These results are
valid below a critical density of $n_{\rm cr}\approx 10^{12}$~cm$^{-3}$
(Lepp \& Shull~1984), and are still commonly used in the literature.
At densities larger than $n_{\rm cr}$, the radiative cooling rate
is provided by Coppola, Lodi \& Tennyson~(2011) who computed updated
potential energy and dipole moment curves.  As for LiH$^+$, whose
abundance is slightly smaller than that of LiH, only the LTE cooling
rate is available (Coppola, Lodi \& Tennyson~2011). Its value is a
factor of 10--100 smaller than that of LiH, making LiH$^+$ a
negligible coolant in cosmological conditions.

\subsubsection{Minor species.}

Cooling by H$_3^+$ has been the subject of a detailed analysis by
Glover \& Savin~(2009) and Miller et al.~(2010).  The former find
that H$_3^+$ is the third important coolant after H$_2$ and HD in
dense ($n\approx 10^7$--$10^9$~cm$^{-3}$) primordial gas. In this
density regime, the consequences of the large uncertainty on the
rate of radiative association of H$_3^+$ discussed in
Section~\ref{criticalrates} are not as strong as those on the
expanding medium, because other channels dominate the formation of
this ion.  As for H$_2^+$ and HD$^+$, Glover \& Savin~(2009) give
cooling functions both in the low- and high-density regime. In the
latter case, however, Coppola, Lodi \& Tennyson~(2011) find a HD$^+$
cooling rate larger by orders of magnitude than that of Glover \&
Savin~(2009) at temperatures $\lesssim 500$~K. Finally, the LTE
cooling rate of HeH$^+$ and its isotopologues has been computed by
Coppola, Lodi \& Tennyson~(2011) (see also Miyake \& Stancil~2007).
Due to is large dipole moment (1.66~debyes, Pavanello et al.~2005),
the LTE cooling rate is estimated to be about ten orders of magnitude
grater than that of H$_2$. No calculations are available for the
low-density regime.

\subsection{Chemical and Thermal Evolution of Collapsing Primordial Clouds}

\begin{figure}[t!] 
\begin{center}
\resizebox{10cm}{!}{\includegraphics{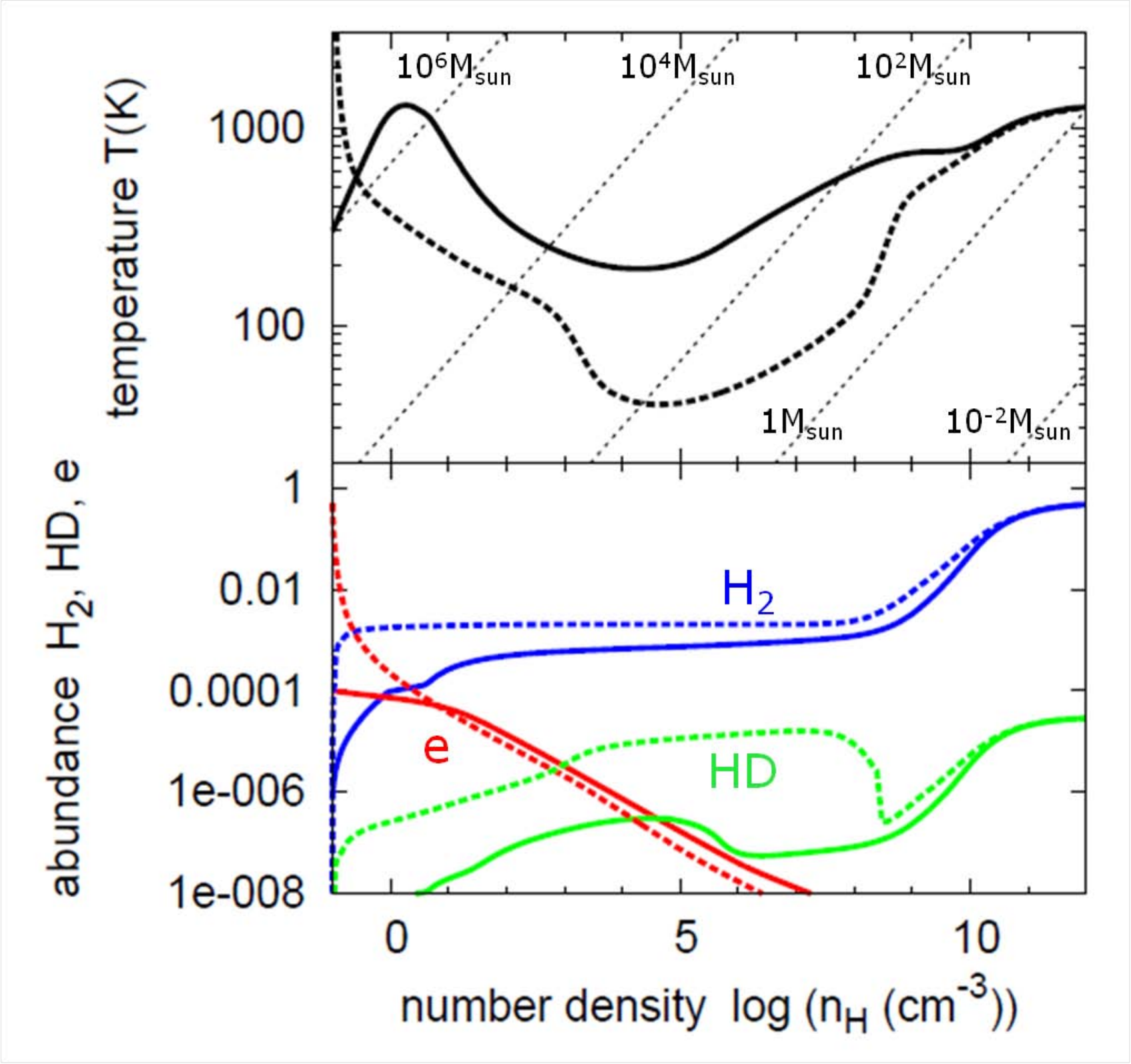}} 
\caption{Thermal ({\em upper}\/) and chemical ({\em lower}\/) evolution of  
collapsing primordial gas clouds. The solid lines refer to the case of  
collapse in a predominantly neutral minihalo, while the dashed lines  
display the evolution of gas which is pre-ionized and cools from high  
temperatures. Lines of constant Jeans mass are shown as dotted lines  
in the upper panel. (Courtesy of K.~Omukai).}
\end{center} 
\label{fig_kazu} 
\end{figure}

The well constrained initial conditions provided by the standard
$\Lambda$CDM model (see Section~\ref{cosmo}) and the knowledge of
the atomic and molecular processes in the primordial gas allow the
investigation of the evolution of dark matter structure and the
formation of the first objects. Current consensus is that the first
generation of stars started to take place at redshifts $z\approx
20$--30 in dark matter halos of mass $\sim 10^6$~M$_\odot$ (called
minihalos) and preceded the formation of the first galaxies (e.g.
Bromm 2012). In these minihalos, star formation could proceed thanks
to the cooling provided by H$_2$ molecules (Couchman \& Rees 1986;
Tegmark et al. 1997) in a predominantly atomic gas and without the
complications of physical processes (viz., turbulence, magnetic
fields, etc.) that characterize the environment of present-day
molecular clouds (however, for a different view see, e.g., Schleicher
et al.  2009; Clark et al.  2011; Turk et al. 2012).

Along with the prevailing minihalos, the $\Lambda$CDM model predicts
the presence of more rare, massive halos with mass $\sim 10^8$~M$_\odot$
and characteristic temperature of $\sim 10^4$~K, too hot for the
survival of molecular hydrogen. In this case, it is believed that
the gas can cool via atomic hydrogen (Oh \& Haiman~2002), but
fragmentation is effectively suppressed with the possible formation
of massive/supermassive black holes (Bromm \& Loeb~2003; Hosokawa
et al.~2012).  However, a situation in which molecular cooling
becomes relevant is when the gas is reionized and recombines out
of equilibrium, as in expanding H{\sc ii} regions and intergalactic
shocks (e.g. Shapiro \& Kang~1987).  The resulting enhanced ionization
fraction promotes the efficient formation of H$_2$ via the H$^-$
channel and HD molecules through D$^+$ via reaction~(\ref{d2}).  If
the HD abundance becomes larger than a critical value, the gas can
cool below the floor temperature of $\sim 200$~K produced by H$_2$,
and even approach the minimum value set by the CMB (Johnson \&
Bromm~2006; Yoshida, Omukai \& Hernquist~2007; McGreer \& Bryan~2008).

The thermal and chemical evolution of the collapsing primordial gas
can be understood following the results of simplified 1-D calculations
that capture the essential microphysics.  {\bf Figure~6} illustrates
the two cases discussed above: a predominantly neutral medium in a
minihalo and a pre-ionized medium.

\subsubsection{Formation of first stars in minihalos.}

As shown in Figure~6 (solid curves), due to the low initial fractional
H$_2$ abundance ($f_{{\rm H}_2} \lesssim 10^{-6}$), the gas collapses
almost adiabatically and H$_2$ formation proceeds mainly via the
H$^-$ channel. At relatively low densities ($n \approx 10$~cm$^{-3}$),
the fractional abundance rises to $f_{{\rm H}_2}\approx 10^{-3}$
which is sufficient to greatly enhance the cooling efficiency via
rovibrational transitions that effectively contrasts the gravitational
compression of the gas and produces the marked drop in temperature
shown in Figure~6 up to densities $n\approx 10^4$~cm$^{-3}$. The
minimum temperature of $ \sim 200$~K sets a characteristic mass for
fragmentation of
\be
M_J \approx 1800\, {\rm M}_\odot \left(\frac{n}{10^4~\rm{cm}^{-3}}\right)^{-1/2} 
\left(T_{\rm g} \over{200~\rm{K}}\right)^{3/2}.
\label{jeansmass}
\ee
This mass scale is a robust result, rather insensitive to variations
in the initial conditions.  At higher densities, however, H$_2$
molecules reach LTE conditions and line cooling saturates. The gas
temperature rises slowly with density, while the fractional H$_2$
and HD abundances remain basically constant at a value of $f_{{\rm
H}_2}\approx 10^{-3}$ and $f_{\rm HD}\approx 10^{-7}$.  The major
change occurs at densities higher than $\sim 10^8$~cm$^{-3}$ when
three body reactions begin to convert the prevailing atomic hydrogen
into H$_2$ molecules, which partly offset the rise in temperature.
The increased H$_2$ abundance drives the accelerated formation of
HD molecules. At densities higher than $\sim 10^{12}$~cm~$^{-3}$,
the cooling is governed by H$_2$ collision-induced emission and
dissociation. Once H$_2$ molecules are almost completely destroyed,
the gas temperature increases adiabatically and pressure forces
eventually succeed in effectively halting gravitational collapse
and a protostellar core is formed that accretes rapidly matter from
the surrounding cloud.

The final outcome of the collapse and the typical masses of the
first stars are still subject to uncertainty. While for a certain
time it seemed that the first stars were predominantly very massive
with characteristic masses $\gtrsim 100$~M$_\odot$ (e.g., Abel et
al. 2002; Bromm, Coppi \& Larson~2002; Yoshida et al. 2006), more
recent calculations including feedback effects have shown that the
growth of protostars at the very high accretion rates ($\dot{\rm
M}\gtrsim 10^{-3}$~M$_\odot$~yr$^{-1}$) expected due to the lack
of coolants can be limited to $\lesssim 50$~M$_\odot$ for a wide
range of initial conditions (McKee \& Tan 2008; Hosokawa et al.
2011). High-resolution 3-D simulations have also shown that multiple
fragments can develop during collapse and circumstellar disk
formation, thus leading to the formation of binary or small multiple
systems or even small clusters of stars of typically low mass (Clark,
Glover \& Klessen~2008; Clark, Glover \& Smith~2011; Turk, Abel \& 
O'Shea~2009; Stacy, Greif \& Bromm~2010). Considering the sensitivity of the accretion
history on the thermal  and dynamical state of the parent gas cloud
and on the physical characteristics of the minihalos, it is likely
that collapse and fragmentation of the primordial gas will produce
a first generation of stars with a substantial distribution in mass.

\subsubsection{Formation of stars from pre-ionized gas.}

The thermo-chemical evolution of primordial gas (see Figure~6,
dashed curves) differs significantly from that described above,
especially in the initial phases of rapid cooling and enhanced H$_2$
formation that results from out-of-equilibrium recombination of
electrons with protons. In these conditions, the higher ionization
fraction promotes the rapid formation of H$_2$ molecules that reach
fractional abundances of about $5\times 10^{-3}$ at densities
$\lesssim 1$~cm$^{-3}$, a factor of $\gtrsim 50$ higher than in the
neutral case.  As a response, the gas drops to temperatures well
below $\sim 150$~K, the threshold value for HD formation (Omukai
et al.~2005).  More HD molecules are readily formed, up to $f_{\rm
HD}\approx 10^{-5}$, while the gas cools to $\sim 30$~K at densities
$\sim 10^4$~cm$^{-3}$, so that the characteristic mass given by
eq.~(\ref{jeansmass}) becomes $\sim 100$~M$_\odot$. The evolution
at higher densities is similar to what described before. The main
difference is the predicted lower masses that would result from the
fragmentation of primordial gas into smaller clouds of typical
solar/subsolar mass (e.g., Nakamura \& Umemura~2002; Nagakura \&
Omukai~2005; Johnson \& Bromm~2006).

\section{INTERACTIONS OF MOLECULES WITH THE CMB}
\label{interactions}

The study of the interaction of primordial molecules with the CMB
is important in view of the possibility to observationally detect
spectral distortions or spatial anisotropies imprinted by molecular
transitions on CMB photons as they travel from the last scattering
surface to a present-day observer. Before the COBE mission (Smoot
et al.~1992), much of the interest about primordial molecules as a
source of opacity was motivated by the lack of evidence for spatial
anisotropies in the CMB, leaving the possibility that an absorbing
layer of molecules extending over a significant range of redshifts
could smear out any primary anisotropy produced at higher redshift
and possibly generate detectable secondary spatial/spectral features
(Dubrovich~1977,1994; Maoli, Melchiorri \& Tosti~1994; Maoli et
al.~1996; Puy \& Signore~2002).

Starting from the late 70s, the main effect investigated was resonant
scattering by molecules in protogalactic clouds moving at some
peculiar velocity with respect to the Hubble flow (Dubrovich~1977),
a process analogous to the Thomson scattering of CMB photons by
free electrons (Sunyaev \& Zel'dovich~1970). Resonant scattering
by molecules, like the Sunyaev-Zel'dovich effect, is an elastic
process in which a low energy photon ($h\nu\ll m_e c^2$) is absorbed
and then re-emitted at the same frequency but in a different
direction, isotropically in the rest frame of the absorber but
anisotropically in the observer's frame. If the scattering particles
have a density $n_{\rm s}$ and a peculiar velocity $v_{\rm pec}$ along
the line of sight (positive away from the observer), the induced
spectral/spatial anisotropy in the CMB temperature to the lowest
order in $v_{\rm pec}/c$ is
\be
\frac{\Delta T_{\rm r}}{T_{\rm r}} = - 
\frac{v_{\rm pec}}{c}(1-e^{-\tau_\nu})
\label{thom}
\ee
where 
\be
\tau_\nu=\int n_{\rm s}\sigma_\nu\, d\ell,
\label{tau}
\ee
is the optical depth of the protogalactic cloud.  For free electrons,
$\sigma_\nu$ is independent of frequency and equal to the Thomson
cross section $\sigma_{\rm T}=8\pi e^4/(3m_e^2 c^4)$, whereas for
electrons bound in molecules the scattering is frequency-dependent
and resonant at the frequency of a molecular transition: for a
single transition with frequency $\nu_{ul}$ between two levels with
density populations $n_{\rm l}$, $n_{\rm ui}$, statistical weights
$g_{\rm l}$, $g_{\rm u}$ and Einstein coefficient $A_{\rm ul}$, the
absorption cross section is
\be
\sigma_\nu=\frac{c^2}{8\pi\nu^2} 
A_{\rm ul}\left(1-\frac{g_{\rm l}n_{\rm u}}{g_{\rm u}n_{\rm l}}\right)
\frac{g_{\rm u}}{g_{\rm l}}\phi(\nu-\nu_{\rm ul})
\ee
where $\phi(\nu-\nu_{\rm ul})$ is the line profile. The important
fact here is that the absorption cross section of electrons bound
in molecules is several orders of magnitude larger than the Thomson
cross section of free electrons. Resonant scattering can also be
greatly enhanced by luminescence, i.e. the absorption of a high
energy photon corresponding to a rovibrational transition followed
by the re-emission of several lower energy photons (Dubrovich \&
Lipovka~1995). Since at $z\approx 300$--500 the frequencies of
rotational transitions fall in the Rayleigh-Jeans region of the CMB
whereas the frequencies of rovibrational transitions fall around
the peak, where the photon density is maximum, luminescence can
result in an increase of $\Delta T_{\rm r}/T_{\rm r}$ at low
frequencies by orders of magnitudes for molecules possessing
low-energy rotational transitions like LiH and H$_2$D$^+$
(Dubrovich~1997).

The limits of integrations in equation~(\ref{tau}) depend on the
size of the protogalactic cloud where molecules are located. In the
case of a uniform distribution of molecules in the primordial gas,
with the abundance depending on redshift as described in the previous
sections, is convenient to transform the integration over space in
an integration over redshift, $d\ell=c\, dz/H(z)$, where $H(z)$ is
the Hubble parameter defined by equation~(\ref{hz}).  In practice, the
absorption (or emission) of photons observed today with frequency
$\nu$ occurs in a small range of redshifts $\Delta z_{\rm i} \approx
(1+z_{\rm i})\Delta\nu_{\rm th}/\nu_{\rm ul}$ around the interaction
redshift $z_{\rm i}$ defined by the condition $\nu(1+z_{\rm
i})=\nu_{\rm ul}$. Since spectral lines expected to arise in the
dark ages are narrow ($\Delta\nu_{\rm th}/\nu_{\rm ul}\ll 1$), the
region contributing to the optical depth is also small ($\Delta
z_{\rm i}/z_{\rm i}\ll 1$) and can be approximated by a Dirac delta
in evaluating the integral in equation~(\ref{tau}) (for details, see
e.g. Bougleux \& Galli~1997).

\subsection{Optical Depth}
\label{opticaldepth}

In a cold gas made mainly of H and He very few atomic transitions
can be excited at low temperatures. A notable exception is the
hyperfine 21~cm transition of neutral hydrogen, with an energy
difference between the $F=1$ and $F=0$ levels corresponding to
0.068~K. This spin-flip transition is expected to have resonantly
absorbed CMB photons in the post-recombination era, producing a
broad absorption feature at frequencies of $\sim 10$~MHz (Varshalovich
\& Khersonskii~1977, see also Field~1959). However, the detection
of this feature is severely hindered, at least from the ground, by
ionospheric distortion and reflection of radio waves and by the
Galactic synchrotron emission that dominates the CMB by several
orders of magnitude in this frequency range (Jester \& Falcke~2009).
The sensitivity of this transition to a variety of processes that
can affect the temperature evolution of the gas before and during
reionization, when the 21~cm signal (either in absorption or in
emission) is redshifted to more convenient frequency bands around
$\sim 100$~MHz, has made this line a prime target for the new
generation of radio interferometers like LOFAR and SKA (for a recent
review of 21~cm line cosmology, see Pritchard \& Loeb~2012).

In addition to the 21~cm line of neutral H, the 6708~\AA\ resonance
transition between the ground state $^2S$ and the first excited
state $^2P$ of neutral Li can also provide a source of opacity of
the primordial gas, as first proposed by Loeb~(2001).  Stancil et
al.~(2002) found that this process could result in a possibly
detectable suppression of the power spectrum amplitude of the CMB.
However they did not consider the dramatic impact of non-thermal
photons on the recombination history of Li, as discussed in
Section~{lithiumrecombination}.  When this effect is taken into
account, the abundance of neutral Li decreases by orders of magnitude,
and the optical depth of the Li 6708~\AA\ line is reduced to only
a few $10^{-5}$ (Switzer \& Hirata~2005).

Unlike atoms, molecules possess rovibrational transitions that can
be excited at the low temperatures characteristic of the pregalactic
gas. Early calculations of the contribution of primordial molecules
to the opacity of the universe were heavily affected by large
uncertainties on the fractional abundance of the species considered.
Dubrovich~(1977) computed the optical depth of the lowest rotational
transitions of HD$^+$, LiH, para- and ortho-H$_2$D$^+$ and found
that the latter could reach unity at cm wavelengths if ${\rm H}_2{\rm
D}^+/{\rm H}\approx 10^{-6}$. Special attention was given to LiH,
because of its high dipole moment (5.89~debyes, Wharton, Gold \&
Klemperer~1960), and therefore its strong rotational and rovibrational
transitions (see, e.g., Zemke \& Stwalley~1980; Bellini et al.~1994;
Gianturco et al.~1996). In addition, because of the lack of accurate
quantal rates for the radiative association of LiH, during the 1980s
and early 1990s the conversion of Li into LiH was expected to be
almost complete, as described in Section~\ref{lithiumchemistry}.

Recently, Black~(2006) has computed the opacity associated with
bound-free photodetachment transitions of the negative hydrogen ion
H$^-$ on the CMB, and found an optical depth of more than $10^{-5}$
at mm wavelengths, which would have observable effects on the CMB.
Following Black's suggestion, Schleicher et al.~(2008) analyzed
both the bound-free and the free-free absorption of H$^-$, including
the corrections for stimulated and spontaneous emission, and found
that the optical depth is dominated by the free-free process for
frequencies below $\sim 100$~GHz, with a $\nu^{-2}$ frequency
dependence, and by the bound-free process at higher frequencies.
The value of the optical depth however is only $\sim 10^{-9}$ at
mm wavelengths, much lower than the estimate by Black~(2006). In
any case, the negative hydrogen ion remains the main source of
opacity in the dark ages at frequencies below $\sim 30$~GHz, whereas
at higher frequencies the optical depth is dominated by the
contribution of rotational transitions of HeH$^+$ and the $^2S$--$^2P$
transition of Li.

{\bf Figure~7} shows the optical depth as function of frequency of
H (21~cm line), Li (6708~\AA\ line), H$^-$, HeH$^+$, HD$^+$, LiH
and H$_2$D$^+$, computed with the abundances shown in Figures~3a--d.
From lower to higher frequencies, the optical depth is dominated
by pure rotational transitions and rovibrational transitions
(free-free and bound-free transitions, respectively, in the case
of H$^-$). For LiH and H$_2$D$^+$ only the contribution of pure
rotational transitions is present in this range of frequencies.

\begin{figure}[t!] 
\begin{center}
\resizebox{10cm}{!}{\includegraphics{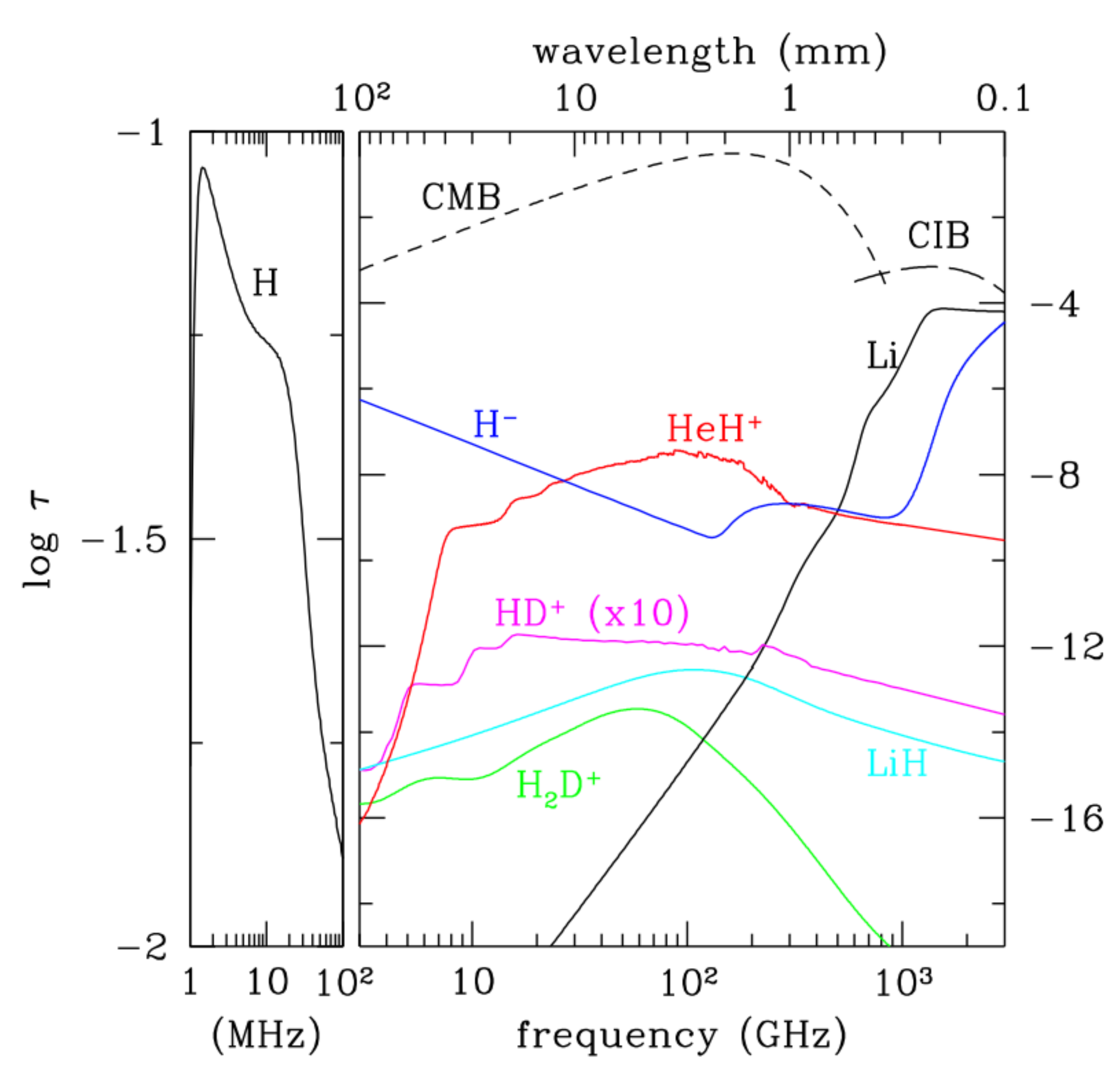}} 
\caption{Redshift-integrated optical depth for several primordial atoms,
molecules and ions as function of frequency/wavelength. For clarity,
the optical depth due to the 21~cm line of neutral hydrogen is shown
on a separate panel, and the contribution of HD$^+$ has been displaced
upwards by a factor of 10. For comparison, the intensity of the CMB and
the Cosmic Infrared Backround (CIB) are also shown in arbitrary units.
Details of the calculations for each species can be found in Palla, Galli
\& Silk~(1995), Bougleux \& Galli~(1997), Palla \& Galli~(2000), Switzer \&
Hirata~(2005), Schleicher et al.~(2008). }
\end{center} 
\label{fig_tau} 
\end{figure}

\subsection{Spectral Distortions}
\label{spectraldistortions}

Besides the smallness of the optical depth of the main molecular
species, another limitation to the detection of spectral distortions
in the CMB induced by resonant molecular scattering is the low value
of the peculiar velocity $v_{\rm pec}$ with respect to the sound
speed. The magnitude of the peculiar velocity has a maximum value
of $\sim 900$~km~s$^{-1}$ for protogalactic clouds of masses
$10^9$--$10^{11}$~M$_\odot$ and is lower for larger or smaller
masses (see e.g. Maoli et al.~1996). It also decreases with redshift
as $(1+z)^{-1/2}$. Thus, $v_{\rm pec}/c \lesssim 10^{-3}/(1+z)^{1/2}$.
Contributions to $v _{\rm pec}$ can also come from internal
hydrodynamical flows, like gravitational collapse. The maximum
collapse velocity is in this case of the order of the free-fall
speed, or about 100--300~km~s$^{-1}$ in the same mass range.  Finally,
the observation of spectral distortions induced by resonant scattering
is limited to regions of relatively low densities: above some
critical density line emission will start to dominate scattering,
erasing any decrement of the CMB brightness temperature produced
by scattering protoclouds (Basu~2007). For example, densities of
about $0.1$-1~cm$^{-3}$ are sufficient to erase the scattering
signature produced by the lowest rotational transitions of HD and
HD$^+$.

The intensity and angular size of emission/absorption lines produced by
various molecules during the evolution of primordial density perturbations
with masses ranging from $10^9$ to $10^{12}$~$M_\odot$ were computed
by Maoli, Melchiorri \& Tosti~(1994), Maoli et al.~(1996). They found that the very initial phases
of nonlinear collapse provide the best conditions for the detection of
resonant scattering.  In fact, at the moment when the cloud collapse
is exactly compensated by the cosmological expansion (turnaround), the
line width reaches the minimum (i.e.  thermal) value, and the absorption
cross section is the highest (Zel'dovich~1978).  The signal thus changes
from a relatively low-intensity broad line in the linear phase to an
intermediate-intensity line of growing broadening during nonlinear
collapse, passing through a high-intensity thin peak at turnaround.

These studies prompted a few attempts to detect spectral signatures of
primordial molecules. The first one, performed at mm wavelengths
with the IRAM radiotelescope, consisted in a search for redshifted
rovibrational transitions of LiH emitted by protogalaxies at $z\approx
100$--200 (de Bernardis et al.~1993). These observations gave upper
limits of $\sim 20$~mK on the intensity of LiH lines, only slightly
above the expected intensity of LiH for optical depths of order
unity, a value that was later found largely overestimated.  Using
a chemical network based on the reaction rates for Li chemistry
known at the time, Palla, Galli \& Silk~(1995) found LiH abundances of the
order of  $10^{-14}$--$10^{-13}$ and a value of only $\sim 10^{-5}$
for the optical depth of LiH at mm wavelengths. Bellini et al.~(1994)
suggested to extend the search at cm wavelengths, where the optical
depth of redshifted pure rotational transitions of the $v=0$ level
would be about two orders of magnitude larger. However, no results
were obtained by Gosachinskij et al.~(2002) in their search for
spectral signatures of LiH at a wavelength of 6.2~cm using the
RATAN-600 radio telescope.  Using more reliable abundances, the
optical depth of LiH was found to be of the order of only $\sim
10^{-11}$ and $\sim 10^{-13}$ at cm and mm wavelengths, respectively
(Bougleux \& Galli~1997).  The search for HD lines appears more
promising, as the optical depths of the three lowest pure rotational
lines are much higher than for LiH, although their intensities are
still below the sensitivities of current facilities like GMT/LMT,
CARMA and ALMA (N\'u\~nez-L\'opez, Lipovka \& Avila-Reese~2006).  Recently, Persson
et al.~(2010) have reported the results of broad band, blind line
surveys with the {\it Odin} satellite to search for the lowest
rotational transitions of H$_2$, as well as higher transitions of
HD and HeH$^+$ redshifted from $z\approx 20$--30, the epoch of the
first structure formation.  Although no lines were detected, the
search set some interesting limits on the amplitudes of the possible
resonant lines and can serve as a pilot study for current and future
radiotelescopes.

\subsection{Spatial Anisotropies}
\label{spatialanisotropies}

In addition to spectral features, resonant scattering from primordial
molecules can also produce secondary spatial anisotropies in the
CMB, leading to changes in the expansion coefficients $C_l$ of the
CMB power spectrum
\be
\frac{\Delta C_l}{C_l}\approx -2\tau_\nu,
\ee 
for $\tau_\nu \ll 1$ and $l\gg 1$ (Maoli, Melchiorri \& Tosti~1994; Basu,
Hern\'andez-Monteagudo \& Sunyaev~2004; Basu~2007).  Thus, a promising
way to detect secondary anisotropies due to resonant scattering is
to compare the expansion coefficients of the angular power spectrum
$C_l$ at different wavelengths to detect signals of $\sim 0.1$~$\mu$K
or less, exploiting the sensitivity of satellites like {\it Planck},
and other ground-based or balloon-borne instruments like the Atacama
Cosmological Telescope (ACT), Atacama Pathfinder Experiment Telescope
(APEX) and the South Pole Telescope (SPT).  However, since the
angular scales of the expected anisotropies are small, of the order
of a few arcmin at best (Maoli et al.~1996), only multipole
coefficients with $l\gtrsim 1000$ will be affected, making the
detection difficult.  A possible experimental setup for the Multi-Beam
Solar Radio Telescope (MSRT, Tuorla, Finland) is described in
Dubrovich, Bajkova \& Khaikin~(2008).

\subsection{Emission by Molecular Formation}

In analogy with H and He recombination, molecule formation also
occurs in excited states that then decay to the fundamental level
emitting photons that produce an excess over the pure Planck
distribution of the CMB. For example, H$_2$ is formed by associative
detachment in vibrationally excited states that quickly decay to
the ground state (see, Section~\ref{hydrogenchemistry}). Since the
maximum production of H$_2$ occurs at redshifts $z\lesssim100$, the
emission of rovibrational transitions with $\lambda \approx 2$~$\mu$m
is redshifted today at $\lambda\lesssim 200$~$\mu$m in the Wien
part of the CMB.  Coincidentally, this is the same wavelength range
where photons produced by Ly-$\alpha$ and two-photon decay of H and
He during recombination are also redshifted (Section~\ref{recombination}).
The H$_2$ emission has been computed by Coppola et al.~(2012, see
also Khersonskii~1982 and Shchekinov \& Entel~1984) for multiquantum
transitions up to $\Delta v=4$. As in the case of H and He recombination
photons, a direct detection of the H$_2$ formation emission in this
wavelength range appears challenging, owing to the presence of an
infrared background (both Galactic and extragalactic) several orders
of magnitude brighter than the CMB (see e.g.  discussion in Wong, Seager 
\& Scott~2006).  However, the excess photons produced by H$_2$ formation
could significantly affect a number of photodestruction rates at
high redshift.

\subsection{Absorption by Thermal Nonequilibrium}

At redshift below $z\approx 500$ matter and radiation are no longer
efficiently coupled by Compton scattering, and the temperatures of
baryon and photons decrease approximately according to the adiabatic
laws $T_{\rm r}\propto (1+z)^{-1}$ and $T_{\rm g}\propto (1+z)^{-2}$,
respectively (see Section~\ref{recombination}).  In these circumstances,
molecules act as heat pumps, transferring energy from the hotter
radiation to the colder gas (Khersonskii~1986; Puy et al.~1993).
The effect on the CMB temperature is of the order of
\be 
\frac{\Delta T_{\rm r}}{T_{\rm r}}\approx 
-\left(1-\frac{T_{\rm ex}}{T_{\rm r}}\right)\tau_\nu,
\ee 
where $T_{\rm ex}$ is the excitation temperature of the transition,
defined by the condition $g_{\rm l}n_{\rm u}/g_{\rm u}n_{\rm
l}=\exp(-h\nu_{\rm ul}/kT_{\rm ex})$. In a collapsing cloud, when
the density becomes larger than a critical density for thermal
equilibrium with collisions, $T_{\rm ex}$ becomes equal to $T_{\rm
g}$,  and the latter rapidly increases above $T_{\rm r}$. In this
case line emission start to dominate (e.g., Mizusawa et al.~2005,
Basu~2007). On the other hand, in the expanding primordial gas after
recombination, when radiative effects still largely dominate over
collisional excitation/de-excitation, the excitation temperature
of a molecular transition is very close to the temperature of the
CMB, $T_{\rm g} < T_{\rm ex}\lesssim T_{\rm r}$.  This explains the
limited absorbing power of a layer of primordial molecules and the
major obstacle for their detection: those species with high dipole
moments, and thus high absorption cross sections, have $T_{\rm
ex}\approx T_{\rm r}$ , while species with low dipole moments have
$\tau_\nu \ll 1$. Puy et al.~(1993) considered only pure rotational
transitions in the $v=0$ level of H$_2$, HD and LiH and found
negligible effects of molecular absorption on the CMB: $\Delta
T_{\rm r}/T_{\rm r}\lesssim 10^{-9}$ at $\nu\sim 40$--$80$~GHz for
rotational transitions of H$_2$ and even less for rotational
transitions of HD and LiH.  Also in the case of H$^-$, the relative
change in the CMB temperature due to the bound-free and free-free
processes discussed above is small, because the absorption and the
emission processes are close to equilibrium, making the net effect
of the order of only $\sim 10^{-10}$ over a frequency range from
100 to 1000~GHz (Schleicher et al.~2008).  As already mentioned in
Section~\ref{opticaldepth}, the 21~cm hyperfine transition of neutral
hydrogen represents an exception to this behavior.  Because of its
extremely small Einstein coefficient ($A_{10}=2.85\times
10^{-15}$~s$^{-1}$), collisions are able to couple $T_{\rm ex}$
(called in this case ``spin temperature'') to $T_{\rm g}$ down to
$z\approx 100$, when the Hubble time becomes longer than the lifetime
of the $F=1$ level and $T_{\rm ex}$ approaches $T_{\rm r}$ until
reionization occurs (Purcell \& Field~1956; Pritchard \& Loeb~2008).

\section{SUMMARY POINTS}
\label{summary}

\begin{itemize}

\item{Assuming that cosmology and standard Big Bang nucleosynthesis
have entered the precision era, the dawn of chemistry can now be
accurately followed from the recombination epoch to the end of the
dark ages.  Chemical networks containing more than $\sim 250$
reactions for $\sim 30$ species have been developed to make
quantitative predictions on their abundances. Most of the remaining
uncertainty can be attributed to the partial knowledge of cross
sections and rate coefficients at the relevant energies and
temperatures.}

\item{Due to the lack of dust grains, primordial chemistry is driven
by slow gas-phase reactions promoted by collisions with electrons
and protons left over from recombination at a level of one part
in $\sim 10^4$.  As a result, only simple diatomic and triatomic
molecules and molecular ions containing H, D, He, and Li are formed
in trace amounts.}

\item{ The low density of the expanding medium and the harsh radiation
field due to the cosmic background limit the maximum production of
each species. While H$_2$ and HD reach freeze-out abundances of
$\lesssim 10^{-6}$ and $\gtrsim 10^{-10}$, respectively, at $z\approx
40$, all other molecules/ions have abundances much smaller than
$10^{-10}$.}

\item{The evolution of H$_2$, H$_2^+$, H$_3^+$ and their deuterated
isotopologues depends to a large extent on the additional contribution
to the cosmic radiation field of photons produced by the recombination
of H and He.  These photons also reduce the abundance of H$^-$ and
Li: the former limits the freeze-out fraction of H$_2$, the latter
never recombines.}

\item{Primordial molecules allow the gas to cool, contract, and
fragment. While the answer to fundamental questions regarding the
formation of the first structures and stars (fragmentation scale,
mass spectrum, maximum mass, etc.) does not depend on chemistry
alone, a full understanding of the thermodynamical properties of
the baryonic matter is a fundamental ingredient in all numerical
simulations addressing these issues. It is rewarding that detailed
chemical networks and accurate cooling functions are now integral
parts of state-of-the-art simulations.}

\item{The interaction of the CMB with the primordial gas is probably
the best probe to test the predictions of primordial chemistry and
provides a unique way to peer into the dark ages prior to structure
formation.  The effects of absorption, emission and resonant
scattering on the temperature and power spectrum of the CMB have
been fully explored theoretically.}

\item{The most promising results are obtained for the negative
hydrogen ion H$^-$ and the HeH$^+$ molecular ion.  The free-free
process of H$^-$ leads to a frequency-dependent change in the power
spectrum of the CMB of the order of $10^{-7}$ at 30~GHz. HeH$^+$
can efficiently scatter CMB photons and smear out primordial
fluctuations in the frequency range 30--300~GHz, leading to a change
in the power spectrum of the order of $10^{-8}$.  Finally, resonant
scatter by neutral Li dominates the optical depth of the universe
in the far-infrared.}

\end{itemize}

\section{FUTURE ISSUES}
\label{future}

\begin{itemize}

\item{Major progress has been achieved by the inclusion of
state-resolved chemical reaction rates for H$_2$ and H$_2^+$, that
are formed out of equilibrium in highly-excited vibrational levels
and then decay radiatively to the fundamental state. It is desirable
to extend this approach to other species like HD, HD$^+$ and HeH$^+$.
Further improvement comes from recent fully-quantal ab initio
calculations of reaction cross sections for He and Li chemistry
specifically performed for applications to the early universe.
Extending these calculations to other species/processes is an obvious
need. It is necessary to complement these theoretical studies with
laboratory experiments, especially at sub-eV energies.}

\item{While excellent agreement between theory and experiments has
been reached for some key reactions, most notably the dissociative
attachment of H$^-$ leading to H$_2$ formation and the dissociative
recombination of H$_3^+$, a number of other important reactions
still suffer from either poor knowledge or inconsistencies.  For
example, in the high density phase of collapse within dark matter
minihalos, the conversion to fully molecular gas by three-body
reactions is still affected by significant uncertainties. We have
provided an assessment for this and several other critical
reactions.}

\item{Searches for signals from primordial molecules and for their
imprint on the CMB are very challenging, in part owing to the extreme
weakness of the expected lines and also because of the poorly
constrained redshift of emission. However, the exploration of the
dark ages using molecular and atomic transitions is already under way with 
the largest available facilities
like ALMA and LOFAR, and dedicated satellites like {\it Planck}.
We can hope that with the advent of next-generation telescopes
(e.g., JWST, SKA, ELTs, etc.) some fundamental breakthrough can be
achieved that will shed light on the dawn of chemistry and its
impact on the birth of the first structures.}

\end{itemize}

\section{ACKNOWLEDGMENTS}

It is a pleasure to thank Stefano Bovino, Carla Maria Coppola,
Fabrizio Esposito, Franco A. Gianturco, Savino Longo, Dominik
Schleicher who have contributed so much over the years to a fruitful
and enjoyable collaboration.  We also acknowledge exciting conversations
on primordial chemistry and star formation with John Black, Simon
Glover, Takashi Hosokawa, Ralf Klessen, Kazuyuki Omukai, Daniel
Savin, and Raffaella Schneider. We thank Fabio Iocco and Ofelia
Pisanti for computing BBN abundances. Finally, we are grateful to
Ewine van Dishoeck, Amiel Sternberg and Malcolm Walmsley for
critically reading an earlier version of this manuscript and providing
comments that improved it.


\begin{thebibliography}{}

\bibitem[]{} Abel T, Anninos P, Zhang Y, Norman, ML. 1997. {\em New Astron.} 2:181 
\bibitem[]{} Abel T, Bryan GL, Norman, ML. 2002. {\em Science} 295:93
\bibitem[]{} Ali-Ha{\"i}moud Y, Hirata CM. 2010. {\em Phys. Rev. D} 82:063521 
\bibitem[]{} Ali-Ha{\"i}moud Y, Hirata CM. 2011. {\em Phys. Rev. D} 83:043513 
\bibitem[]{} Alizadeh E, Hirata C. 2011. {\em Phys. Rev. D} 84:083011
\bibitem[]{} Asplund M, Lambert DL, Nissen P, Primas F, Smith VV. 2006. {\em Ap. J.} 644:229
\bibitem[]{} Aver E, Olive KA, Skillman ED. 2010. {\em J. Cosmol. Astrop. Phys.} 5:3
\bibitem[]{} Aver E, Olive KA, Skillman ED. 2012. {\em J. Cosmol. Astrop. Phys.} 4:4
\bibitem[]{} Ayouz M, Lopes R, Raoult M, Dulieu O, Kokoouline, V. 2011. {\em Phys. Rev. Lett.} 83:052712
\bibitem[]{} Bacchus-Montabonel MC, Talbi, D. 1999. {\em J. Molec. Struc.} 463:91
\bibitem[]{} Balakrishnan N, Qu{\'e}m{\'e}ner G, Forrey RC, Hinde RJ, Stancil PC. 2011. {\em J. Chem. Phys.} 134:014301 
\bibitem[]{} Bania TM, Rood RT, Balser DS. 2010. In {\em Light Elements in the Universe}, eds. C Charbonnel et. al., IAU Symp. 268, p. 81
\bibitem[]{} Basu D, Barua AK. 1984. {\em J. Phys. B} 17:1537 
\bibitem[]{} Basu K, Hern{\'a}ndez-Monteagudo C, Sunyaev RA. 2004. {\em Astron. Astrophys.} 416:447
\bibitem[]{} Basu K. 2007. {\em New Astr. Rev.} 51:431
\bibitem[]{} Bates DR. 1951. {\em MNRAS} 111:303
\bibitem[]{} Bellini M, de Natale P, Inguscio M, Fink E, Galli D, Palla F. 1994. {\em Ap. J.} 424:507
\bibitem[]{} Bennett OJ, Dickinson AS, Leininger T, Gad{\'e}a FX. 2003. {\em MNRAS} 341:361 (Erratum in 2008. {\em MNRAS} 384:1743)
\bibitem[]{} Black, JH. 1990. In {\em Molecular Astrophysics}, ed. TW Hartquist (Cambridge: University Press), p.473
\bibitem[]{} Black, JH. 2006. {\em Faraday Discuss.} 133:27  
\bibitem[]{} Bodo E, Gianturco FA, Martinazzo, R. 2003. {\em Phys. Rep.} 384:85
\bibitem[]{} Bonifacio P, Sbordone L, Caffau E, Ludvig HG, Spite M, et al. 2012. {\em Astron. Astrophys.} 542:A87
\bibitem[]{} Bougleux E, Galli D. 1997. {\em MNRAS} 288:638
\bibitem[]{} Bovino S, Wernli M, Gianturco, FA. 2009. {\em Ap. J.} 699:383 
\bibitem[]{} Bovino S, Stoecklin T, Gianturco FA. 2010a. {\em Ap. J.} 708:1560 (Erratum in 2010. {\em Ap. J.} 713:711)
\bibitem[]{} Bovino S, Tacconi M, Gianturco FA, Stoecklin T. 2010. {\em Ap. J.} 724:126
\bibitem[]{} Bovino S, Tacconi M, Gianturco FA. 2011a. {\em Ap. J.} 740:101 (Erratum in 2012. {\em Ap. J.} 748:150)
\bibitem[]{} Bovino S, Tacconi M, Gianturco FA. 2011b. {\em Phys. Scr.} 84:028103 
\bibitem[]{} Bovino S, Tacconi M, Gianturco FA, Galli D. 2011c. {\em Astron. Astrophys.} 529:A140 
\bibitem[]{} Bovino S, Tacconi M, Gianturco FA, Galli D, Palla F. 2011d. {\em Ap. J.} 731:107 
\bibitem[]{} Bovino S, {\v C}ur\'{\i}k R, Galli D, Tacconi M, Gianturco FA. 2012. {\em Ap. J.} 752:19 
\bibitem[]{} Bromm V. 2012. {\em arXiv1203.3824}
\bibitem[]{} Bromm V, Loeb A. 2003. {\em Nature} 425:812
\bibitem[]{} Bromm V, Loeb A. 2003. {\em Ap. J.} 596:34
\bibitem[]{} Bromm V, Yoshida N. 2011. {\em Annu. Rev.}{\em Astron. Astrophys.} 49:373
\bibitem[]{} Bromm V, Coppi PS, Larson RB. 2002. {\em Ap. J.} 564:23
\bibitem[]{} Bromm V, Yoshida N, Hernquist L, McKee CF. 2009. {\em Nature} 459:49 
\bibitem[]{} Bruhns H, Kreckel H, Miller KA, Urbain X, Savin DW. 2010. {\em Phys. Rev. A} 82:042708-1
\bibitem[]{} Caffau E, Bonifacio P, Fran{\c c}ois P, Spite M., Spite F., et al. 2011. {\em Nature} 477:67 
\bibitem[]{} Caffau E, Bonifacio P, Fran{\c c}ois P, Spite M., Spite F., et al. 2012. {\em Astron. Astrophys.} 542:A51
\bibitem[]{} Carlberg RG. 1981. {\em MNRAS} 197:1021
\bibitem[]{} Cassisi S, Castellani V. 1993. {\em Ap. J.}S 88:509
\bibitem[]{} Chandrasekhar S. 1958. {\em Ap. J.} 128:114
\bibitem[]{} Charutz DM, Last I, Baer M. 1997. {\em J. Chem. Phys.} 106:7654
\bibitem[]{} Chluba J, Thomas RM. 2011. {\em MNRAS} 412:748
\bibitem[]{} Chluba J, Vasil GM, Dursi LJ. 2010. {\em MNRAS} 407:599
\bibitem[]{} Chluba J, Fung J, Switzer ER. 2012. {\em MNRAS} 423:3227
\bibitem[]{} Christlieb N, Bessell MS, Beers TC., Gustafsson B, Korn A, et al. 2002. {\em Nature} 419:904
\bibitem[]{} C\'{\i}zek M, Hor\'acek J, Domcke W. 1998. {\em J. Phys. B} 31:2571
\bibitem[]{} Clark PC, Glover SCO, Klessen RS. 2008. {\em Ap. J.} 672:757
\bibitem[]{} Clark PC, Glover SCO, Klessen RS, Bromm V. 2011. {\em Ap. J.} 727:110 
\bibitem[]{} Clark PC, Glover SCO, Smith RJ. 2011. {\em Science} 331:1040
\bibitem[]{} Coc A, Goriely S, Xu Y, Saimpert M, Vangioni E. 2012. {\em Ap. J.} 744:158 
\bibitem[]{} Combes F, Wiklind T. 1998. {\em Astron. Astrophys.} 334:L81 
\bibitem[]{} Cooke R, Pettini M, Steidel CC. 2011. {\em MNRAS} 417:1534
\bibitem[]{} Coppola CM, Lodi L, Tennyson J. 2011. {\em MNRAS} 415:487
\bibitem[]{} Coppola CM, Longo S, Capitelli M, Palla F, Galli D. 2011. {\em Ap. J. Suppl.} 193:7 
\bibitem[]{} Coppola CM, D'Introno R, Galli D, Tennyson J, Longo S. 2012. {\em Ap. J.}S 199:16 
\bibitem[]{} Couchman HMP, Rees MJ. 1986. {\em MNRAS} 221:53
\bibitem[]{} {\v C}ur\'{\i}k R, Greene CH. 2007. {\em Phys. Rev. Lett.} 98:173201
\bibitem[]{} {\v C}ur\'{\i}k R, Greene CH. 2008. {\em J. of Phys. Conf. Ser.} 115:012016 
\bibitem[]{} de Bernardis P, Dubrovich VK, Encrenaz P, Maoli R, Masi S, et al. 1993. {\em Astron. Astrophys.} 269:1
\bibitem[]{} Dalgarno A, Lepp S. 1987. In {\em Astrochemistry}, eds. MS Vardya \& SP Tarafdar (Dordrecht: Reidel), p. 109
\bibitem[]{} Dalgarno A, Kirby K, Stancil PC. 1996. {\em Ap. J.} 458:397
\bibitem[]{} Dalgarno A, van der Loo MPJ. 2006. {\em Ap. J.} 646:L91
\bibitem[]{} Dalgarno A, Weisheit JC, Black JH. 1973. {\em Astrophys. Lett.} 14:77
\bibitem[]{} Datz S, Larsson M, Stromholm C, Sundstr\"om G, Zengin V, et al. 1995. {\em Phys. Rev. A} 52:2901 
\bibitem[]{} Diep P, Johnson JK. 2000. {\em J. Chem. Phys.} 112:4465 
\bibitem[]{} Dinerstein HL, Geballe TR. 2001. {\em Ap. J.} 562:515
\bibitem[]{} Dubrovich VK. 1977. {\em Sov. Astr. Lett.}. 3:181 (orig. 1977. {\em Pis'ma Astron. Zh.} 3:339)
\bibitem[]{} Dubrovich VK. 1993. {\em Sov. Astr. Lett.}. 19:53 (orig. 1993. {\em Pis'ma Astron. Zh.} 19:132)
\bibitem[]{} Dubrovich VK. 1994. {\em Astron. Astrophys.} Trans. 5:57
\bibitem[]{} Dubrovich VK. 1997. {\em Astron. Astrophys.} 324:27
\bibitem[]{} Dubrovich VK, Lipovka AA. 1995. {\em Astron. Astrophys.} 296:301
\bibitem[]{} Dubrovich V, Bajkova A, Khaikin VB. 2008. {\em New Astron.} 13:28 
\bibitem[]{} Dumitriu I, Saenz A. 2009. J. Phys. Conf. Ser. 194:152026 
\bibitem[]{} Dunn GH. 1968. {\em Phys. Rev.} 172:1 
\bibitem[]{} Ekstr\"om S, Coc A, Descouvemont P, Meynet G., Olive KA, et al. 2010. {\em Astron. Astrophys.} 514:A62
\bibitem[]{} Esposito F, Capitelli M. 2009. {\em J. Chem. Phys.} A 113:15307
\bibitem[]{} Fehsenfeld FC, Howard CJ, Ferguson EE. 1973. {\em J. Chem. Phys.} 58:5841
\bibitem[]{} Fendt WA, Chluba J, Rubi{\~n}o-Mart{\'{\i}}n JA, Wandelt BD. 2009. {\em Ap. J.}S 181:627
\bibitem[]{} Field GB. 1959. {\em Ap. J.} 129:563
\bibitem[]{} Fields BD. 2011. {\em Ann. Rev. of Nucl. Part. Sci.} 61:47 
\bibitem[]{} Fixsen DJ. 2009. {\em Ap. J.} 707:916 
\bibitem[]{} Flower DR, Roueff E. 1979. {\em Astron. Astrophys.} 72:361 
\bibitem[]{} Flower DR, Harris GJ. 2007. {\em MNRAS} 377:705 
\bibitem[]{} Frebel A, Johnson JL, Bromm V. 2007. {\em MNRAS} 380:L40
\bibitem[]{} Frebel A, Collet R, Eriksson K, Christlieb N, Aoki W. 2008. {\em Ap. J.} 684:588
\bibitem[]{} Friedel DN, Kemball A, Fields BD. 2011. {\em Ap. J.} 738:37 
\bibitem[]{} Fumagalli M, O'Meara JM, Prochaska JX. 2011. {\em Science} 334:1245 
\bibitem[]{} Galli D, Palla F. 1989. In {\em Astronomy, Cosmology and Fundamental Physics}, eds. M Caffo et al. (Dordrecht: Kluwer), p. 421
\bibitem[]{} Galli D, Palla F. 1998. {\em Astron. Astrophys.} 335:403
\bibitem[]{} Galli D, Palla F. 2000. In {\em The First Stars}, eds. A Weiss et al. (Berlin: Springer), p. 229
\bibitem[]{} Galli D, Palla F. 2002. {\em Planet. Sp. Sci.} 50:1197
\bibitem[]{} Garc\'{\i}a P\'erez AE, Aoki W., Inoue S, Ryan SG, Suzuki TK, Chiba M. 2009. {\em Astron. Astrophys.} 504:213
\bibitem[]{} Gay CD, Stancil PC, Lepp S, Dalgarno A. 2011. {\em Ap. J.} 737:44
\bibitem[]{} Gerlich D, 1982. In {\em Symposium on Atomic \& Surface Physics}, eds. W Lindinger et al. (Dordrecht: Kluwer), p. 304
\bibitem[]{} Gerlich D, Horning S. 1992. {\em Chem. Rev.} 92:1509
\bibitem[]{} Gerlich D, Jusko P, Rou{\v c}ka, {\v S}, Zymak I, Pla{\v s}il R, Glos\'{\i}k J. 2012. {\em Ap. J.} 749:22
\bibitem[]{} Gianturco FA, Gori~Giorgi P, Berriche H, Gadea FX. 1996. {\em Astron. Astrophys.}S 117:377
\bibitem[]{} Gianturco FA, Gori~Giorgi P. 1996. {\em Phys. Rev. A} 54:1
\bibitem[]{} Gianturco FA, Gori Giorgi P. 1997. {\em Ap. J.} 479:560
\bibitem[]{} Glassgold AE, Galli D, Padovani M. 2012. {\em Ap. J.} 756:157
\bibitem[]{} Glos\'{\i}k J, Korolov I, Pla{\v s}il R, Novotny O, Kotrik T, et al. 2008. {\em J. Phys. B}. 41:191001
\bibitem[]{} Glover SCO, Abel T. 2008. {\em MNRAS} 388:1627
\bibitem[]{} Glover SCO, Savin DW. 2009. {\em MNRAS} 393:911 
\bibitem[]{} Glover SCO, Savin DW, Jappsen AK. 2006. {\em Ap. J.} 640:553
\bibitem[]{} Gosachinskij IV, Dubrovich VK, Zhelenkov SR, Il'in GN, Prozorov VA. 2002. {\em Astr. Rep.} 46:543 
\bibitem[]{} Grin D, Hirata CM. 2010. {\em Phys. Rev. D} 81:083005 
\bibitem[]{} Guberman SL. 1994. {\em Phys. Rev. A} 49:4277 
\bibitem[]{} Holliday MG, Muckerman JT, Friedman L. 1971. {\em J. Chem. Phys.} 54:1058 
\bibitem[]{} Hirasawa T, Aizu K, Taketani M. 1969. {\em Prog. Theor. Phys.} 41:835
\bibitem[]{} Hirata CM, Padmanabhan N. 2006. {\em MNRAS} 372:1175 
\bibitem[]{} Hirata CM, Switzer ER. 2008. {\em Phys. Rev. D} 77:083007
\bibitem[]{} Hosokawa T, Omukai K, Yoshida N, Yorke HW. 2011. {\em Science} 334:1250
\bibitem[]{} Hosokawa T, Omukai K, Yorke HW. 2012. {\em Ap. J.} 756:93
\bibitem[]{} Howk JC, Lehenr N, Fields BD, Mathews GJ. 2012. {\em Nature} 489:121
\bibitem[]{} Hutchins JB. 1976. {\em Ap. J.} 191:375
\bibitem[]{} Iocco F, Mangano G, Miele G, Pisanti O, Serpico PD. 2009. {\em Phys. Rep.} 472:1 
\bibitem[]{} Izotov Y, Thuan TX. 2010. {\em Ap. J.} 710:L67
\bibitem[]{} Izotov Y, Thuan TX, Stasinska G. 2007. {\em Ap. J.} 662:15
\bibitem[]{} Janev RK, Langer WD, Evans K. 1987. {\em Springer Series on Atoms and Plasmas} (Berlin: Springer)
\bibitem[]{} Jester S, Falcke H. 2009. {\em New Astron.} 53:1
\bibitem[]{} Johnson JL, Bromm V. 2006, {\em MNRAS} 366:247
\bibitem[]{} Jones BJT, Wise RFG. 1985. {\em Astron. Astrophys.} 149:144
\bibitem[]{} Ju{\v r}ek M, {\v S}pirko V, Kraemer WP. 1995. {\em Chem. Phys.} 193:287
\bibitem[]{} Karpas Z, Anicich V, Huntress WT. 1979. {\em J. Chem. Phys.} 70:2877 
\bibitem[]{} Keisler R, Reichardt CL, Aird KA. 2011. {\em Ap. J.}, 743:28
\bibitem[]{} Khersonskii VK. 1982. {\em Ap. Space Sci.} 88:21 
\bibitem[]{} Khersonskii VK. 1986. {\em Astrophys.} 24:114 (orig. 1986. {\em Astrofizika} 24:191)
\bibitem[]{} Khersonskii VK, Lipovka AA. 1993. {\em Bull. Spec. Astrophys. Observ.} 36:88
\bibitem[]{} Kholupenko EE, Ivanchik AV, Balashev SA, Varshalovich DA. 2011. {\em MNRAS} 417:2417 
\bibitem[]{} Kimura M, Lane NF, Dalgarno A, Dixson RG. 1993. {\em Ap. J.} 405:801 
\bibitem[]{} Kirkman D, Tytler D, Suzuki N, O'Meara JM, Lubin D. 2003. {\em Ap. J. Suppl.} 149:1
\bibitem[]{} Klessen R, Glover SCO, Clark PC. 2012. {\em MNRAS} 421:3217
\bibitem[]{} Komatsu E, Smith KM, Dunkley J, Bennett CL, Gold B, et al. 2011. {\em Ap. J. Suppl.} 192:18 
\bibitem[]{} Kreckel H, Motsch M, Mikosch J., Glos\'{\i}k J, Pla{\v s}il R, et al. 2005. {\em Phys. Rev. Lett.} 95:3201
\bibitem[]{} Kreckel H, Bruhns H, C{\'{\i}}zek M., Glover SCO, Miller KA, et al. 2010. {\em Science} 329:69
\bibitem[]{} Krolik JH. 1990. {\em Ap. J.} 353:21
\bibitem[]{} Krsti\'c PS. 2002. {\em Phys. Rev. A} 66:042717 
\bibitem[]{} Larsson M, Lepp S, Dalgarno A, Str\"omholm C, Sundstr\"om G, et al. 1996. {\em Astron. Astrophys.} 309:L1
\bibitem[]{} Latter WB. 1989. {\em PhD Thesis}, Univ. of Arizona
\bibitem[]{} Latter WB, Black JH. 1991. {\em Ap. J.} 372:161
\bibitem[]{} Lebedev VS, Presnyakov LP, Sobel'Man II. 2000. {\em Astr. Rep.} 44:338 
\bibitem[]{} Lee TG, Balakrishnan N, Forrey RC, Stancil PC, Shaw G., et al. 2008. {\em Ap. J.} 689:1105 
\bibitem[]{} Lepp S, Shull JM. 1984. {\em Ap. J.} 270:578 
\bibitem[]{} Lepp S, Stancil PC, Dalgarno A. 2002. {\em J. Phys. B} 35:57 
\bibitem[]{} Linder F, Janev RK, Botero J. 1995. {\em Atomic and Molecular Processes in Fusion Edge Plasmas}, ed. RK Janev (New York: Plenum Press), p. 397
\bibitem[]{} Lipovka A, N\'u\~nez-L\'opez R, Avila-Reese V. 2005. {\em MNRAS} 361:850
\bibitem[]{} Liu XW, Barlow MJ, Dalgarno A, Tennyson J., Lim T., et al. 1997. {\em MNRAS} 290:L71 
\bibitem[]{} Loeb A. 2001. {\em Ap. J.} 555:L1 
\bibitem[]{} Loeb A. 2010. {\em How did the First Stars and Galaxies Form?} (Princeton: Princeton Univ. Press) 
\bibitem[]{} Longo S, Coppola CM, Galli D, Palla F, Capitelli M. 2011. {\em Rend. Fis. Acc. Lincei} 22:1
\bibitem[]{} Maoli R, Melchiorri F, Tosti D. 1994. {\em Ap. J.} 425:372
\bibitem[]{} Maoli R, Ferrucci V, Melchiorri F, Signore M, Tosti D. 1996. {\em Ap. J.} 457:1
\bibitem[]{} Martinez O, Yang Z, Betts NB, Snow TP, Bierbaum VM. 2009. {\em Ap. J.} 705:L172
\bibitem[]{} Matsuda T, Sat{\-o} H, Takeda H. 1969. {\em Prog. Theor. Phys.} 42:219 
\bibitem[]{} Matsuda T, Sat{\-o} H, Takeda H. 1971. {\em Prog. Theor. Phys.} 46:416
\bibitem[]{} McCall BJ, Huneycutt AJ, Saykally RJ, Djuric N, Dunn GH, et al. 2004. {\em Phys. Rev. A} 70:052716
\bibitem[]{} McCrea WH. 1960. {Proc. Roy. Soc.} A 256:245
\bibitem[]{} McCrea WH, McNally D. 1960. {\em MNRAS} 121:238 
\bibitem[]{} McDowell MRC. 1961. {\em Observatory} 81:240
\bibitem[]{} McGreer ID, Bryan GL. 2008. {\em Ap. J.} 685:8
\bibitem[]{} McKee CF, Tan JC. 2008. {\em Ap. J.} 681:771
\bibitem[]{} Michael JV, Fisher JR. 1990. {\em J. Phys. Chem.} 93:3318
\bibitem[]{} Mielke SL, Lynch GC, Truhlar DG, Schwenke DW. 1994. {\em J. Phys. Chem.} 98:8000
\bibitem[]{} Miller KA, Bruhns H, Eli\'ai{\v s}ek J, {\v C}\'{\i}{\v z}ek M, Kreckel H, et al. 2011. {\em Phys. Rev. A} 84:e2709 
\bibitem[]{} Miller S, Stallard T, Melin H, Tennyson J. 2010. {\em Faraday Discuss.} 147:283 
\bibitem[]{} Mizusawa H, Omukai K, Nishi R. 2005. {\em Publ. Astron. Soc. Japan} 57:951 
\bibitem[]{} Mitchell DN, LeRoy DJ. 1973. {\em J. Chem. Phys.} 58:3449
\bibitem[]{} Miyake S, Stancil PC. 2007. {\em Bull. AAS} 39:139.09 
\bibitem[]{} Miyake S, Gay CD, Stancil PC. 2011. {\em Ap. J.} 735:21
\bibitem[]{} Moorhead JM, Lowe RP, Wehlau WH, Maillard JP, Bernath PF. 1988. {\em Ap. J.} 326:899 
\bibitem[]{} Moseley J, Aberth W, Peterson JR. 1970. {\em Phys. Rev. Lett.} 24:435 
\bibitem[]{} Nagakura T, Omukai K. 2005. {\em MNRAS} 364:1378
\bibitem[]{} Nakamura F, Umemura N. 2002. {\em Ap. J.} 569:549
\bibitem[]{} Nolte JL, Stancil PC, Lee TG, Balakrishnan N, Forrey RC. 2012. {\em Ap. J.} 744:62 
\bibitem[]{} Norris JE, Christlieb N, Korn AJ, Eriksson K, Bessell MS. 2007. {\em Ap. J.} 670:774
\bibitem[]{} Noterdaeme P, L\'opez S, Dumont V, Ledoux C, Molaro P, Petitjean P. 2012. {\em Astron. Astrophys.} 542:L33
\bibitem[]{} Novikov I, Zel'dovich YB. 1967. {\em Annu. Rev.}{\em Astron. Astrophys.} 5:627
\bibitem[]{} N{\'u}{\~n}ez-L{\'o}pez R, Lipovka A, Avila-Reese V. 2006. {\em MNRAS} 369:2005 
\bibitem[]{} Oh SP, Haiman Z. 2002. {\em Ap. J.} 569:558
\bibitem[]{} Olive K, Skillman ED. 2004. {\em Ap. J.} 617:29
\bibitem[]{} Olive KA, Petitjean P, Vangioni E, Silk J. 2012. {\em MNRAS}, 426:1427
\bibitem[]{} Orel AE. 1987. {\em J. Chem. Phys.} 87:314
\bibitem[]{} Pagel BEJ. 1959. {\em MNRAS} 119:609
\bibitem[]{} Palla F, Salpeter EE, Stahler SW. 1983. {\em Ap. J.} 271:632
\bibitem[]{} Palla F, Galli D, Silk J. 1995. {\em Ap. J.} 451:44 
\bibitem[]{} Palla F, Galli D. 2000. In {\em H$_2$ in Space}, eds. F Combes \& G Pineau Des F\^orets (Cambridge: Cambridge Univ. Press), p. 119
\bibitem[]{} Pavanello M, Bubin S, Molski M, Adamowicz L. 2005. {\em J. Chem. Phys.} 123:104306 
\bibitem[]{} Peart B, Hayton DA. 1992. {\em J. Phys. B} 25:5109
\bibitem[]{} Pedersen HB, Altevogt S, Jordon-Thaden B, Heber O, Lammich L., et al. 2007. {\em Phys. Rev. Lett.} 98:223202 
\bibitem[]{} Peebles PJE. 1968. {\em Ap. J.} 153:1
\bibitem[]{} Peebles PJE. 1993. {\em Principles of Physical Cosmology} (Princeton: Princeton Univ. Press), Chap.~6
\bibitem[]{} Peebles PJE, Dicke RH. 1968. {\em Ap. J.} 154:891
\bibitem[]{} Persson CM, Maoli R, Encrenaz P, Hjalmarson \AA, Olberg M, et al. 2010. {\em Astron. Astrophys.} 515:A72 
\bibitem[]{} Pettini M, Zych BJ, Murphy MT, Lewis A, Steidel CC. 2008. {\em MNRAS} 391:1499
\bibitem[]{} Pettini M, Cooke R. 2012 {\em MNRAS} 425:2477 
\bibitem[]{} Pritchard JR, Loeb A. 2008. {\em Phys. Rev. Lett.} 78:103511
\bibitem[]{} Pritchard JR, Loeb A. 2012. {\em Rep. Progr. Phys.} 75:086901
\bibitem[]{} Purcell EM, Field GB. 1956. {\em Ap. J.} 124:542 
\bibitem[]{} Puy D, Alecian G, Le Bourlot J, Leorat J, Pineau Des For\^ets G. 1993. {\em Astron. Astrophys.} 267:337 
\bibitem[]{} Puy D, Signore M. 2002. {\em New Astron.} Rev. 46:709
\bibitem[]{} Puy D. Signore M. 2007. {\em New Astron.} 51:411
\bibitem[]{} Puy D, Dubrovich V, Lipovka A, Talbi D, Vonlanthen P. 2007. {\em Astron. Astrophys.} 476:685
\bibitem[]{} Rafelsky M, Wolfe AM, Prochaska JX, Neeleman M, Mendez AJ. 2012. {\em Ap. J.} 755:89
\bibitem[]{} Ramaker DE, Peek JM. 1979. {\em J. Chem. Phys.} 71:1844 
\bibitem[]{} Ripamonti E, Mapelli M, Ferrara A. 2007. {\em MNRAS} 375:1399
\bibitem[]{} Roberge W, Dalgarno A. 1982. {\em Ap. J.} 255:489 
\bibitem[]{} Rubi{\~n}o-Mart{\'{\i}}n JA, Chluba J, Fendt WA, Wandelt BD. 2010. {\em MNRAS} 403:439 
\bibitem[]{} Rutherford JA, Vroom DA. 1973. {\em J. Chem. Phys.} 58:4076
\bibitem[]{} Saha S, Datta KK, Barua AK. 1978. {\em J. Phys. B} 11:3349 
\bibitem[]{} Sakimoto K. 1989. {\em Chem. Phys. Lett.} 164:294
\bibitem[]{} Saslaw WC, Zipoy D. 1967. {\em Nature}, 216:976
\bibitem[]{} Sasaki S, Takahara F. 1993. {\em Publ. Astron. Soc. Japan} 45:655 
\bibitem[]{} Savin DW. 2002. {\em Ap. J.} 566:599
\bibitem[]{} Savin DW, Brickhouse NS, Cowan JJ, Drake RP, Federman SR, et al. 2012. {\em Rep. Progr. Phys.} 75:036901 
\bibitem[]{} Savin DW. 2013. In {\it First Stars IV}, eds. M Umemura \& K Omukai, AIP 1840, in press
\bibitem[]{} Savin DW, Krsti\'c PS, Haiman Z, Stancil PC. 2004. {\em Ap. J.} 606:L167 (Erratum 2004. {\em Ap. J.} 607:L147) 
\bibitem[]{} Sbordone L, Bonifacio P, Caffau E, Ludvig HG, Behara NT, et al. 2010. {\em Astron. Astrophys.} 522:A26
\bibitem[]{} Schleicher DRG, Galli D, Palla F, Camenzind M, Klessen RS, et al. 2008. {\em Astron. Astrophys.} 490:521 
\bibitem[]{} Schleicher DRG, Galli D, Glover SCO, Banerjee R, Palla F, et al. 2009. {\em Ap. J.} 703:1096
\bibitem[]{} Schmeltekopf AL, Fehsenfeld FF, Ferguson EE. 1967. {\em Ap. J.} 148:L155
\bibitem[]{} Schneider IF, Dulieu O, Giusti-Suzor A, Roueff E. 1994. {\em Ap. J.} 424:983 (Erratum 1997. {\em Ap. J.} 486:580)
\bibitem[]{} Schneider IF, Suzor-Weiner A. 2002. {\em Contr. to Plasma Physics} 42:578
\bibitem[]{} Schneider R, Omukai K, Limongi M, Ferrara A, Salvaterra R, Chieffi A., Bianchi S. 2012. {\em MNRAS} 423:L60
\bibitem[]{} Scott D, Moss A. 2009. {\em MNRAS} 397:445 
\bibitem[]{} Seager S, Sasselov SS, Scott D. 1999. {\em Ap. J.} 523:L1
\bibitem[]{} Seager S, Sasselov SS, Scott D. 2000. {\em Ap. J.} 128:407
\bibitem[]{} Serpico PD, Esposito S, Iocco F. 2004. {\em J. Cosmol. Astropart. Phys.} 12:10
\bibitem[]{} Shapiro PR, Kang H. 1987. {\em Ap. J.} 318:32 
\bibitem[]{} Shavitt I. 1959. {\em J. Chem. Phys.} 31:1359
\bibitem[]{} Shchekinov YA, Entel MB. 1984. {\em Sov. Astr.} 28:270 (orig. 1984. {\em Astron. Zh.} 61:460)
\bibitem[]{} Silk J. 1977. {\em Ap. J.} 214:718
\bibitem[]{} Smith D, {\v S}panel P. 1993. {\em Int. J. Mass Spectr. Ion. Proc.} 129:163
\bibitem[]{} Smoot GF, Bennett CL, Kogut A., Wright EL, Aymont J, et al. 1992. {\em Ap. J.}, 396:L1 
\bibitem[]{} Sodoga K, Loreau J, Lauvergnat D, Justum Y, Vaeck N, et al. 2009. {\em Phys. Rev. A} 80:033417
\bibitem[]{} Spergel DN, Verde L, Peiris HV, Komatsu E, Nolta MR, et al. 2003. {\em Ap. J. Suppl.} 148:175 
\bibitem[]{} Stacy A, Greif T, Bromm V. 2010. {\em MNRAS} 403:45
\bibitem[]{} Stancil PC, Babb JF, Dalgarno A. 1993. {\em Ap. J.} 414:672
\bibitem[]{} Stancil PC, Lepp S, Dalgarno A. 1996, {\em Ap. J.} 458:401 
\bibitem[]{} Stancil PC, Lepp S, Dalgarno A. 1998. {\em Ap. J.} 509:1 
\bibitem[]{} Stancil PC, Loeb A, Zaldarriaga M, Dalgarno A, Lepp S. 2002. {\em Ap. J.} 580:29 
\bibitem[]{} St\"arck J, Meyer W. 1993. {\em Chem. Phys.} 176:83
\bibitem[]{} Steigman G. 2007. {\em Annu. Rev. Nucl. Part. Sci.} 57:463 
\bibitem[]{} Stenrup M, Larson A, Elander N. 2009. {\em Phys. Rev. A} 79:012713
\bibitem[]{} Sundstr\"om G, Mowat JR, Danared H. 1994. {\em Science} 263:785
\bibitem[]{} Sunyaev RA, Zel'dovich YB. 1970. {\em Ap. Space Sci.} 7:3 
\bibitem[]{} Switzer ER, Hirata CM. 2005. {\em Phys. Rev. D} 72:083002.1 
\bibitem[]{} Switzer ER, Hirata CM. 2008a. {\em Phys. Rev. D} 77:083006
\bibitem[]{} Switzer ER, Hirata CM. 2008b. {\em Phys. Rev. D} 77:083008
\bibitem[]{} Szucs S, Karemera M, Terao M,  Brouillard F. 1984. {\em J. Phys. B} 17:1613
\bibitem[]{} Takagi T. 2002. {\em Phys. Scr.}, 96:52 
\bibitem[]{} Takayanagi N, Nishimura S. 1960. {\em Publ. Astron. Soc. Japan} 12:77
\bibitem[]{} Takeda H, Sato H, Matsuda T. 1969. {\em Progr. Theor. Phys.} 41:840
\bibitem[]{} Talbi D, Bacchus-Montabonel MC. 1998. {\em Chem. Phys.} 232 267
\bibitem[]{} Tegmark M, Silk J, Rees MJ, Blanchard A, Abel T, Palla F. 1997. {\em Ap. J.} 474:1
\bibitem[]{} Turk MJ, Abel T, O'Shea B. 2009. {\em Science} 325:601
\bibitem[]{} Turk MJ, Clark PC, Glover SCO, Greif TH, Abel T, Klessen R, Bromm V. 2011. {\em Ap. J.} 726:55 
\bibitem[]{} Turk MJ, Oishi JS, Abel T, Bryan GL. 2012. {\em Ap. J.} 745:154
\bibitem[]{} Umemura M, Omukai K. 2013. {\em First Stars IV}, AIP Conf. Proc. 1480
\bibitem[]{} Urbain X. 2010. {\em APS Div. At. Mol. Opt. Phys. Meeting Abstracts}, p. 1003
\bibitem[]{} Varshalovich DA, Khersonskii VK. 1977. {\em Sov. Astr. Lett.}. 3:155 (orig. 1977. {\em Pis'ma Astron. Zh.} 3:291)
\bibitem[]{} Vonlanthen P, Rauscher T, Winteler C, Puy D, Signore M, Dubrovich V. 2009. {\em Astron. Astrophys.} 503:47 
\bibitem[]{} Wharton L, Gold LP, Klemperer W. 1960. {\em J. Chem. Phys.} 33:1255 
\bibitem[]{} Watson WD. 1976. {\em Rev. Mod. Phys.} 48:513
\bibitem[]{} Watson WD, Christensen RB, Deissler RJ. 1978. {\em Astron. Astrophys.} 65:159
\bibitem[]{} Wolfe AM, Gawiser E, Prochaska JX. 2005. {\em Annu. Rev. Astron. Astrophys.} 43:861
\bibitem[]{} Wong WY, Seager S, Scott D. 2006. {\em MNRAS} 367:1666 
\bibitem[]{} Wong WY, Moss A, Scott D. 2008. {\em MNRAS} 386:1023 
\bibitem[]{} Wrathmall SA, Gusdorf A, Flower DR. 2007. {\em MNRAS} 382:133 
\bibitem[]{} Yoneyama T. 1970. {\em Publ. Astron. Soc. Japan} 24:87
\bibitem[]{} Yong D, Norris JE, Bessell MS, Christlieb N, Asplund M, et al. 2012. {\em arXiv1208.3016v1}
\bibitem[]{} Yoon JS, Kim YW, Kwon DC, Song MY, Chang WS, et al. 2010. {\em Rep. Progr. Phys.} 73:116401 
\bibitem[]{} Yoshida N, Omukai K, Hernquist L, Abel T. 2006. {\em Ap. J.} 652:6
\bibitem[]{} Yoshida N, Omukai K, Hernquist L. 2007. {\em Ap. J.} 667:L117
\bibitem[]{} Zel'dovich YB, Kurt VG, Sunyaev RA. 1969. {\em Soviet Phys.-JETP Lett.} 28:146 (orig. 1968. {\em Zh. Eksp. Theor. Fiz.} 55:278)
\bibitem[]{} Zel'dovich YB. 1978. {\em Sov. Astr. Lett.} 4:88 (orig. 1978. {\em Pis'ma Astron. Zh.} 4:165)
\bibitem[]{} Zemke WT, Stwalley WC. 1980. {\em J. Chem. Phys.} 73:5584
\bibitem[]{} Zinchenko I, Dubrovich V, Henkel C. 2011. {\em MNRAS} 415:L78 
\bibitem[]{} Zygelman B, Stancil PC, Dalgarno, A. 1998. {\em Ap. J.} 508:151 
\end{thebibliography}
\end{document}